\documentclass[11pt,english]{article}
\usepackage{lmodern}
\usepackage{helvet}
\usepackage{courier}

\usepackage[T1]{fontenc}
\usepackage[latin9]{inputenc}
\usepackage{geometry}
\usepackage{color}
\usepackage{babel}
\usepackage{array}
\usepackage{longtable}
\usepackage{float}
\usepackage{booktabs}
\usepackage{units}
\usepackage{textcomp}
\usepackage{multirow}
\usepackage{amsmath}
\usepackage{amssymb}
\usepackage{graphicx}
\usepackage{setspace}
\usepackage{esint}
\usepackage[unicode=true,pdfusetitle,
 bookmarks=true,bookmarksnumbered=false,bookmarksopen=false,
 breaklinks=false,pdfborder={0 0 0},pdfborderstyle={},backref=false,colorlinks=true]
 {hyperref}

\makeatletter

\providecommand{\tabularnewline}{\\}

\newcommand{\lyxaddress}[1]{
	\par {\raggedright #1
	\vspace{1.4em}
	\noindent\par}
}

\usepackage{setspace}
\usepackage{etoolbox}
\patchcmd{\env@cases}{1.2}{1.8}{}{}

\makeatother

\begin{document}
\title{Description of seismic sources in underground mines:\\
 Dynamic stress fracturing around tunnels and strainbursting}
\author{Dmitriy Malovichko \& Alex Rigby}
\maketitle

\lyxaddress{Institute of Mine Seismology, 10 Church Street, Kingston, 7050, Australia}
\begin{abstract}
This paper considers the dynamic fracturing of the rockmass surrounding
a tunnel statically loaded by compressional stress as a possible source
of seismic events in underground mines. This begins with two-dimensional
dynamic modelling of failure for six plausible scenarios that span
various loadings, tunnel profiles, rockmass parameters, and methods
of event initiation. In each case, the seismic source derived from
these models has significant negative isotropic (implosive) and negative
compensated linear vector dipole (pancake-shape) components as well
as a P-axis that is approximately aligned with the direction of maximum
compressional principal stress. These features indicate that at wavelengths
larger than the diameter of the tunnel and the extent of damage along
it, seismic radiation is controlled by the elastic convergence of
the surrounding rockmass rather than by rock fracturing. To aid in
the analysis of such events, an analytical approximation of the source
mechanism is suggested {[}Equation (\ref{eq:Definition=0000233d}){]}
that is based solely on mechanical and geometric properties: the magnitudes
$\sigma_{\mathrm{max}}$ and $\sigma_{\mathrm{min}}$ of the maximum
and minimum principal stresses orthogonal to the tunnel's axis, the
Poisson's ratio $\nu$ of the rockmass, the length $L_{3}$ of dynamic
fracturing along the tunnel, the effective tunnel dimension $\overline{L_{A}}$,
and the increase in depth of failure $\triangle d_{f}^{A}$ in the
direction of $\sigma_{\mathrm{min}}$. Furthermore, it is shown that
the scalar seismic moment can be approximated as $\left|\mathbf{M}\right|\approx2[(1-\nu)/(1-2\nu)]\left|\sigma_{max}\right|L_{3}\overline{L_{A}}\triangle d_{f}^{A}$.
The suggested approximations are considered in the context of seismic
data from a real underground mine. It is shown that many mechanisms
inverted from observed waveforms are consistent with the suggested
model. Furthermore, it is demonstrated that the proposed source mechanism
approximation can be used for the forensic analysis of damaging seismic
events and quantitative monitoring of the evolution of fractured zones
around tunnels.
\end{abstract}

\section{Introduction\label{sec:Introduction}}

In mines, episodes of sudden inelastic deformation are often induced
or triggered by the excavation of rocks, with examples including slip
along a fault adjacent to a mined-out stope or the failure of a pillar
between two tunnels. As discussed in \cite{Malovichko-2020}, these
nearby excavations can have a significant effect on the radiation
of seismic waves and should be taken into account in the elastodynamic
Green's function adopted in the modelling or inversion of waveforms.
An alternative approach is to consider the excavations as part of
the seismic source. Expressions describing such point sources are
suggested in \cite{Malovichko-2020} and are appropriate when seismic
wavelengths exceed the combined size of the volume of sudden inelastic
deformation and nearby excavations.

\begin{figure}[H]
\begin{centering}
\includegraphics[width=0.8\textwidth]{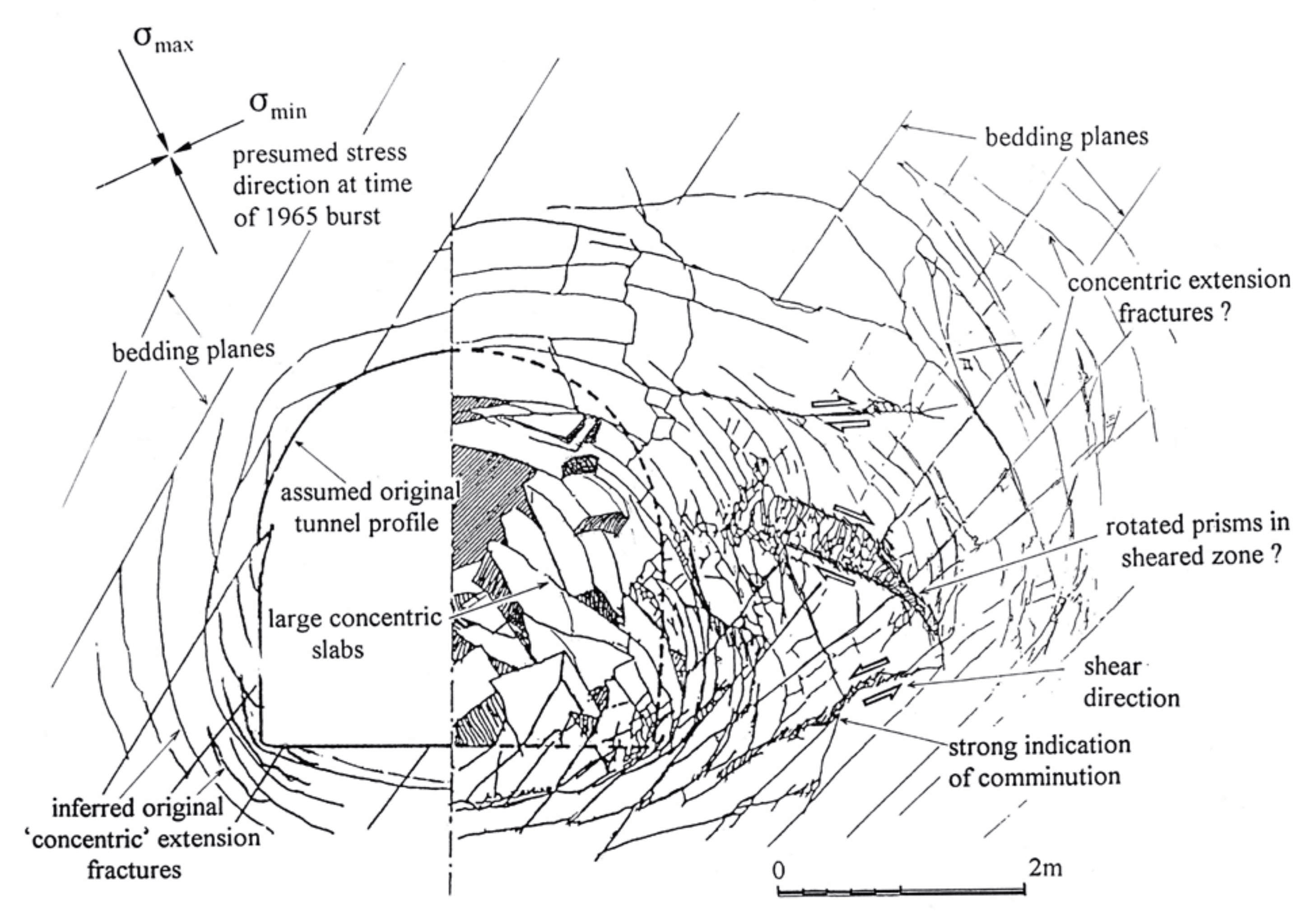}
\par\end{centering}
\caption{\label{fig:Ortlepp-sketch}Interpretation of fracturing around a rockburst-damaged
tunnel in a deep gold mine in South Africa \cite{Ortlepp-1997}. The
left half of the sketch describes conditions expected to exist before
the rockburst, and the right half was drawn based on a photograph
of damage.}
\end{figure}

The focus of this paper is the application of these expressions to
a particular type of dynamic process in underground mines: violent
stress fracturing around tunnels. If this process results in observable
damage to the excavation (as is shown in Figure \ref{fig:Ortlepp-sketch}),
then it is referred to as a strainburst. Strainbursts are of great
concern in the mining industry as they compromise safety and disrupt
production plans. Typically, instances of violent damage to excavations
are thoroughly investigated to identify causes and contributing factors.
Such forensic investigations can be aided by the use of seismic data,
whether it be the waveforms of the seismic event associated with the
strainburst or a catalog of seismic events recorded around the strainburst's
location. Note that dynamic stress fracturing around a tunnel need
not be accompanied by damage to excavations: the ground support system
(rock bolts, mesh, shotcrete, etc.) may accommodate the deformation
of fractured rock and prevent damage. Even in such cases, the fracturing
process can be intensive and fast enough to radiate detectable seismic
waves, and the analysis of these signals is beneficial from the perspective
of understanding the deformation around tunnels and the assessment
of consumption of ground support capacity.

As is shown in Figure \ref{fig:Ortlepp-sketch}, the overall process
of dynamic stress fracturing may be decomposed into smaller-scale
episodes of shear or tensile rupturing of the rockmass, the seismic
radiation of which can be approximately described by conventional
double-couple or tensile-crack point sources (although modelling the
medium as unbounded space may not be appropriate \cite{Sileny-2001}).
Given the potentially sub-meter length scales of these constituent
processes, observation of the radiated waves would require a dense
seismic array consisting of acoustic emission sensors capable of recording
in the multi-$\unit{kHz}$ frequency range. However, typical seismic
systems in underground mines have sparse arrays of sensors adequate
for the recording of signals with wavelengths on the order of tens
or hundreds of meters. Such waves do not provide enough resolution
to distinguish the details of fracturing as shown in Figure \ref{fig:Ortlepp-sketch},
and only the overall pattern of fracturing can be inferred. In such
a case, it is convenient to consider the tunnel as part of the seismic
source, which makes it possible to utilize a simple elastodynamic
Green's function for homogeneous, isotropic space that does not take
the presence of the tunnel into account.

To the best knowledge of the authors, there are no established quantitative
point-source models describing the low-frequency content of seismic
radiation produced by strainbursts (or dynamic stress fracturing around
excavations in general). However, some characteristics of such sources
have been postulated. For example, \cite{Ortlepp-1997} states that
strainbursts have an implosive seismic signature (first motion towards
the source from the seismic records) and a local magnitude ranging
from $-0.2$ to $0.0$. In \cite{McGarr-1992b}, an example of a seismic
event is considered whose source presumably involves the closure of
an excavation (stope). It is suggested that the isotropic component
of the moment tensor $M_{ij}$ for such events can be interpreted
in terms of a coseismic volumetric change $\triangle V=\mathrm{tr}(M_{ij})/(3\lambda+2\mu)$,
where $\lambda$ and $\mu$ are the Lamé moduli for the rockmass.
This interpretation is based on Equation (3.35) of \cite{Aki-Richards-2009},
which describes the transformational expansion (or contraction) of
a spherical volume in an isotropic medium.

Seismic sources associated with inelastic deformation around tunnels
were briefly considered in a previous work \cite{Malovichko-2020},
where two simple cases were analyzed. In the first case, a circular
tunnel in an elastic-brittle-plastic Mohr-Coulomb material was loaded
hydrostatically. Analytical solutions for stress and displacement
\cite{Fritz-1984} were then used to evaluate a corresponding seismic
source mechanism that took the effect of the tunnel into account.
This mechanism had a significant isotropic implosive component and
a deviatoric component in the form of a negative (largest dipole compressional)
compensated linear vector dipole (CLVD), which is hereafter referred
to as a ``pancake-shape'' CLVD. The second case considered was the
nonhydrostatic loading of a rectangular tunnel. In the absence of
analytical expressions for stress and displacement, a finite-difference
analysis was performed. This involved modelling deformation of limited
extent along the tunnel's axis on two of its sides. Again, the corresponding
seismic source mechanism, which accounted for the tunnel, consisted
mainly of implosive and pancake-shape CLVD components.

In this paper, we first extend these previous results in Section \ref{sec:Modelling},
where six cases of nonhydrostatically loaded two-dimensional tunnels
are considered that differ in loading, tunnel profile, rockmass parameters,
and method of event initiation. For each of these cases, numerical
modelling of deformation around the tunnel is performed. Using these
modelling results, two different approaches to calculating seismic
mechanisms are outlined, and the main features of these mechanisms
are discussed. While such modelling can be a useful instrument in
the forensic analysis of strainbursts, such an approach may not always
be computationally feasible. To address this limitation, Section \ref{sec:Approximation}
suggests an analytical approximation that provides a quantitative
relation between the geometric and mechanical characteristics of the
fracturing process around a tunnel (tunnel dimension, properties of
loading, etc.) and the parameters of an equivalent seismic point source
(mechanism and scalar moment). This approximate model is verified
against the results of Section \ref{sec:Modelling}, with a good agreement
between the two approaches being demonstrated. Section \ref{sec:Examples}
presents examples of real data that can be interpreted in terms of
the discussed source model and its suggested approximation. The paper
is concluded in Section \ref{sec:Discussion-and-Conclusions}.

\section{\label{sec:Modelling}Numerical modelling of sources}

\subsection{Conceptual overview\label{subsec:Conceptual-model}}

\begin{figure}
\begin{centering}
\includegraphics[width=0.45\textwidth]{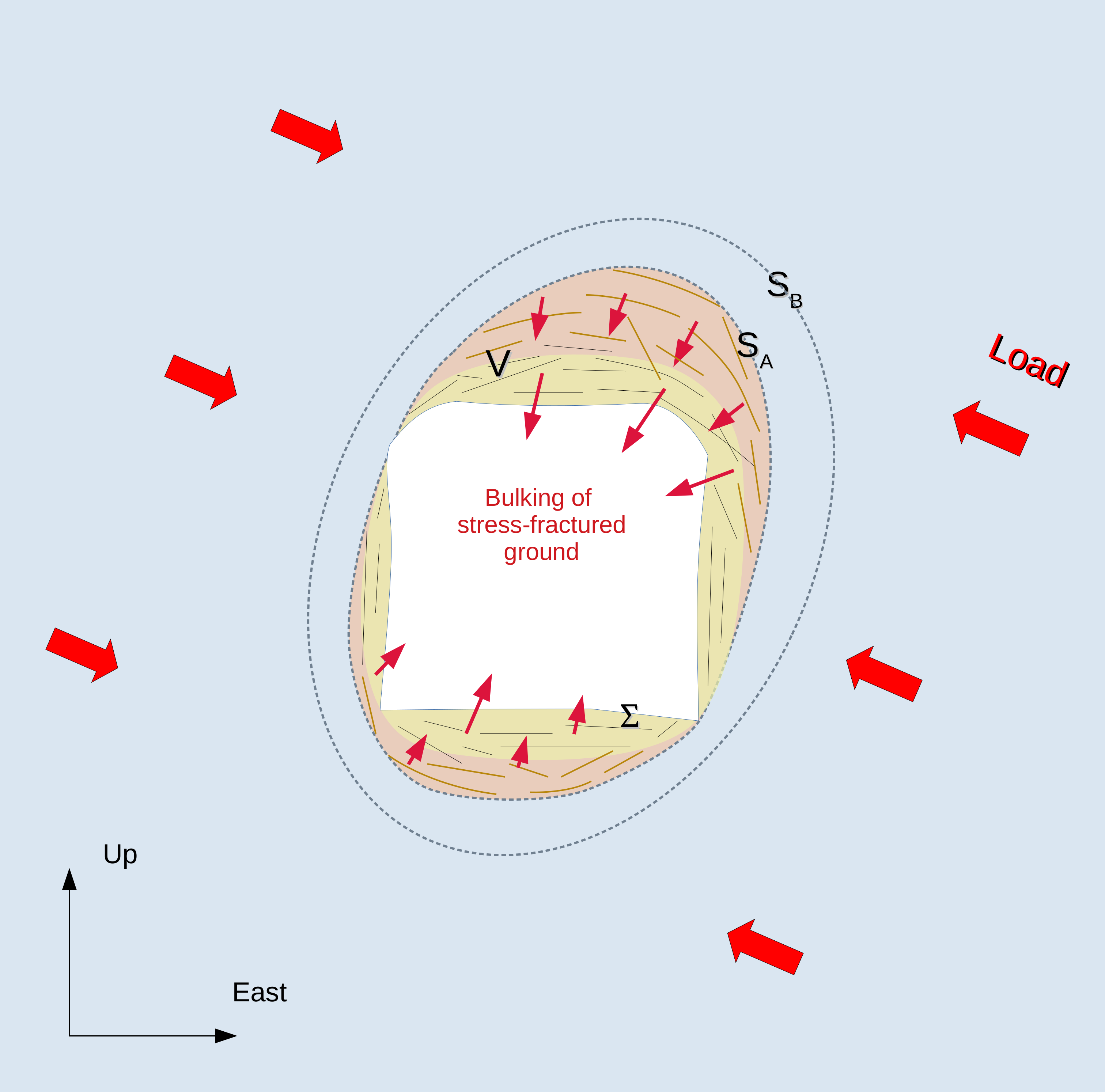}
\par\end{centering}
\caption{\label{fig:Conceptual-model}A conceptual model of fracturing around
a tunnel. It is assumed that the growth of damage zone from yellow
to red happens suddenly (within a fraction of second). Several geometrical
characteristics used in the calculation of seismic source parameters
are marked: the volume $V$ containing damaged rock, the tunnel surface
$\Sigma$, the external boundary $S_{A}$ of damaged rock, and a surface
$S_{B}$ enclosing both the tunnel and the damaged region.}
\end{figure}

In this section and Section \ref{sec:Approximation}, we consider
horizontal tunnels, like the one shown in Figure \ref{fig:Conceptual-model},
that are located in a homogeneous rockmass with no other nearby excavations.
For notational simplicity, we take these tunnels to be north-south
oriented. Nonhydrostatic loading is considered, with maximum compressional
stress orthogonal to the tunnel being either horizontal, vertical,
or plunging to the east (as is depicted in Figure \ref{fig:Conceptual-model}).
In the latter case, such loading facilitates rockmass damage in the
eastern shoulder (upper corner) and bottom western corner of the tunnel
as shown in yellow in Figure \ref{fig:Conceptual-model}.

Suppose that some process (for example, a slow increase in loading,
the stress wave from a distant blast, etc.) triggers dynamic fracturing,
which can be described as an expansion of the damaged zone (shown
in red in Figure \ref{fig:Conceptual-model}). This will typically
be accompanied by a bulking of the rockmass (increase of volume due
to geometrical inconsistencies of rock fragments) and its overall
movement into the excavation. If the described processes of fracturing
and convergence occur suddenly (within a fraction of a second), then
this can result in the radiation of seismic waves through the surrounding
elastic zone. These waves can be strong enough to be detected by seismic
sensors that are typically used for the monitoring of underground
mines. As such, the considered process constitutes the source of a
seismic event.

The sudden unidirectional deformation of the rockmass moving into
the excavation contributes to seismic radiation and can be described
by a single-force point source \cite{Takei-Kumazawa-1994}. While
this contribution to total radiation is expected to be minor, proper
quantification is required and deserves a separate study. The major
portion of radiation is anticipated to come from the next term in
the polynomial decomposition of equivalent forces: the seismic moment
tensor (or, more correctly, the stress glut tensor) \cite{Backus-Mulcahy-1976a}.
This tensor's components can be calculated using either the Kirchhoff-type
or adjusted conventional expressions suggested in \cite{Malovichko-2020}:

\begin{singlespace}
\begin{equation}
M_{ij}=M_{ij}^{T}+M_{ij}^{U}=\iintop_{S}\Delta T_{i}(\mathbf{\mathbf{x}})(x_{j}-x_{j}^{(0)})\:dS(\mathbf{x})-\iintop_{S}c_{ijkl}(\mathbf{x})\Delta u_{k}(\mathbf{x})n_{l}(\mathbf{x})\:dS(\mathbf{x}),\label{eq:kirchhoff}
\end{equation}

\begin{equation}
M_{ij}=M_{ij}^{e}+M_{ij}^{u}=\iiintop_{V}c_{ijkl}(\mathbf{x})\triangle e_{kl}(\mathbf{\mathbf{x}})\:dV(\mathbf{x})-\iintop_{\Sigma}c_{ijkl}(\mathbf{x})\Delta u_{k}(\mathbf{x})n_{l}(\mathbf{x})\:d\Sigma(\mathbf{x}),\label{eq:adjusted-conventional}
\end{equation}
where $\mathbf{\mathbf{\triangle T}=\mathbf{T}^{\mathrm{(after)}}-\mathbf{T}^{\mathrm{(before)}}}$
and $\mathbf{\triangle u}=\mathbf{u}^{\mathrm{(after})}-\mathbf{u}^{\mathrm{(before})}$
are the differences in traction and displacement, respectively, before
and after expansion of the damaged region to volume $V$, $\mathbf{\Delta e}$
is the stress-free strain tensor, $S$ is a surface in the elastic
region enclosing the damage, $\varSigma$ is the tunnel's surface,
$\mathbf{x}^{(0)}$ is an arbitrary point, $\mathbf{n}$ is an inward
unit normal to $\varSigma$ or $S$, and $c_{ijkl}$ is the stiffness
tensor. For an isotropic medium, $c_{ijkl}=\lambda\delta_{ij}\delta_{kl}+\mu(\delta_{ik}\delta_{jl}+\delta_{il}\delta_{jk})$,
where $\delta_{ij}$ is the Kronecker delta function. The stress-free
strain is mathematically equivalent to the increment of plastic strain
$\mathbf{\Delta}\boldsymbol{\varepsilon}^{p}$ according to the theory
of plastic flow \cite{Kachanov-2004}. $\varSigma$, $V$, and two
possible selections of the surface $S$ are shown in Figure \ref{fig:Conceptual-model}.
As discussed in \cite{Malovichko-2020}, the dimension of $S$ must
be kept smaller than the wavelengths of interest, meaning that it
must cut through the tunnel to the north and south of the extent of
damage in this case (this is valid provided the excluded parts of
the excavation experience small displacements during the period of
inelastic deformation of $V$).
\end{singlespace}

Each of the stated expressions is split naturally into two terms:
the first into traction $\mathbf{M^{T}}$ and displacement $\mathbf{M^{U}}$
components, and the second into strain $\mathbf{M^{e}}$ and displacement
$\mathbf{M^{u}}$ components. The contribution of these terms to the
total mechanism will be compared using the conventional norm $\left|\mathbf{M}\right|=\sum_{ij}\sqrt{M_{ij}^{2}/2}$
defining a scalar moment \cite{Silver-Jordan-1982}. Note that while
$\left|\mathbf{M^{e}}+\mathbf{M^{U}}\right|\leq\left|\mathbf{M^{e}}\right|+\left|\mathbf{M^{U}}\right|$
and $\left|\mathbf{M^{e}}+\mathbf{M^{u}}\right|\leq\left|\mathbf{M^{e}}\right|+\left|\mathbf{M^{u}}\right|$,
equality need not hold in either case.

\subsection{Cases\label{subsec:Cases}}

\begin{figure}[H]
\centering{}\includegraphics[width=0.3\textwidth]{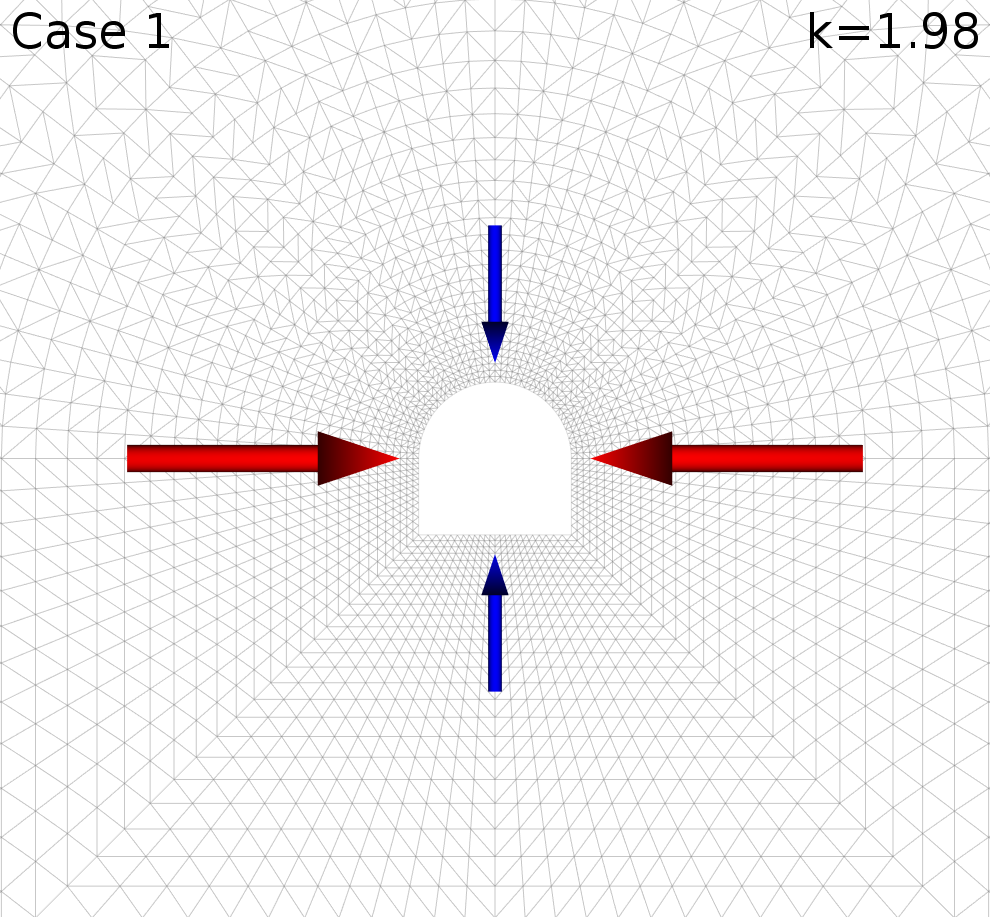}\hspace{0.02\textwidth}\includegraphics[width=0.3\textwidth]{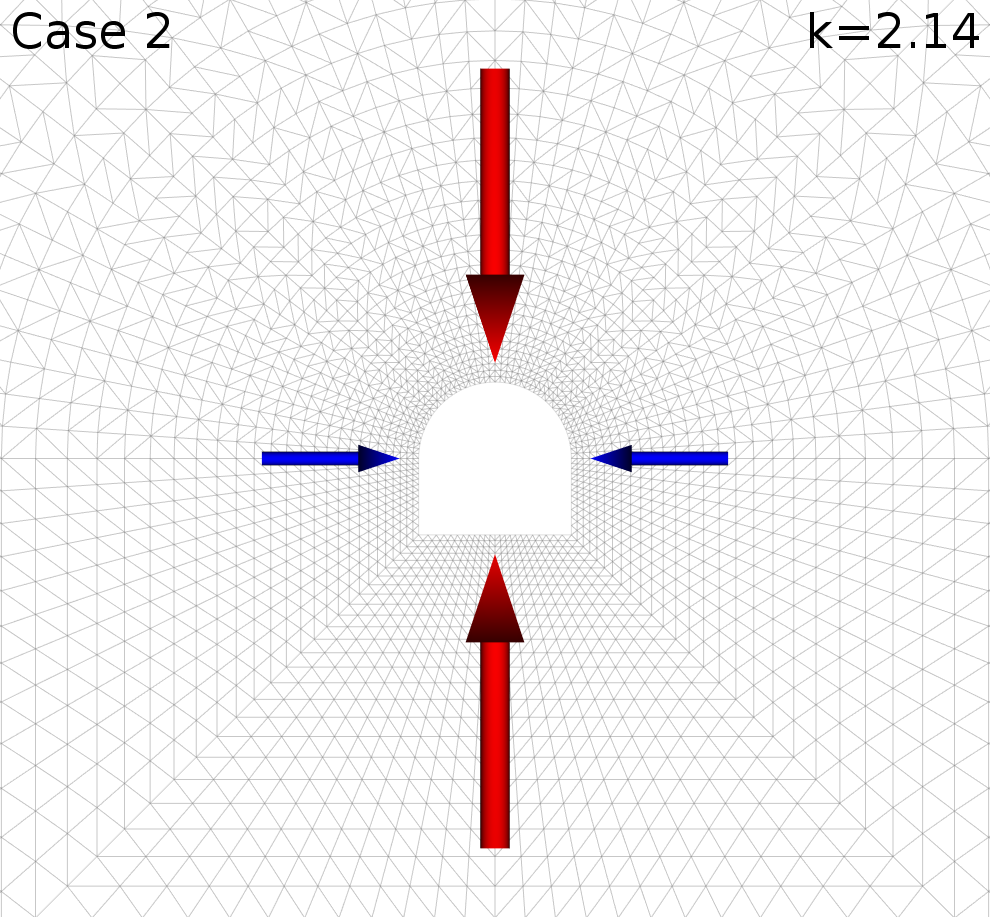}\hspace{0.02\textwidth}\includegraphics[width=0.3\textwidth]{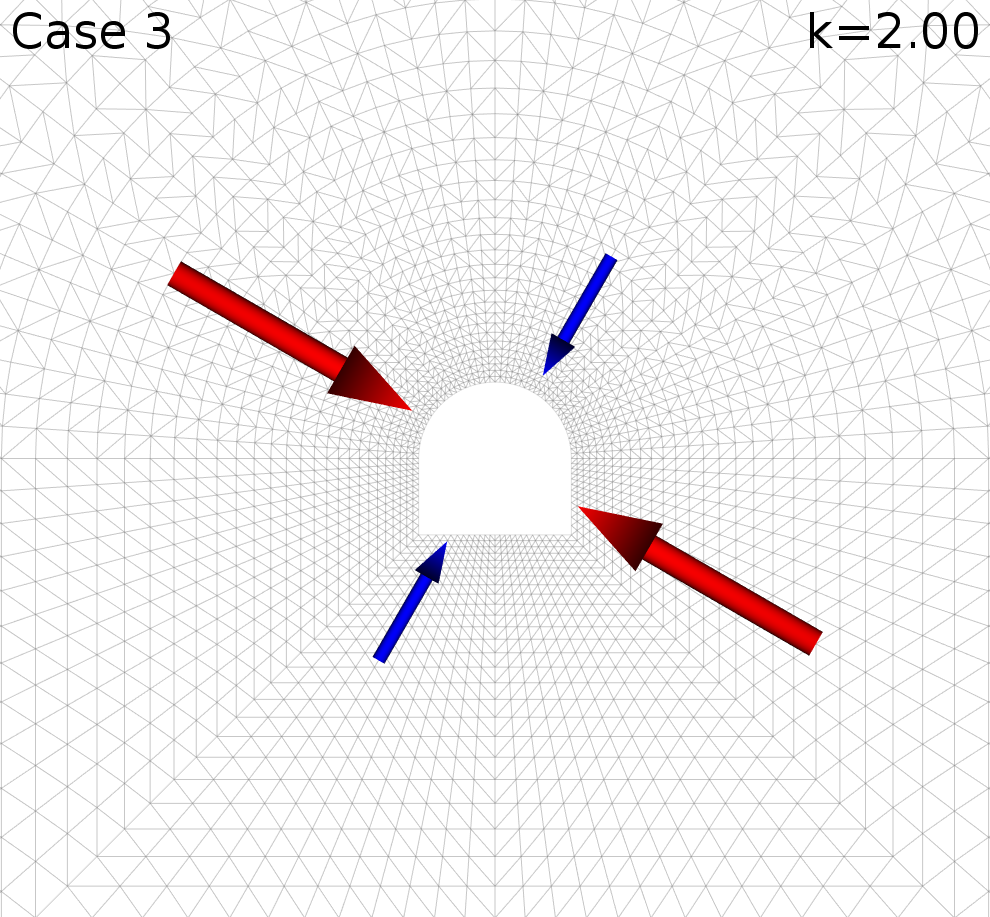}\vspace{0.02\textwidth}
\includegraphics[width=0.3\textwidth]{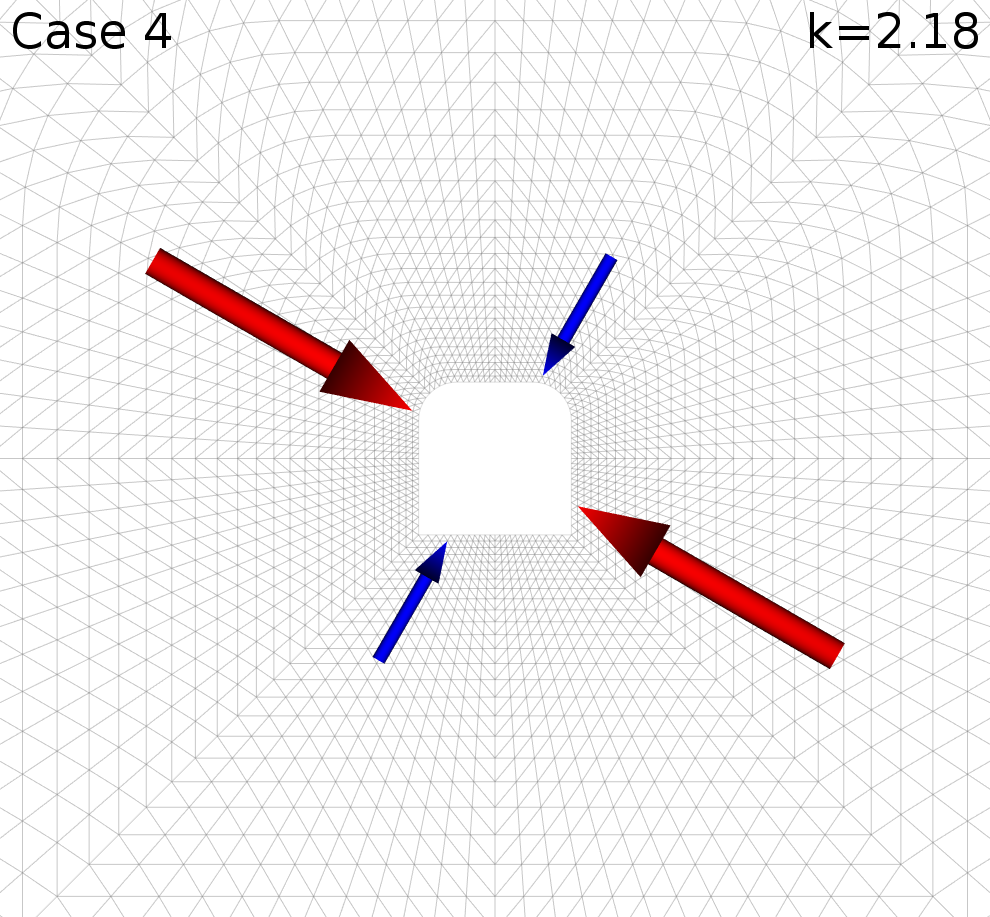}\hspace{0.02\textwidth}\includegraphics[width=0.3\textwidth]{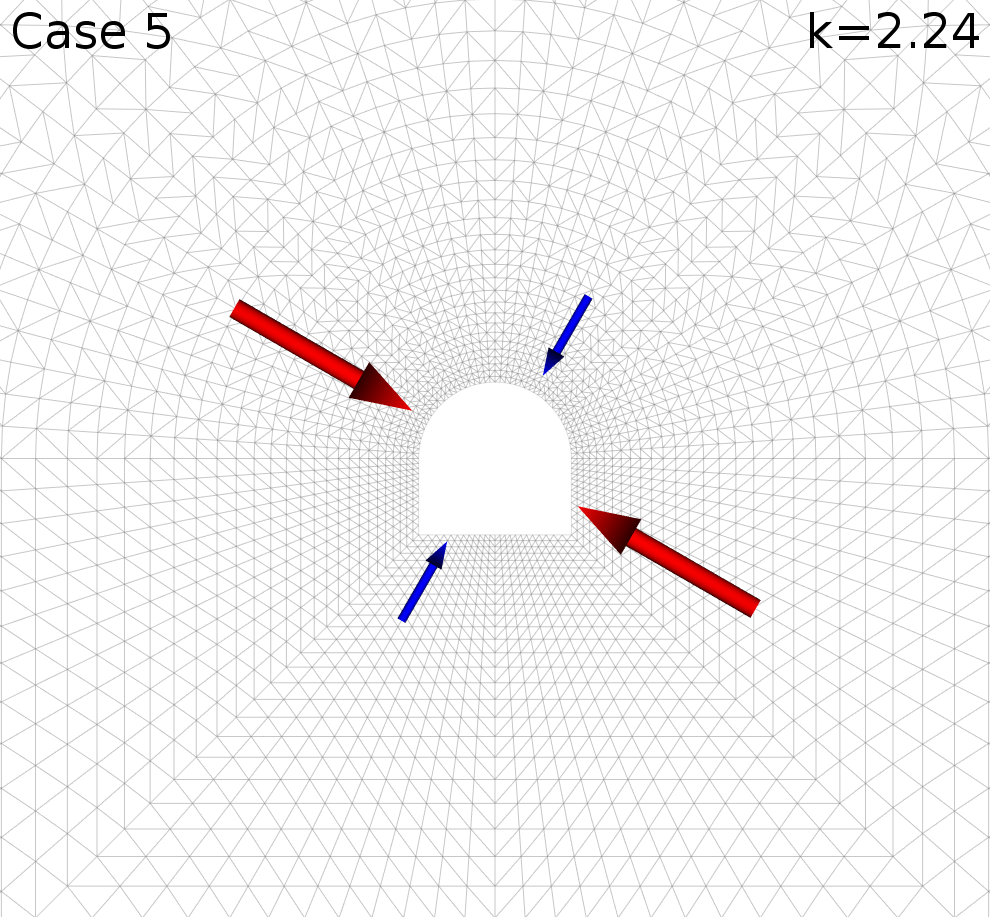}\hspace{0.02\textwidth}\includegraphics[width=0.3\textwidth]{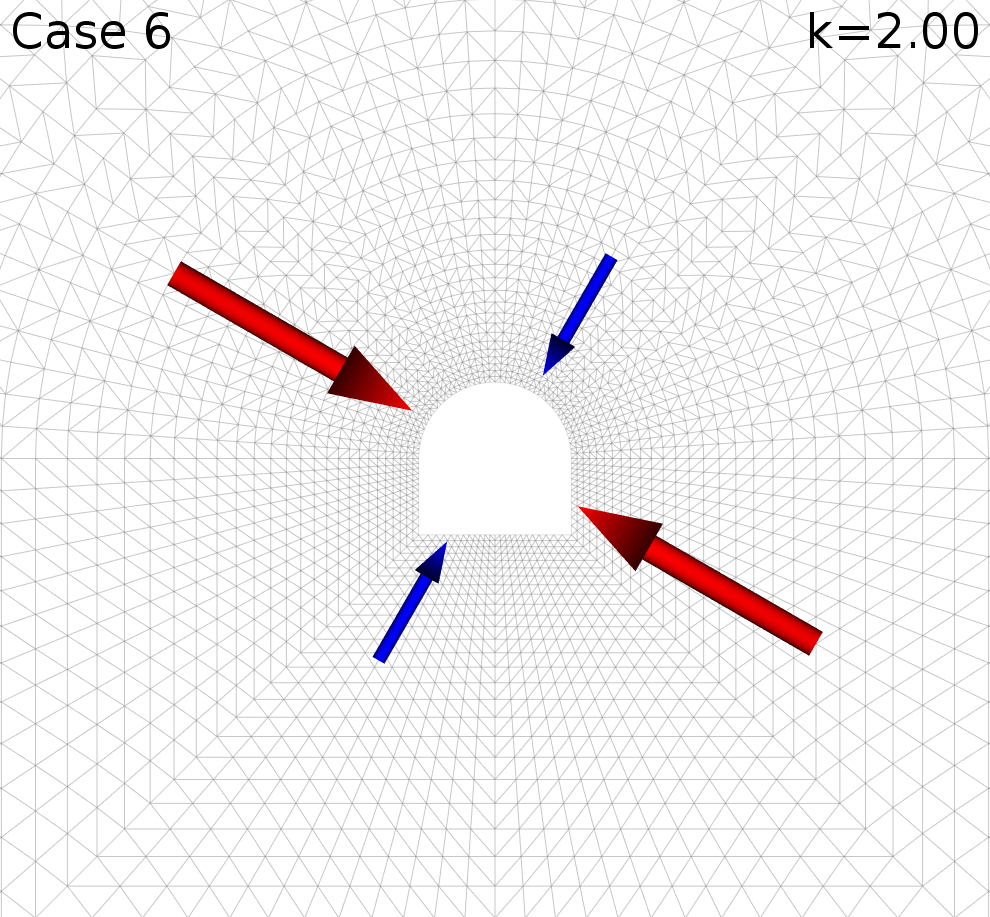}\caption{\label{fig:case-summary}Graphical summary of the six cases outlined
in Subsection \ref{subsec:Cases}. The arrows show the orientation
of the maximum (red) and minimum (blue) in-plane loading, the ratio
of which is listed in the top right.}
\end{figure}

As depicted in Figure \ref{fig:case-summary}, six cases based on
the general setup described in Subsection \ref{subsec:Conceptual-model}
are considered in this paper. To simplify analysis, we approximate
the dynamic stress fracturing along a finite extent of a tunnel by
considering a finite-length ($L_{3}=\unit[5]{m}$) slice of an infinite-length
tunnel (oriented north-south), which reduces the problem to two dimensions
under the assumption of plane strain. The tunnel cross section has
both a height and width of $\unit[5]{m}$, square bottom corners and
rounded shoulders; for Cases 1-4 and 6 the radius of curvature of
this rounding is $R=\unit[2.5]{m}$ (that is, the top half of the
tunnel is circular), while for Case 5, we take $R=\unit[1.25]{m}$
(resulting in a geometry similar to that shown in Figure \ref{fig:Conceptual-model}).

As noted in Subsection \ref{subsec:Conceptual-model}, nonhydrostatic
loading is considered. The minor in-plane loading is $\sigma_{\mathrm{min}}=-\unit[20]{MPa}$
for Case 5 and $\unit[-30]{MPa}$ for the remaining cases. Maximum
in-plane loading is $\sigma_{\max}=k\sigma_{\min}$, where the stress
ratio $k$ for each case can be found in Figure \ref{fig:case-summary}.
Major loading is horizontal for Case 1, vertical for Case 2, and plunging
$30^{\circ}$ to the east for the remaining cases.

A cohesion-weakening-friction-strengthening (CWFS) material \cite{HAJIABDOLMAJID2002731,diederichs20072003}
is used for all cases. Details of implementation and verification
of this constitutive model are outlined in Appendix \ref{sec:Appendix-1:-Details}.
In all cases, the Lamé moduli defining the material's elastic behavior
are $\mu=\lambda=\unit[30]{GPa}$ (giving a Poisson's ratio of $\nu=0.25$).
For Cases 1-4 and 6, the material has friction angle $\phi=25^{\circ}$,
cohesion $\unit[c=50]{MPa}$, residual friction angle $\phi_{r}=50^{\circ}$,
residual cohesion $\unit[c_{r}=2]{MPa}$, dilation angle $\psi=20^{\circ}$,
tensile strength $\unit[\sigma_{t}=8]{MPa}$, and a deviatoric plastic
strain limit $\varepsilon_{c}^{p}=\varepsilon_{\phi}^{p}=3\times10^{-3}$.
For Case 5, a weaker material is used with $\phi=15^{\circ}$, $\unit[c=30]{MPa}$,
$\phi_{r}=50^{\circ}$, $\unit[c_{r}=2]{MPa}$, $\unit[\sigma_{t}=5]{MPa}$,
and $\varepsilon_{c}^{p}=\varepsilon_{\phi}^{p}=3\times10^{-3}$.

After reaching equilibrium with the loading and material properties
as described above, expansion of the failure region is triggered by
slowly increasing the stress ratio by $\Delta k=0.01$ for Cases 1-5.
This serves to replicate slow loading of the tunnel by nearby mining.
For Case 6, expansion is induced by dynamically loading the tunnel
with a stress wave. In particular, a half-sine pulse of force density
directed vertically is applied in a $\unit[30\times5]{m}$ region
centered $\unit[20]{m}$ beneath the tunnel with a period of $\unit[0.01]{s}$
and amplitude of $\unit[10^{6}]{N\cdot m^{-3}}$ (this results in
a peak ground velocity of approximately $\unit[0.2]{ms^{-1}}$).

\subsection{\label{subsec:Results}Results}

Dynamic simulation of the outlined cases was performed using the Material
Point Method (MPM). The MPM implementation used \cite{Basson-mpm}
is based on the formulations of \cite{Nairn-2003,Jassim-2013}. Verification
across a number of relevant cases for which analytical solutions are
available has been undertaken and is documented in the supplemental
material. In an MPM simulation, the domain of interest is discretized
as Lagrangian material points, or particles, that interact with each
other via the nodes of a Eulerian background grid (this allows for
convenient handling of large deformations without remeshing). Given
the plane strain assumption, this grid is based on a thin plate that
is meshed by a single layer of tetrahedra. Details on the construction
of this mesh, the initial and boundary conditions employed, and other
aspects of the simulation process can be found in Appendix \ref{sec:Appendix-1:-Details}.
Animations showing the dynamics of the expansion of the failed region
are included in the supplemental material.

\begin{figure}
\begin{centering}
\includegraphics[width=0.3\textwidth]{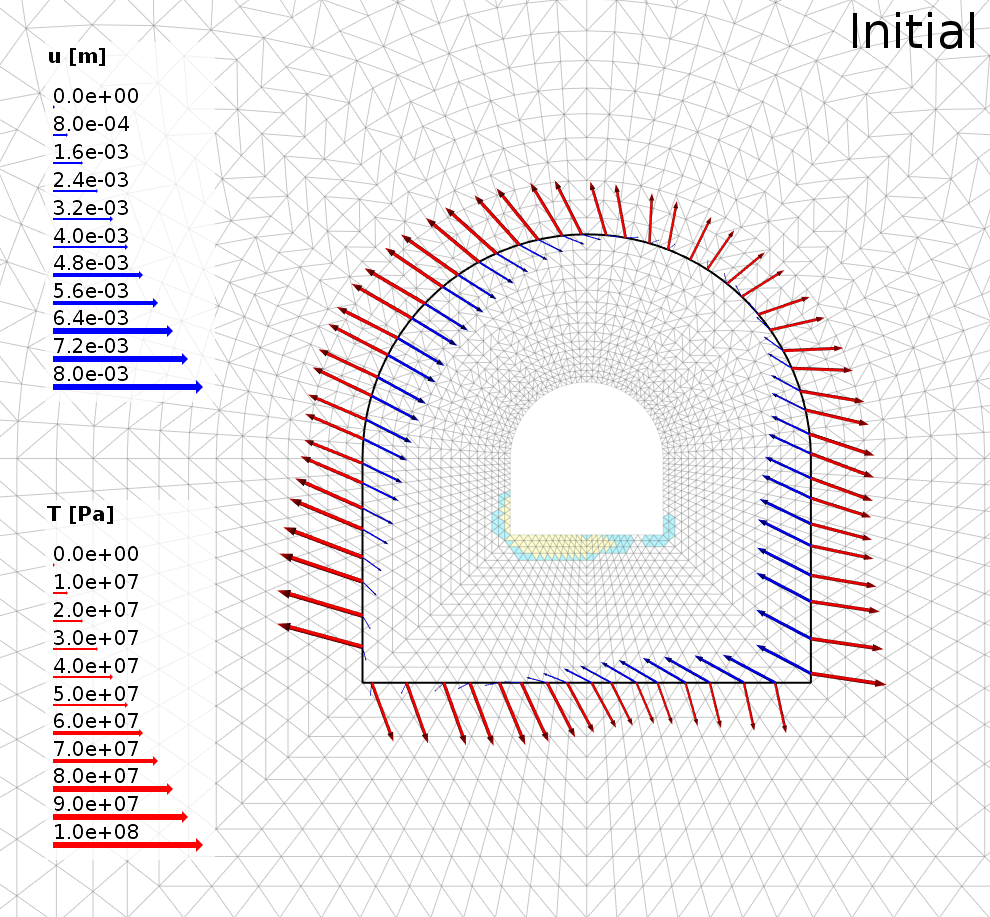}\hspace{0.02\textwidth}\includegraphics[width=0.3\textwidth]{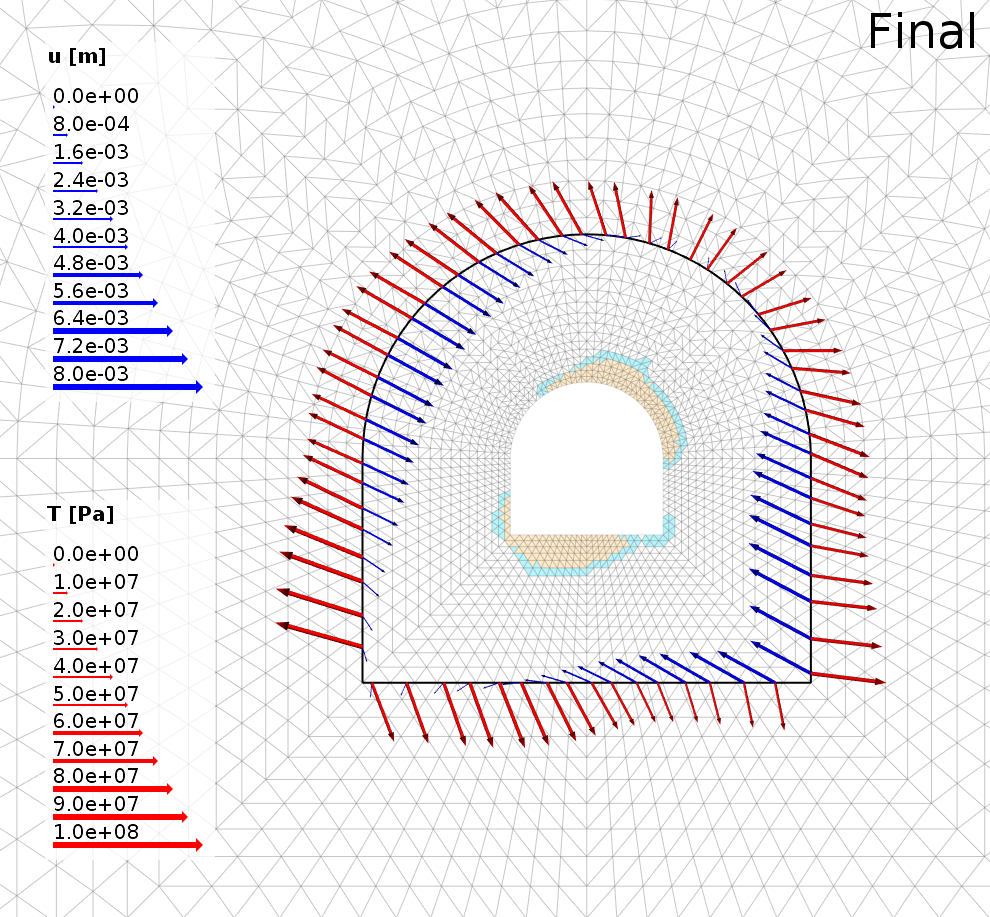}\hspace{0.02\textwidth}\includegraphics[width=0.3\textwidth]{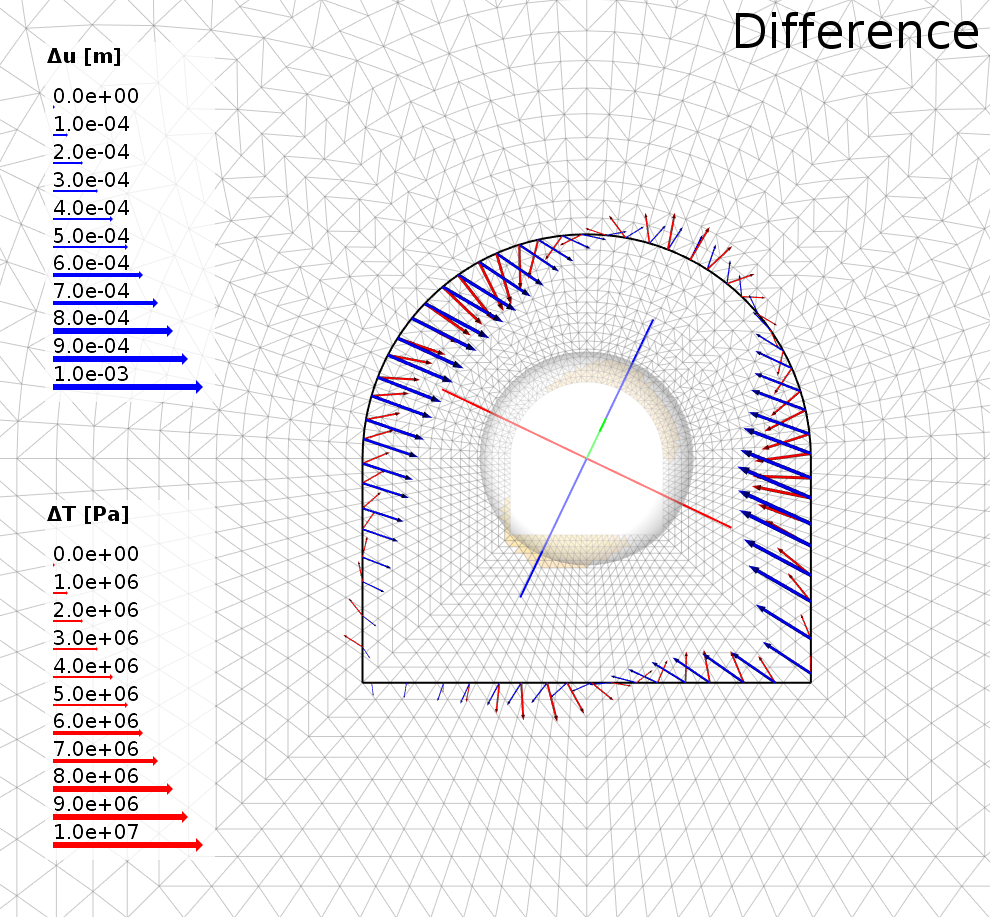}
\par\end{centering}
\begin{centering}
\includegraphics[width=0.3\textwidth]{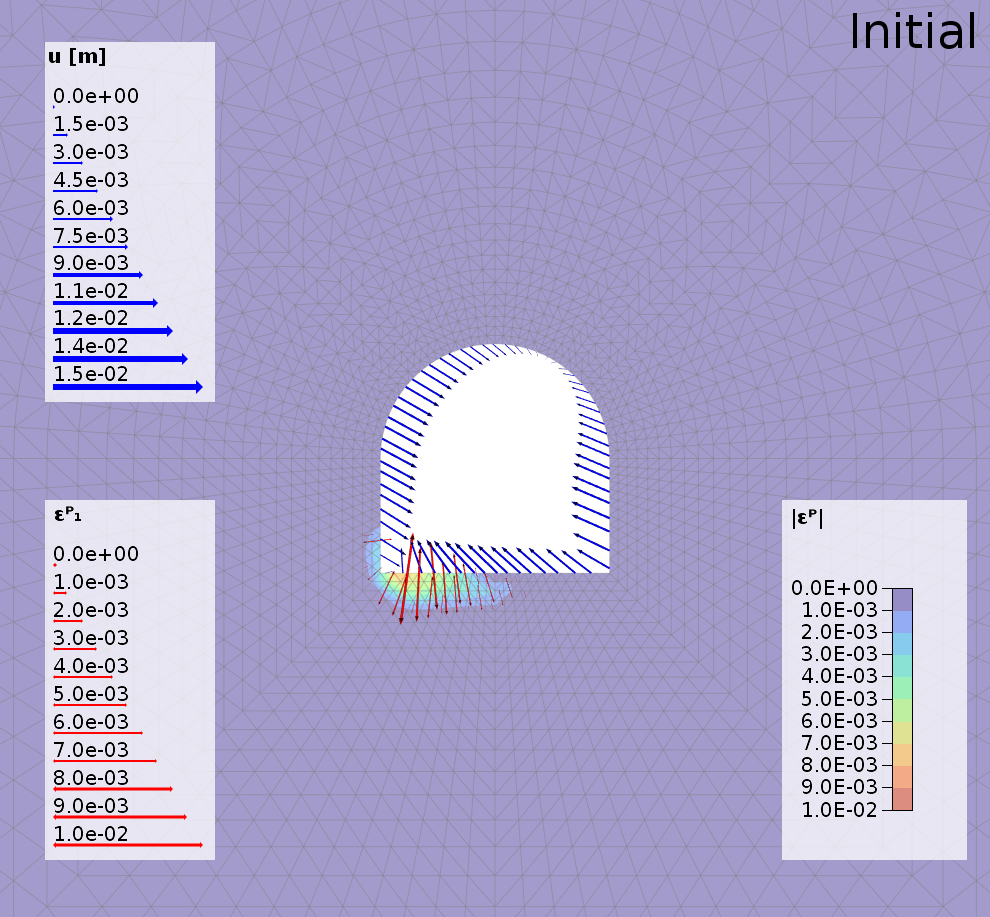}\hspace{0.02\textwidth}\includegraphics[width=0.3\textwidth]{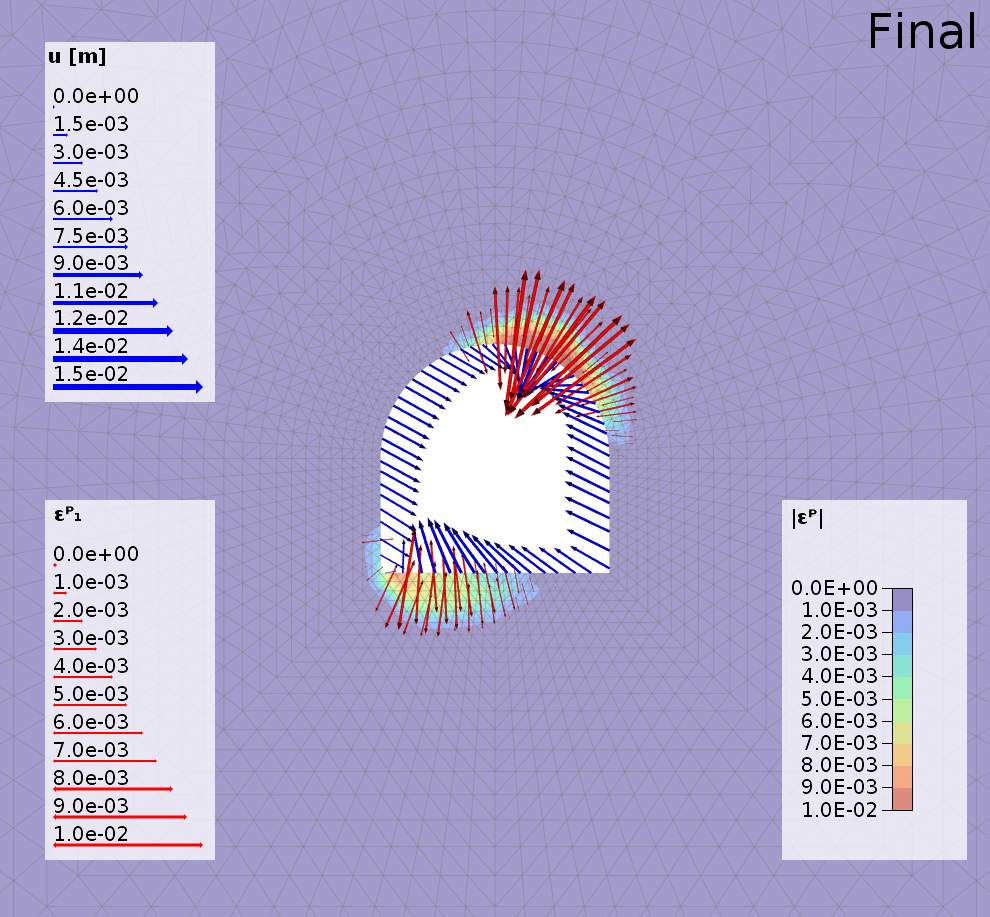}\hspace{0.02\textwidth}\includegraphics[width=0.3\textwidth]{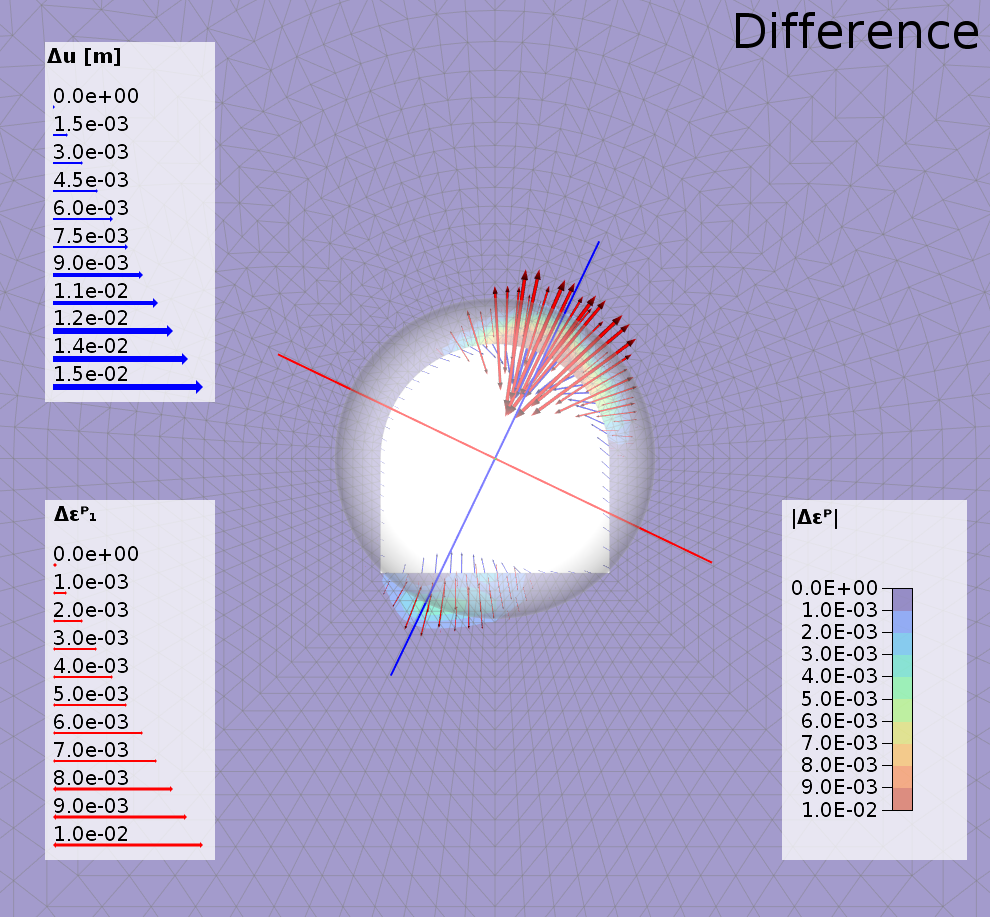}
\par\end{centering}
\caption{\label{fig:Conv-vs-Kirch-case3}Illustration of the inputs in seismic
mechanism calculation for Case 3 using Equations (\ref{eq:kirchhoff})
and (\ref{eq:adjusted-conventional}). Also shown are the resulting
mechanisms as beachballs (P-axis in red, B-axis in green, and T-axis
in blue). Descriptions are provided in the text.}
\end{figure}

Using Equations (\ref{eq:kirchhoff}) and (\ref{eq:adjusted-conventional}),
it is possible to extract source mechanisms from the results of numerical
modelling. We take Case 3 as an example of this procedure, which is
represented graphically in Figure \ref{fig:Conv-vs-Kirch-case3}.
The top left and middle plots show values and directions of traction
$\mathbf{T}$ and displacement $\mathbf{u}$ on a surface $S$ enclosing
the damaged region before and after its expansion. The top right plot
shows their differences ($\mathbf{\Delta T}$ and $\mathbf{\triangle u}$,
respectively), which are used in Equation (\ref{eq:kirchhoff}) to
calculate a seismic mechanism that is shown as a beachball (with the
P-axis in red, B-axis in green, and T-axis in blue). The bottom left
and middle plots depict plastic strain $\boldsymbol{\varepsilon}^{p}$
and tunnel surface displacement $\mathbf{u}$ before and after the
expansion of the damaged region. Their differences ($\mathbf{\Delta}\boldsymbol{\varepsilon}^{p}$
and $\mathbf{\triangle u}$, respectively, shown in the bottom right)
are then used in Equation (\ref{eq:adjusted-conventional}) to determine
the source mechanism shown. In the Kirchhoff case, the surface $S$
was chosen to be composed from the north-south oriented faces lying
on the contour shown in black; however, as shown in Appendices \ref{sec:Appendix-1:-Details}
and \ref{sec:Appendix-2:-Transformations} for all cases, this selection
does not affect the resulting moment (as long as the contour is sufficiently
far from the external boundary of the modelling domain). Note that
this surface does not enclose the damaged region in the north-south
direction. This stems from our two-dimensional approach, which means
that extending $S$ to include faces on the north or south ends of
the mesh would result in intersecting the damaged region. However,
the results of this subsection will demonstrate that this compromise
in surface selection does not result in significant deviation of the
results of Equation (\ref{eq:kirchhoff}) from those of Equation (\ref{eq:adjusted-conventional}).

Figures showing the difference between the pre- and post-expansion
states for all cases are given in Appendix \ref{sec:Appendix-1:-Details}.
To compare the source mechanisms obtained using Equations (\ref{eq:kirchhoff})
and (\ref{eq:adjusted-conventional}) for these cases, three main
characteristics are summarized in Tables \ref{tab:Results-M0} and
\ref{tab:Results-Hudson+stereo}:
\begin{itemize}
\item Source size in terms of scalar \textcolor{black}{scalar seismic moment
}$\left|\mathbf{M}\right|$ (expressed in $\unit{N\cdot m}$) \textcolor{black}{and
Hanks-Kanamori moment magnitude }$m_{\mathrm{HK}}=0.667\log_{10}\left|\mathbf{M}\right|-6.033$
\cite{Hanks-1979} (Table \ref{tab:Results-M0}).
\item Signs and ratios of eigenvalues expressed in terms of $k$ and $T$
parameters as displayed in the source-type plot of \cite{Hudson-1989}
(middle column of Table \ref{tab:Results-Hudson+stereo}).
\item Orientations of principal axes as displayed in a lower-hemisphere
stereonet plot (last column of Table \ref{tab:Results-Hudson+stereo}).
\end{itemize}
\textcolor{black}{The Kirchhoff-type }{[}Equation (\ref{eq:kirchhoff}){]}
and\textcolor{black}{{} adjusted conventional }{[}Equation (\ref{eq:adjusted-conventional}){]}\textcolor{black}{{}
expressions provide similar results with regard to sizes and orientation
of principal P axes: the difference in moment magnitude }$m_{HK}$\textcolor{black}{{}
does not exceed $0.02$ and the difference in the directions of principal
axes is within $1^{\circ}$. Furthermore, the source types are similar
with respect to the sign and size of their isotropic components, with
}the difference in $k$ parameter between Kirchhoff-type and adjusted
conventional solutions being less than $0.03$. There is a larger
disparity in the $T$ parameter (which measures the deviation of the
deviatoric component from a double-couple model); the largest such
difference being $0.36$ for Case 3. However, we note that the region
of the source plot corresponding to significant CLVD and isotropic
components ($T$ above $0.6$ and $k$ below $-0.4$) has a high density
of isolines of constant $T$, which indicates that this seemingly
substantial difference in the $T$ parameter is not necessarily translated
to a significant difference in other metrics (such as the angle $\omega$
explained in Section \ref{sec:Discussion-and-Conclusions}). The difference
in $T$ parameter values may be an artifact of the previously discussed
compromises made in the selection of integration surface for the Kirchhoff-type
expression. The similarity in B- and T-axis eigenvalues for sources
of this type also results in their orientations not being well constrained.
This can be seen for Case 6, where these axes are interchanged for
the two methods of moment tensor calculation.

Overall, the sources derived from modelling have significant negative
isotropic (implosive) and negative compensated linear vector dipole
(pancake-shape) components as well as a P-axis that is aligned with
the direction of maximum compressional principal stress. These features
indicate that at wavelengths larger than the diameter of the tunnel
and the extent of the damage along it, seismic radiation is controlled
by the elastic convergence of the surrounding rockmass rather than
by rock fracturing (which is explosive when considered in isolation).
Furthermore, there is not a strong dependence of the method of triggering
on the resulting source. This is evidenced by the similarity between
the seismic mechanisms for Cases 3 and 6, which differ only in the
method of triggering (quasi-static load increase and transient stress
wave, respectively).

\textcolor{black}{}
\begin{table}
\centering{}\textcolor{black}{\caption{\textcolor{black}{\label{tab:Results-M0}Comparison of the the modelled
source mechanisms in terms of scalar moment }$\left|\mathbf{M}\right|$\textcolor{black}{{}
and Hanks--Kanamori moment magnitude }$m_{\mathrm{HK}}$\textcolor{black}{.
The listed variants are: }$K$\textcolor{black}{-- Kirchhoff-type
expression {[}Equation (\ref{eq:kirchhoff}){]}, }$K-T$\textcolor{black}{{}
-- Kirchhoff-type expression traction term, }$K-D$\textcolor{black}{{}
-- Kirchhoff-type expression displacement term, }$C$\textcolor{black}{{}
-- adjusted conventional expression {[}Equation (\ref{eq:adjusted-conventional}){]},
}$C-S$\textcolor{black}{{} -- adjusted conventional expression strain
term, }$C-D$\textcolor{black}{{} -- adjusted conventional expression
displacement term, }$N$\textcolor{black}{{} -- numerically evaluated
expression for the elliptical cavity described in Subsection \ref{subsec:Numerical-approximation},
$N-T$ -- numerical expression traction term, $N-D$ -- numerical
expression displacement term, }$A$\textcolor{black}{{} -- analytical
approximation for the effective elliptical cavity described in Subsection
\ref{subsec:Analytical-approximation}, $\left|C_{M}\right|$ --
simple approximations of the $A$ case. The ``}$\left|\mathbf{M}\right|$\textcolor{black}{{}
ratio'' rows list moments normalized to the Kirchhoff-type moment
}$K$\textcolor{black}{. The} $\triangle m_{\mathrm{HK}}$\textcolor{black}{{}
rows list difference in moment magnitude compared to the Kirchhoff-type
moment }$K$. Traction and dispalcement terms for Kirchhoff-type and
numerical expressions have been calculated for $\unit[15]{m}$ diameter
contours.}
\medskip{}
}%
\begin{tabular}{>{\centering}p{0.045\columnwidth}>{\centering}p{0.15\columnwidth}>{\centering}p{0.045\columnwidth}>{\centering}p{0.045\columnwidth}>{\centering}p{0.045\columnwidth}>{\centering}p{0.045\columnwidth}>{\centering}p{0.045\columnwidth}>{\centering}p{0.045\columnwidth}>{\centering}p{0.045\columnwidth}>{\centering}p{0.045\columnwidth}>{\centering}p{0.045\columnwidth}>{\centering}p{0.045\columnwidth}>{\centering}p{0.045\columnwidth}}
\toprule 
\textcolor{black}{\scriptsize{}Case} & \textcolor{black}{\scriptsize{}Parameter} & \textbf{\textcolor{black}{\scriptsize{}$K$}} & \textbf{\textcolor{black}{\scriptsize{}$K-T$}} & \textbf{\textcolor{black}{\scriptsize{}$K-D$}} & \textbf{\textcolor{black}{\scriptsize{}$C$}} & \textbf{\textcolor{black}{\scriptsize{}$C-S$}} & \textbf{\textcolor{black}{\scriptsize{}$C-D$}} & \textbf{\textcolor{black}{\scriptsize{}$N$}} & \textbf{\textcolor{black}{\scriptsize{}$N-T$}} & \textbf{\textcolor{black}{\scriptsize{}$N-D$}} & \textbf{\textcolor{black}{\scriptsize{}$A$}} & {\scriptsize{}$\left|C_{M}\right|$}\tabularnewline
\midrule
\multirow{4}{0.045\columnwidth}{\textcolor{black}{\scriptsize{}1}} & \textcolor{black}{\scriptsize{}$\unit[\left|\mathbf{M}\right|]{[10^{9}N\cdot m]}$} & \textbf{\scriptsize{}8.55} & {\scriptsize{}2.62} & {\scriptsize{}6.01} & \textbf{\scriptsize{}8.19} & {\scriptsize{}12.13} & {\scriptsize{}15.41} & \textbf{\scriptsize{}11.17} & {\scriptsize{}3.70} & {\scriptsize{}7.58} & \textbf{\scriptsize{}11.43} & \textbf{\scriptsize{}12.22}\tabularnewline
 & \textcolor{black}{\scriptsize{}$m_{\mathrm{HK}}$} & \textbf{\scriptsize{}0.59} & {\scriptsize{}0.25} & {\scriptsize{}0.49} & \textbf{\scriptsize{}0.58} & {\scriptsize{}0.69} & {\scriptsize{}0.76} & \textbf{\scriptsize{}0.67} & {\scriptsize{}0.35} & {\scriptsize{}0.55} & \textbf{\scriptsize{}0.67} & \textbf{\scriptsize{}0.69}\tabularnewline
 & \textcolor{black}{\scriptsize{}$\left|\mathbf{M}\right|$ ratio} & \textbf{\scriptsize{}-} & {\scriptsize{}-} & {\scriptsize{}-} & \textbf{\scriptsize{}0.96} & {\scriptsize{}-} & {\scriptsize{}-} & \textbf{\scriptsize{}1.31} & {\scriptsize{}-} & {\scriptsize{}-} & \textbf{\scriptsize{}1.34} & \textbf{\scriptsize{}1.43}\tabularnewline
 & \textcolor{black}{\scriptsize{}$\triangle m_{\mathrm{HK}}$} & \textbf{\scriptsize{}-} & {\scriptsize{}-} & {\scriptsize{}-} & \textbf{\scriptsize{}-0.01} & {\scriptsize{}-} & {\scriptsize{}-} & \textbf{\scriptsize{}0.08} & {\scriptsize{}-} & {\scriptsize{}-} & \textbf{\scriptsize{}0.08} & \textbf{\scriptsize{}0.10}\tabularnewline
\midrule
\multirow{4}{0.045\columnwidth}{\textcolor{black}{\scriptsize{}2}} & \textcolor{black}{\scriptsize{}$\unit[\left|\mathbf{M}\right|]{[10^{9}N\cdot m]}$} & \textbf{\scriptsize{}8.06} & {\scriptsize{}2.39} & {\scriptsize{}5.75} & \textbf{\scriptsize{}7.82} & {\scriptsize{}11.42} & {\scriptsize{}15.01} & \textbf{\scriptsize{}9.86} & {\scriptsize{}3.20} & {\scriptsize{}6.76} & \textbf{\scriptsize{}9.72} & \textbf{\scriptsize{}10.10}\tabularnewline
 & \textcolor{black}{\scriptsize{}$m_{\mathrm{HK}}$} & \textbf{\scriptsize{}0.57} & {\scriptsize{}0.22} & {\scriptsize{}0.47} & \textbf{\scriptsize{}0.56} & {\scriptsize{}0.67} & {\scriptsize{}0.75} & \textbf{\scriptsize{}0.63} & {\scriptsize{}0.30} & {\scriptsize{}0.52} & \textbf{\scriptsize{}0.63} & \textbf{\scriptsize{}0.64}\tabularnewline
 & \textcolor{black}{\scriptsize{}$\left|\mathbf{M}\right|$ ratio} & \textbf{\scriptsize{}-} & {\scriptsize{}-} & {\scriptsize{}-} & \textbf{\scriptsize{}0.97} & {\scriptsize{}-} & {\scriptsize{}-} & \textbf{\scriptsize{}1.22} & {\scriptsize{}-} & {\scriptsize{}-} & \textbf{\scriptsize{}1.21} & \textbf{\scriptsize{}1.25}\tabularnewline
 & \textcolor{black}{\scriptsize{}$\triangle m_{\mathrm{HK}}$} & \textbf{\scriptsize{}-} & {\scriptsize{}-} & {\scriptsize{}-} & \textbf{\scriptsize{}-0.01} & {\scriptsize{}-} & {\scriptsize{}-} & \textbf{\scriptsize{}0.06} & {\scriptsize{}-} & {\scriptsize{}-} & \textbf{\scriptsize{}0.05} & \textbf{\scriptsize{}0.07}\tabularnewline
\midrule
\multirow{4}{0.045\columnwidth}{\textcolor{black}{\scriptsize{}3}} & \textcolor{black}{\scriptsize{}$\unit[\left|\mathbf{M}\right|]{[10^{9}N\cdot m]}$} & \textbf{\scriptsize{}4.69} & {\scriptsize{}1.34} & {\scriptsize{}3.39} & \textbf{\scriptsize{}4.33} & {\scriptsize{}5.02} & {\scriptsize{}7.36} & \textbf{\scriptsize{}5.62} & {\scriptsize{}1.85} & {\scriptsize{}3.83} & \textbf{\scriptsize{}5.79} & \textbf{\scriptsize{}6.12}\tabularnewline
 & \textcolor{black}{\scriptsize{}$m_{\mathrm{HK}}$} & \textbf{\scriptsize{}0.41} & {\scriptsize{}0.05} & {\scriptsize{}0.32} & \textbf{\scriptsize{}0.39} & {\scriptsize{}0.43} & {\scriptsize{}0.55} & \textbf{\scriptsize{}0.47} & {\scriptsize{}0.14} & {\scriptsize{}0.36} & \textbf{\scriptsize{}0.48} & \textbf{\scriptsize{}0.49}\tabularnewline
 & \textcolor{black}{\scriptsize{}$\left|\mathbf{M}\right|$ ratio} & \textbf{\scriptsize{}-} & {\scriptsize{}-} & {\scriptsize{}-} & \textbf{\scriptsize{}0.92} & {\scriptsize{}-} & {\scriptsize{}-} & \textbf{\scriptsize{}1.20} & {\scriptsize{}-} & {\scriptsize{}-} & \textbf{\scriptsize{}1.23} & \textbf{\scriptsize{}1.30}\tabularnewline
 & \textcolor{black}{\scriptsize{}$\triangle m_{\mathrm{HK}}$} & \textbf{\scriptsize{}-} & {\scriptsize{}-} & {\scriptsize{}-} & \textbf{\scriptsize{}-0.02} & {\scriptsize{}-} & {\scriptsize{}-} & \textbf{\scriptsize{}0.05} & {\scriptsize{}-} & {\scriptsize{}-} & \textbf{\scriptsize{}0.06} & \textbf{\scriptsize{}0.08}\tabularnewline
\midrule
\multirow{4}{0.045\columnwidth}{\textcolor{black}{\scriptsize{}4}} & \textcolor{black}{\scriptsize{}$\unit[\left|\mathbf{M}\right|]{[10^{9}N\cdot m]}$} & \textbf{\scriptsize{}1.91} & {\scriptsize{}0.41} & {\scriptsize{}1.51} & \textbf{\scriptsize{}1.87} & {\scriptsize{}2.49} & {\scriptsize{}3.33} & \textbf{\scriptsize{}2.42} & {\scriptsize{}0.79} & {\scriptsize{}1.66} & \textbf{\scriptsize{}2.66} & \textbf{\scriptsize{}2.83}\tabularnewline
 & \textcolor{black}{\scriptsize{}$m_{\mathrm{HK}}$} & \textbf{\scriptsize{}0.15} & {\scriptsize{}-0.29} & {\scriptsize{}0.09} & \textbf{\scriptsize{}0.15} & {\scriptsize{}0.23} & {\scriptsize{}0.32} & \textbf{\scriptsize{}0.22} & {\scriptsize{}-0.10} & {\scriptsize{}0.11} & \textbf{\scriptsize{}0.25} & \textbf{\scriptsize{}0.27}\tabularnewline
 & \textcolor{black}{\scriptsize{}$\left|\mathbf{M}\right|$ ratio} & \textbf{\scriptsize{}-} & {\scriptsize{}-} & {\scriptsize{}-} & \textbf{\scriptsize{}0.98} & {\scriptsize{}-} & {\scriptsize{}-} & \textbf{\scriptsize{}1.27} & {\scriptsize{}-} & {\scriptsize{}-} & \textbf{\scriptsize{}1.40} & \textbf{\scriptsize{}1.48}\tabularnewline
 & \textcolor{black}{\scriptsize{}$\triangle m_{\mathrm{HK}}$} & \textbf{\scriptsize{}-} & {\scriptsize{}-} & {\scriptsize{}-} & \textbf{\scriptsize{}-0.01} & {\scriptsize{}-} & {\scriptsize{}-} & \textbf{\scriptsize{}0.07} & {\scriptsize{}-} & {\scriptsize{}-} & \textbf{\scriptsize{}0.10} & \textbf{\scriptsize{}0.11}\tabularnewline
\midrule
\multirow{4}{0.045\columnwidth}{\textcolor{black}{\scriptsize{}5}} & \textcolor{black}{\scriptsize{}$\unit[\left|\mathbf{M}\right|]{[10^{9}N\cdot m]}$} & \textbf{\scriptsize{}1.23} & {\scriptsize{}0.31} & {\scriptsize{}0.93} & \textbf{\scriptsize{}1.18} & {\scriptsize{}1.18} & {\scriptsize{}1.74} & \textbf{\scriptsize{}1.73} & {\scriptsize{}0.58} & {\scriptsize{}1.16} & \textbf{\scriptsize{}1.64} & \textbf{\scriptsize{}1.66}\tabularnewline
 & \textcolor{black}{\scriptsize{}$m_{\mathrm{HK}}$} & \textbf{\scriptsize{}0.03} & {\scriptsize{}-0.37} & {\scriptsize{}-0.05} & \textbf{\scriptsize{}0.02} & {\scriptsize{}0.01} & {\scriptsize{}0.13} & \textbf{\scriptsize{}0.13} & {\scriptsize{}-0.19} & {\scriptsize{}0.01} & \textbf{\scriptsize{}0.11} & \textbf{\scriptsize{}0.11}\tabularnewline
 & \textcolor{black}{\scriptsize{}$\left|\mathbf{M}\right|$ ratio} & \textbf{\scriptsize{}-} & {\scriptsize{}-} & {\scriptsize{}-} & \textbf{\scriptsize{}0.96} & {\scriptsize{}-} & {\scriptsize{}-} & \textbf{\scriptsize{}1.41} & {\scriptsize{}-} & {\scriptsize{}-} & \textbf{\scriptsize{}1.34} & \textbf{\scriptsize{}1.35}\tabularnewline
 & \textcolor{black}{\scriptsize{}$\triangle m_{\mathrm{HK}}$} & \textbf{\scriptsize{}-} & {\scriptsize{}-} & {\scriptsize{}-} & \textbf{\scriptsize{}-0.01} & {\scriptsize{}-} & {\scriptsize{}-} & \textbf{\scriptsize{}0.10} & {\scriptsize{}-} & {\scriptsize{}-} & \textbf{\scriptsize{}0.08} & \textbf{\scriptsize{}0.09}\tabularnewline
\midrule
\multirow{4}{0.045\columnwidth}{{\scriptsize{}6}} & \textcolor{black}{\scriptsize{}$\unit[\left|\mathbf{M}\right|]{[10^{9}N\cdot m]}$} & \textbf{\scriptsize{}4.54} & {\scriptsize{}1.56} & {\scriptsize{}3.02} & \textbf{\scriptsize{}4.20} & {\scriptsize{}5.05} & {\scriptsize{}7.43} & \textbf{\scriptsize{}5.75} & {\scriptsize{}1.89} & {\scriptsize{}3.92} & \textbf{\scriptsize{}5.97} & \textbf{\scriptsize{}6.19}\tabularnewline
 & \textcolor{black}{\scriptsize{}$m_{\mathrm{HK}}$} & \textbf{\scriptsize{}0.40} & {\scriptsize{}0.10} & {\scriptsize{}0.29} & \textbf{\scriptsize{}0.38} & {\scriptsize{}0.44} & {\scriptsize{}0.55} & \textbf{\scriptsize{}0.47} & {\scriptsize{}0.15} & {\scriptsize{}0.36} & \textbf{\scriptsize{}0.48} & \textbf{\scriptsize{}0.49}\tabularnewline
 & \textcolor{black}{\scriptsize{}$\left|\mathbf{M}\right|$ ratio} & \textbf{\scriptsize{}-} & {\scriptsize{}-} & {\scriptsize{}-} & \textbf{\scriptsize{}0.93} & {\scriptsize{}-} & {\scriptsize{}-} & \textbf{\scriptsize{}1.27} & {\scriptsize{}-} & {\scriptsize{}-} & \textbf{\scriptsize{}1.32} & \textbf{\scriptsize{}1.36}\tabularnewline
 & \textcolor{black}{\scriptsize{}$\triangle m_{\mathrm{HK}}$} & \textbf{\scriptsize{}-} & {\scriptsize{}-} & {\scriptsize{}-} & \textbf{\scriptsize{}-0.02} & {\scriptsize{}-} & {\scriptsize{}-} & \textbf{\scriptsize{}0.07} & {\scriptsize{}-} & {\scriptsize{}-} & \textbf{\scriptsize{}0.08} & \textbf{\scriptsize{}0.09}\tabularnewline
\bottomrule
\end{tabular}
\end{table}

\noindent \begin{center}
\begin{longtable}[l]{>{\centering}m{0.05\textwidth}>{\centering}m{0.42\textwidth}>{\centering}m{0.42\textwidth}}
\addlinespace[0.1cm]
\caption{\label{tab:Results-Hudson+stereo}Source mechanism characteristics
for modelled Cases 1-6. Mechanisms are shown in source-type plots
of \cite{Hudson-1989} and in lower-hemisphere stereonet plots. The
labels describe the procedure used to calculate the mechanism:\textcolor{black}{{}
}$K$\textcolor{black}{-- Kirchhoff-type expression {[}Equation (\ref{eq:kirchhoff}){]},
}$C$\textcolor{black}{{} -- adjusted conventional expression {[}Equation
(\ref{eq:adjusted-conventional}){]}, }$N$\textcolor{black}{{} --
numerically evaluated expression for the elliptical cavity described
in Subsection \ref{subsec:Numerical-approximation}, }$A$\textcolor{black}{{}
-- analytical approximation for the effective elliptical cavity described
in Subsection \ref{subsec:Analytical-approximation}. The overlays
in the source-type plots describing different sources are explained
in Subsection \ref{subsec:Classification}.}}
\tabularnewline\addlinespace[0.1cm]
\midrule 
\addlinespace[0.1cm]
\textcolor{black}{\footnotesize{}Case} & {\footnotesize{}Source-type plot} & {\footnotesize{}Stereonet plot of principal axes}\tabularnewline\addlinespace[0.1cm]
\midrule
\endfirsthead
\addlinespace[0.1cm]
\caption{continued.}
\tabularnewline\addlinespace[0.1cm]
\midrule 
\addlinespace[0.1cm]
\textcolor{black}{\footnotesize{}Case} & {\footnotesize{}Source-type plot} & {\footnotesize{}Stereonet plot of principal axes}\tabularnewline\addlinespace[0.1cm]
\midrule
\endhead
\addlinespace[0.07cm]
{\footnotesize{}1} & {\footnotesize{}\includegraphics[width=0.44\textwidth]{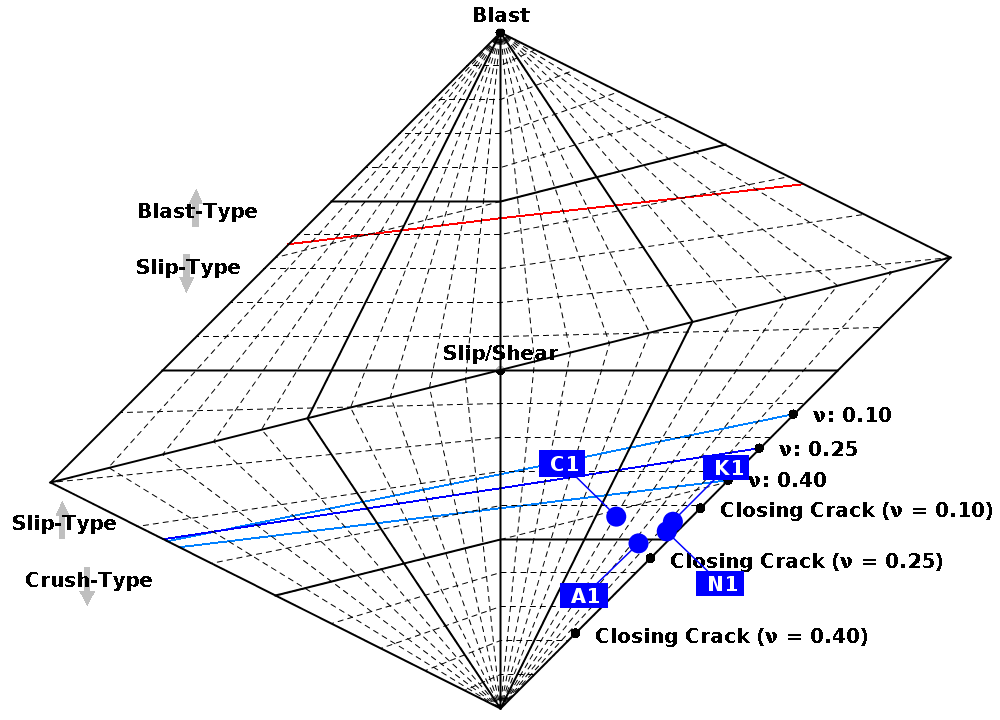}} & {\footnotesize{}\includegraphics[width=0.33\textwidth]{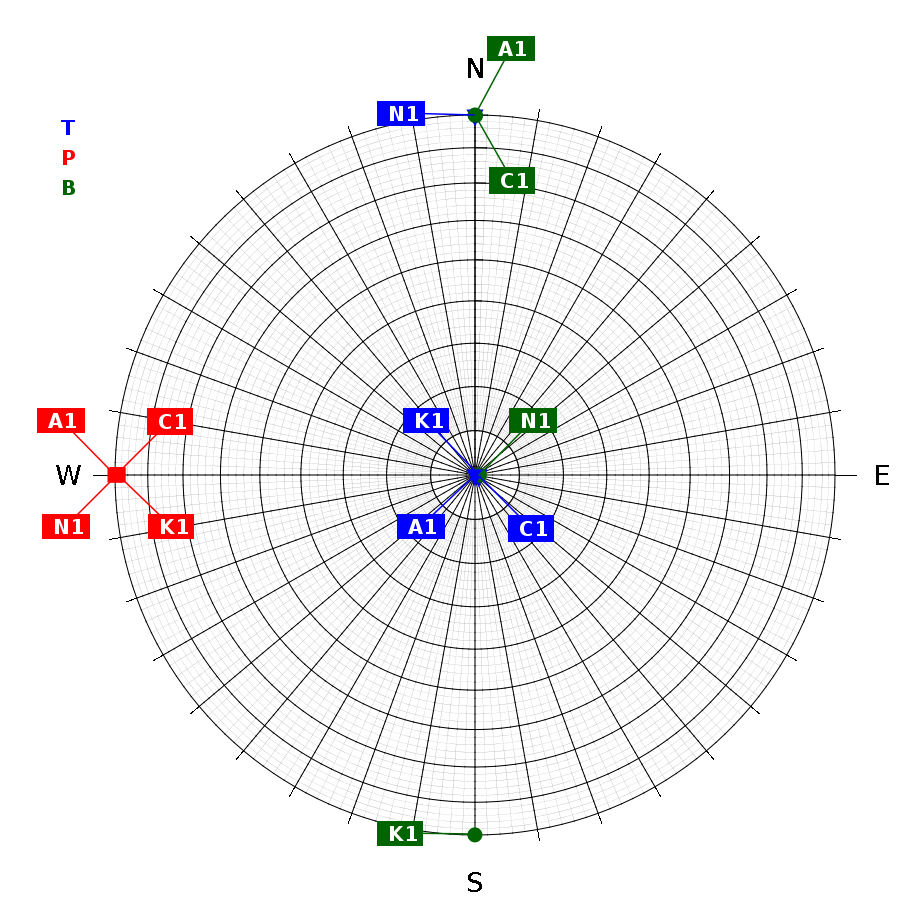}}\tabularnewline\addlinespace[0.07cm]
\midrule
\addlinespace[0.07cm]
{\footnotesize{}2} & {\footnotesize{}\includegraphics[width=0.44\textwidth]{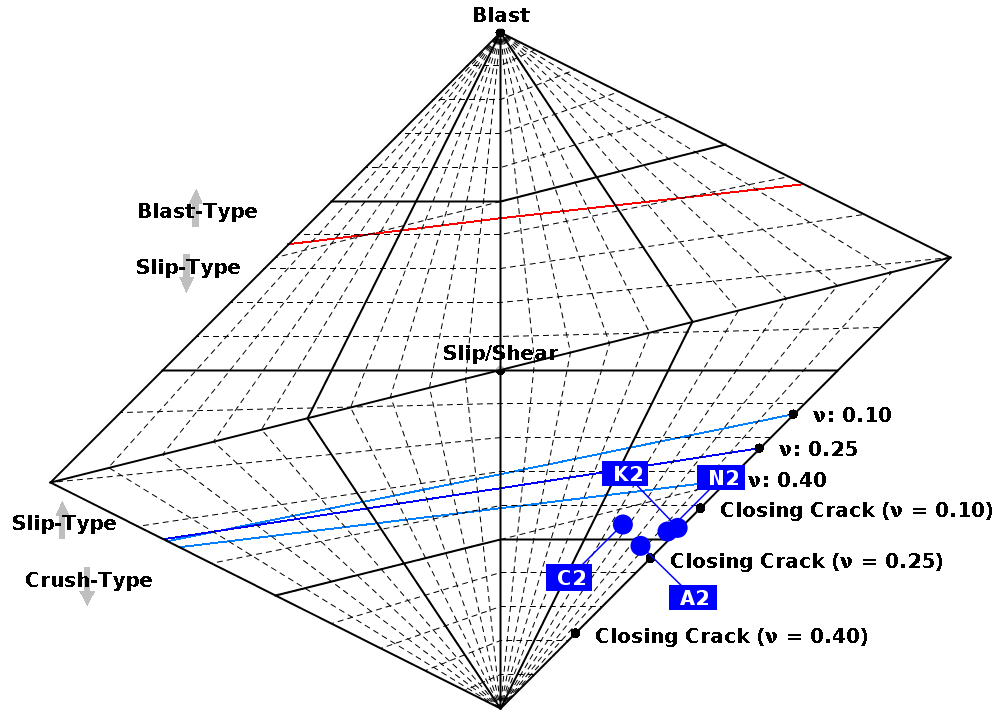}} & {\footnotesize{}\includegraphics[width=0.33\textwidth]{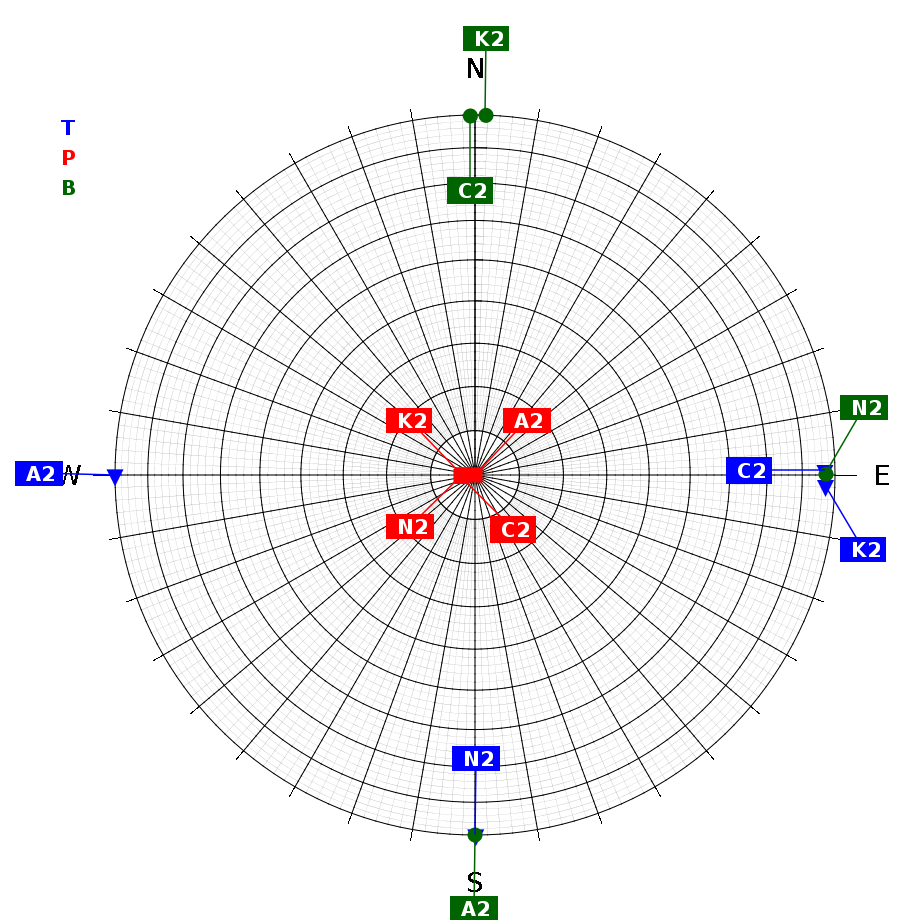}}\tabularnewline\addlinespace[0.07cm]
\midrule
\addlinespace[0.07cm]
{\footnotesize{}3} & {\footnotesize{}\includegraphics[width=0.44\textwidth]{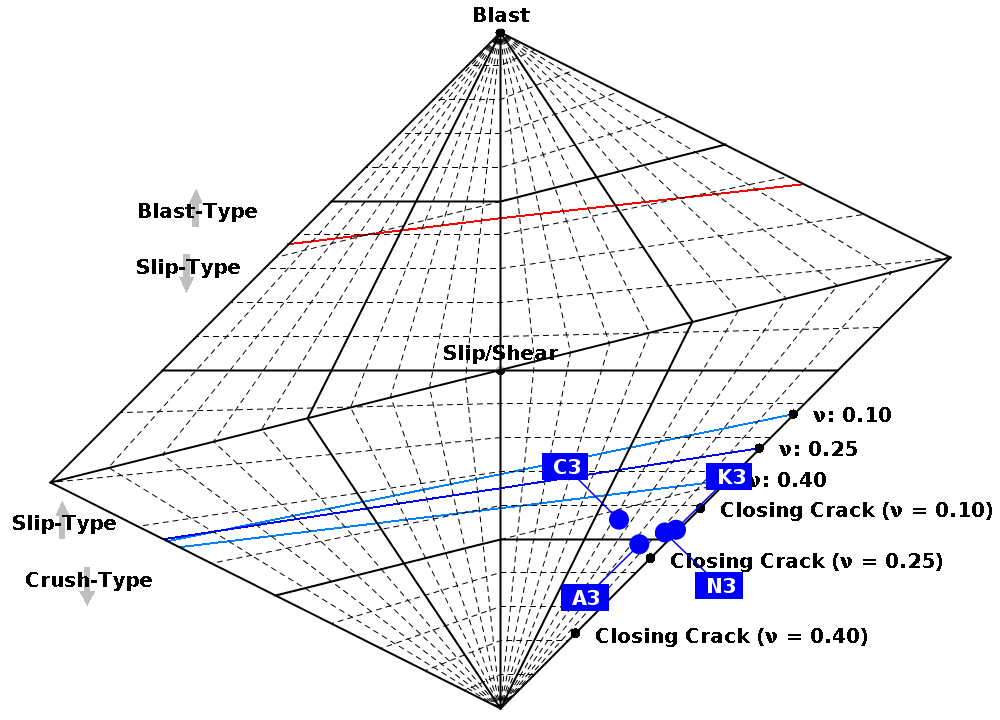}} & {\footnotesize{}\includegraphics[width=0.33\textwidth]{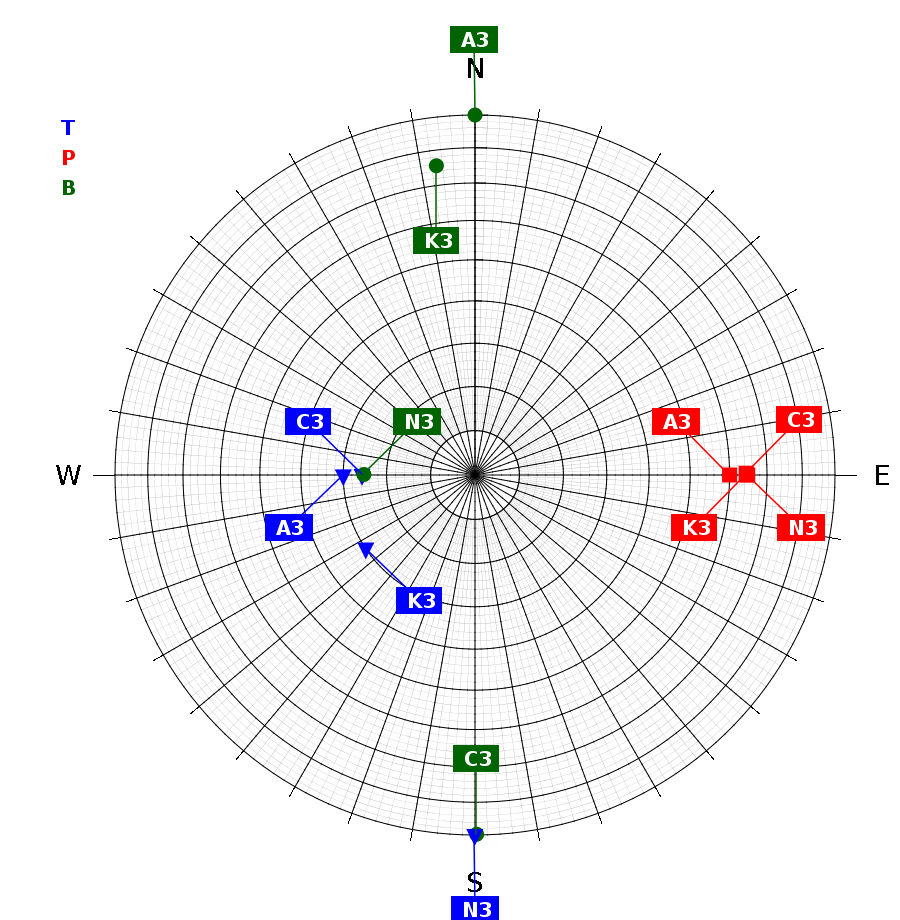}}\tabularnewline\addlinespace[0.07cm]
\midrule
\addlinespace[0.1cm]
{\footnotesize{}4} & {\footnotesize{}\includegraphics[width=0.44\textwidth]{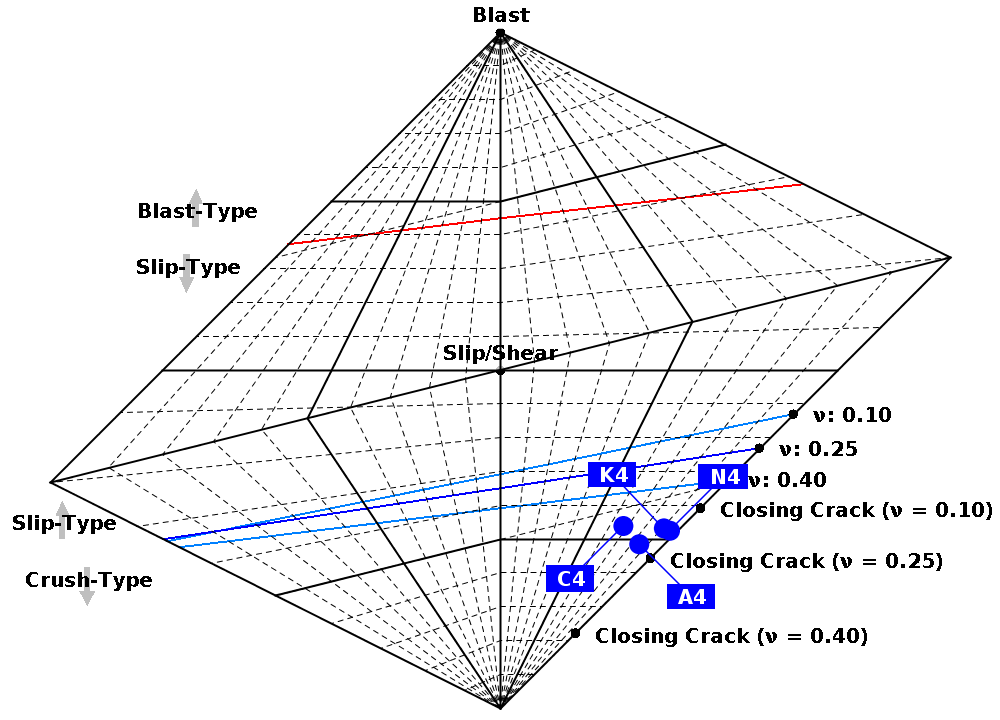}} & {\footnotesize{}\includegraphics[width=0.33\textwidth]{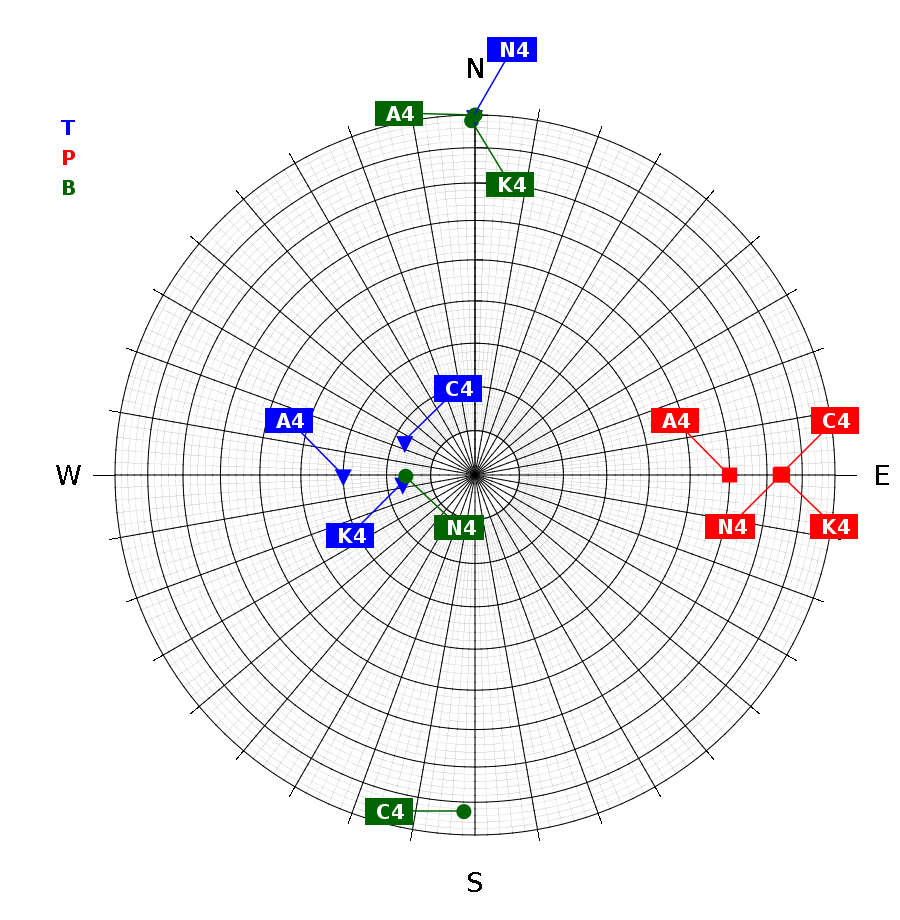}}\tabularnewline\addlinespace[0.1cm]
\midrule
\addlinespace[0.1cm]
{\footnotesize{}5} & {\footnotesize{}\includegraphics[width=0.44\textwidth]{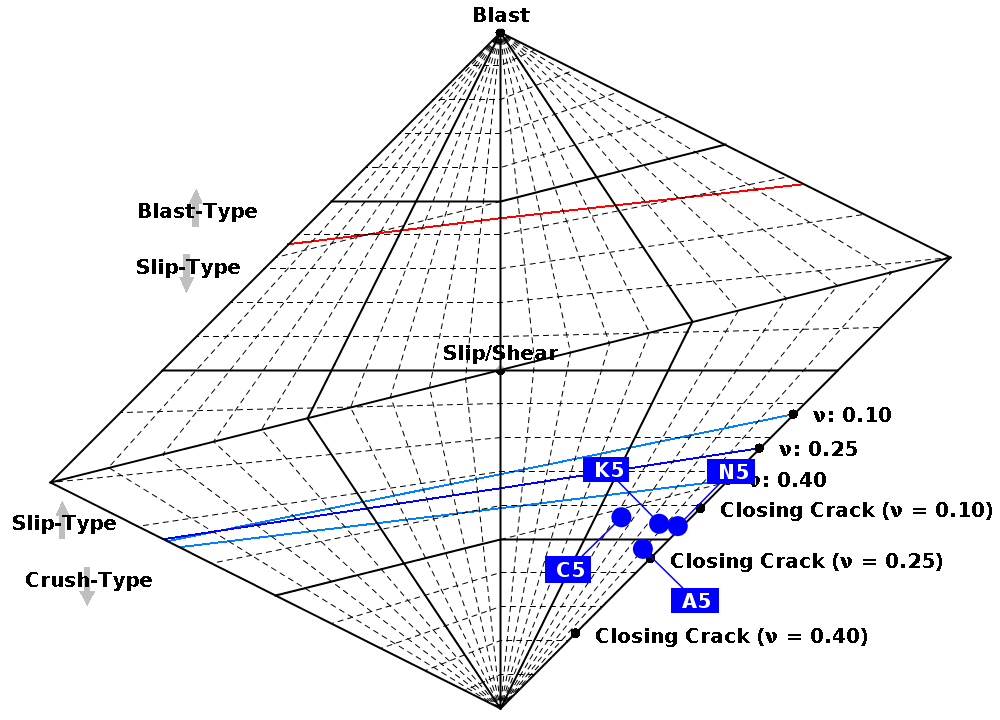}} & {\footnotesize{}\includegraphics[width=0.33\textwidth]{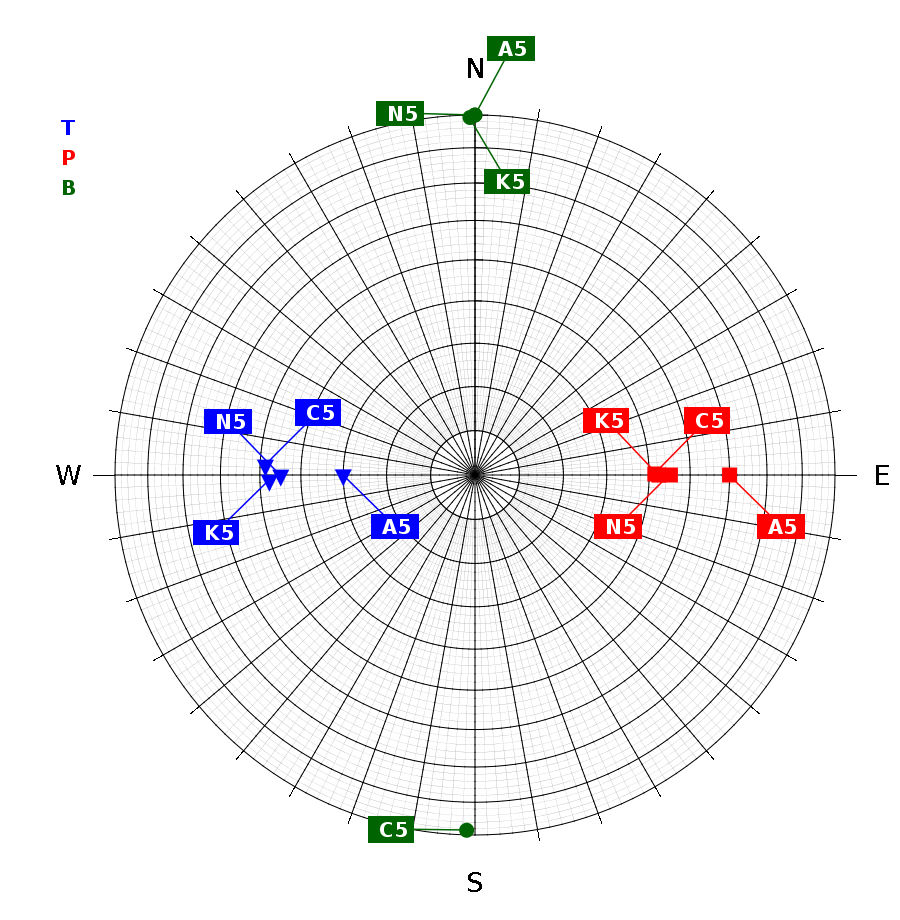}}\tabularnewline\addlinespace[0.1cm]
\midrule
\addlinespace[0.1cm]
{\footnotesize{}6} & {\footnotesize{}\includegraphics[width=0.44\textwidth]{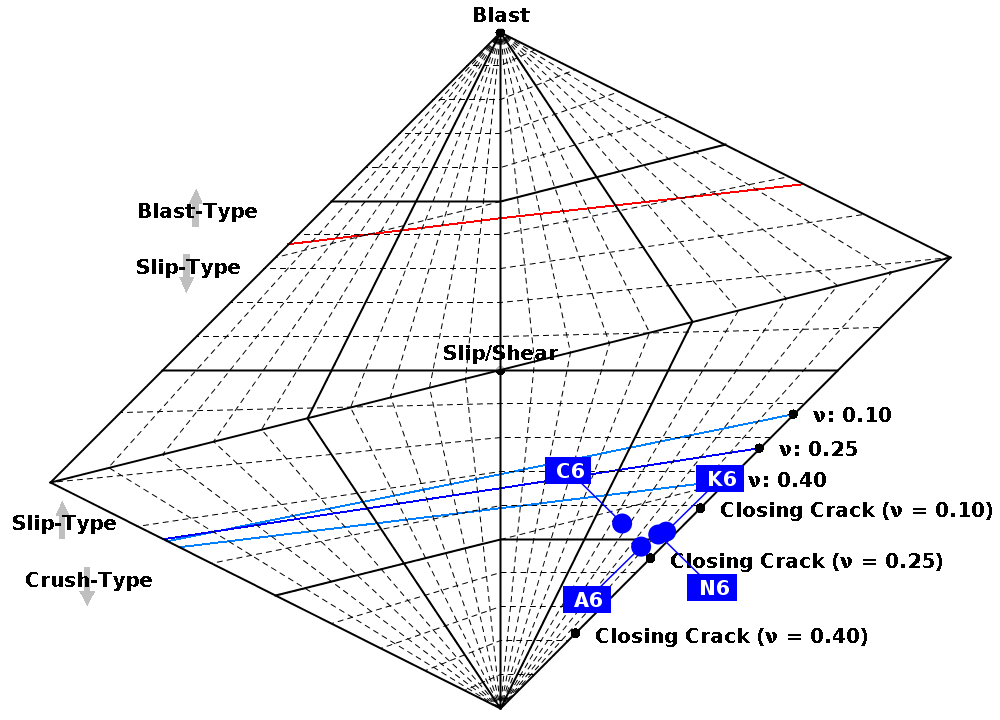}} & {\footnotesize{}\includegraphics[width=0.33\textwidth]{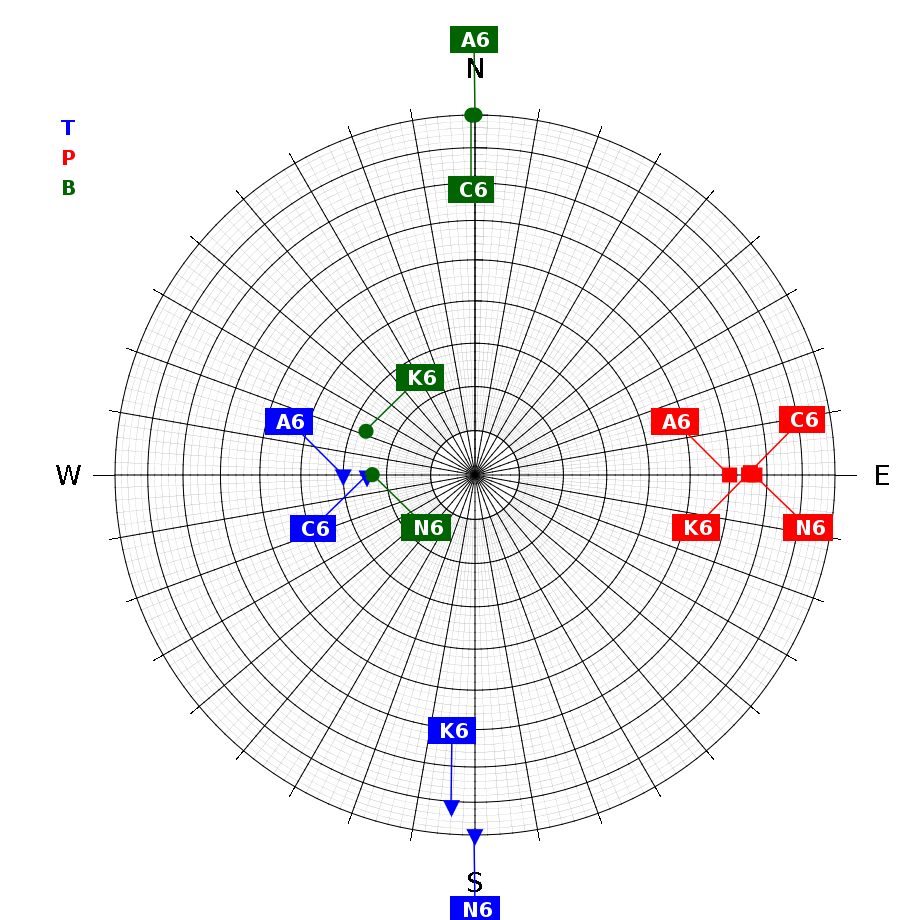}}\tabularnewline\addlinespace[0.1cm]
\bottomrule
\end{longtable}
\par\end{center}

\noindent \newpage Table \ref{tab:Results-M0} also shows the contribution
of the constituent terms of Equations (\ref{eq:kirchhoff}) and (\ref{eq:adjusted-conventional})
to the total solutions. For the Kirchhoff-type expression, it can
be seen that $\left|\mathbf{M^{T}}\right|<\left|\mathbf{M^{U}}\right|<\left|\mathbf{M}\right|$
in all five cases, which is a result of each term contributing similar
content to the overall seismic mechanism; in a sense, $\mathbf{M^{T}}$
and $\mathbf{M^{U}}$ combine constructively to form $\mathbf{M}$.
We note that while $\mathbf{M}=\mathbf{M^{T}}+\mathbf{M^{U}}$ is
independent of the contour selected from integration, the components
are not; however, their constructive nature seems to hold in general
(see Appendix \ref{sec:Appendix-1:-Details}). Behavior is quite different
for the adjusted conventional expression, where the terms $\mathbf{M^{e}}$
and $\mathbf{M^{u}}$ combine ``destructively,'' resulting in $\left|\mathbf{M}\right|\lessapprox\left|\mathbf{M^{e}}\right|<\left|\mathbf{M^{u}}\right|$
in each case: the closure of the tunnel (implosive $\mathbf{M^{u}}$)
needs to overcompensate for the dilation of the fracturing rockmass
(explosive $\mathbf{M^{e}}$). An important feature of the counteracting
nature of $\mathbf{M^{e}}$ and $\mathbf{M^{u}}$ is the potential
for a significant amount of inelastic deformation (large $\left|\mathbf{M^{e}}\right|$)
producing considerable excavation closure (large $\left|\mathbf{M^{u}}\right|$)
but resulting in relatively weak seismic radiation (small $\left|\mathbf{M}\right|$).
Physically, this means that large sudden inelastic deformation may
constitute a small-magnitude seismic source if the deformed volume
has poor mechanical coupling with the surrounding rock mass.

The conventional expression for a moment tensor \{see, for example,
Equation (3.32) in \cite{Aki-Richards-2009}\} includes only the strain
term $\mathbf{M^{e}}$ of Equation (\ref{eq:adjusted-conventional}).
Such a definition suggests interpreting scalar moment in terms of
average plastic strain; that is, $\left|\mathbf{M}\right|\approx kV\overline{\boldsymbol{\triangle\varepsilon^{p}}}$,
where $k$ is the relevant elastic moduli (shear modulus $\mu$ is
typically used) and $V$ is the volume of inelastic deformation. However,
the results presented here (in particular that $\left|\mathbf{M}\right|\lesssim\left|\mathbf{M^{e}}\right|$)
mean that this approach potentially leads to an underestimation of
average plastic strain for events associated with dynamic stress fracturing
around tunnels. It is more appropriate to interpret the scalar moment
in terms of mechanical and geometric parameters describing fracturing
around the tunnel. Such expressions will be presented in the next
section.

\section{\label{sec:Approximation}Approximate description of sources}

The numerical modelling of seismic source mechanisms presented in
Section \ref{sec:Modelling} is a useful instrument in the forensic
analysis of strainbursts or seismic events accompanied by significant
deformation of tunnels. However, building and solving such models
may be computationally prohibitive. Furthermore, several iterations
of building and solving would likely be required to produce a theoretical
source mechanism that matches observations (including scalar moment).
The situation would be complicated further in the case of the observation
of multiple seismic events with the characteristic source mechanism
features presented in the previous section (implosive isotropic component
and pancake-shape CLVD component). To address these complications,
this section discusses an approximate model for the source mechanism
of such events that relies only on geometric and mechanical inputs,
which provides a means for analysis without the use of numerical modelling.

\subsection{\label{subsec:Effective-cavity}Effective cavity expansion}

\begin{figure}[H]
\begin{centering}
\includegraphics[width=0.45\textwidth]{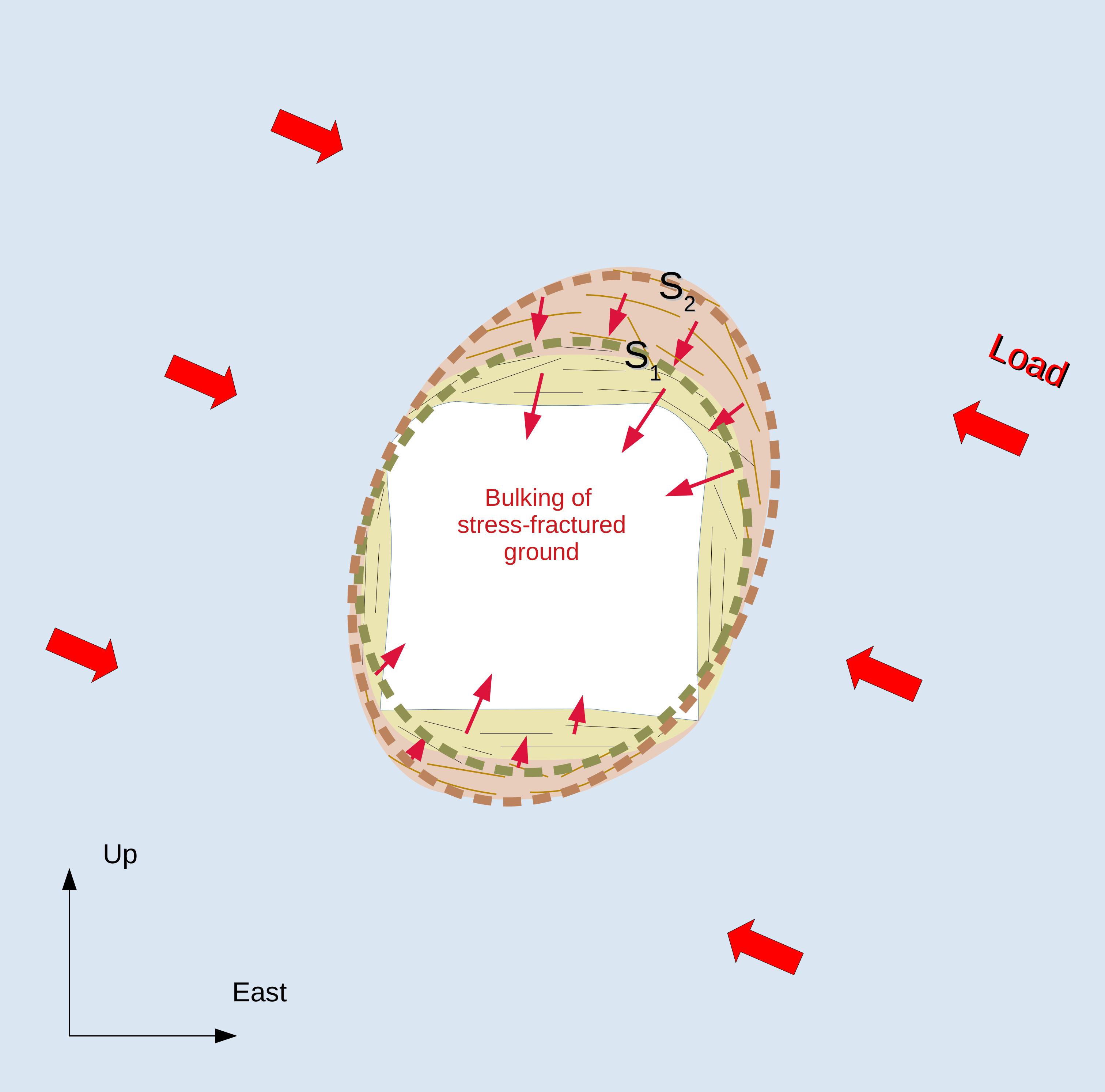}\hspace{1cm}\includegraphics[width=0.45\textwidth]{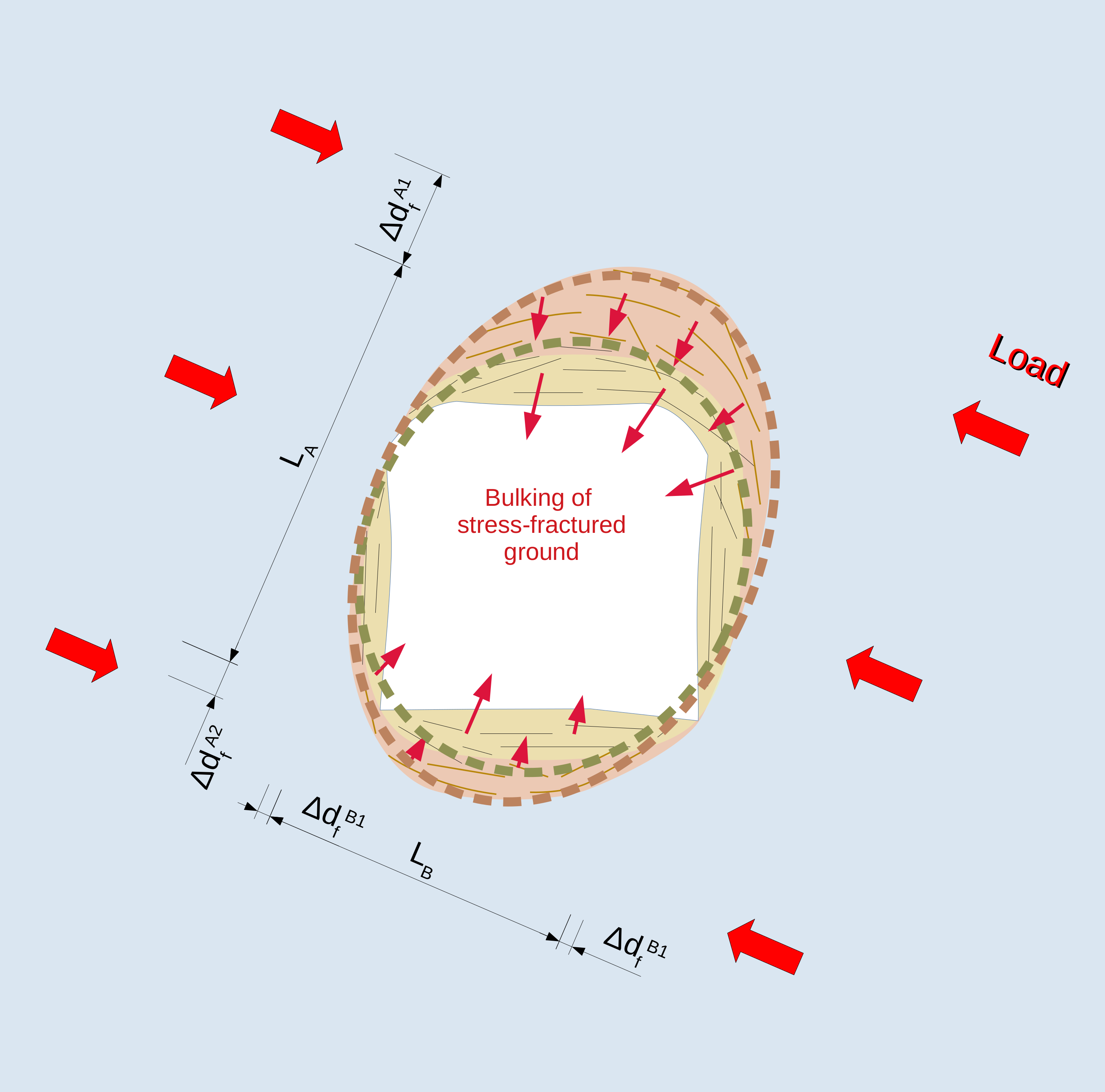}
\par\end{centering}
\caption{\label{fig:Cavity-expansion-and-parametrisation}Approximation of
a tunnel and its damage zone by an elliptical cavity \textit{(left)}.
Geometrical parameters describing the model \textit{(right)}.}
\end{figure}

A tunnel and its associated damaged zone can be approximated by a
stress-free elliptical cavity in an elastic medium. An example of
such an approximation is shown in Figure \ref{fig:Cavity-expansion-and-parametrisation}
for the conceptual model discussed in Subsection \ref{subsec:Conceptual-model}.
An appropriate cavity should, when loaded in the same manner as the
tunnel itself, reproduce the major characteristics of stress and strain
outside the tunnel's damaged zone. Obviously, such a cavity cannot
replicate all details of the actual stress and strain fields; for
example, some parts of the damaged zone or even the interior of the
tunnel can be located outside the cavity as shown in Figure \ref{fig:Cavity-expansion-and-parametrisation},
so modelled stress and strain in these areas will not be accurate.
However, we shall show that such small-scale imperfections do not
prohibit the selection of cavities that yield reasonable approximations
of the seismic source mechanisms described by Equation (\ref{eq:kirchhoff}).

The discussed problem of stress fracturing around tunnels requires
two elliptical cavities: one for the initial state prior to the expansion
of the damaged region (the yellow zone in Figure \ref{fig:Cavity-expansion-and-parametrisation})
and another for the state after this expansion (the red region in
Figure \ref{fig:Cavity-expansion-and-parametrisation}). Analytical
expressions for stress and displacement around a two-dimensional elliptical
cavity are presented in \cite{Maugis-1992} for plane stress and plane
strain (the latter being used here) and can be utilized in the Kirchhoff-type
moment tensor definition of Equation (\ref{eq:kirchhoff}) by selecting
any contour $S$ that encloses both ellipses. Note that we constrain
the north-south extent of the surface $S$ in the same manner as described
in Subsection \ref{subsec:Cases}; that is, we effectively take a
$L_{3}=\unit[5]{m}$ slice of the plane strain solution, which implies
infinite extent of damage along the tunnel's axis.

\subsection{\label{subsec:Numerical-approximation}Numerical approximation}

\begin{figure}
\begin{centering}
\includegraphics[width=0.3\textwidth]{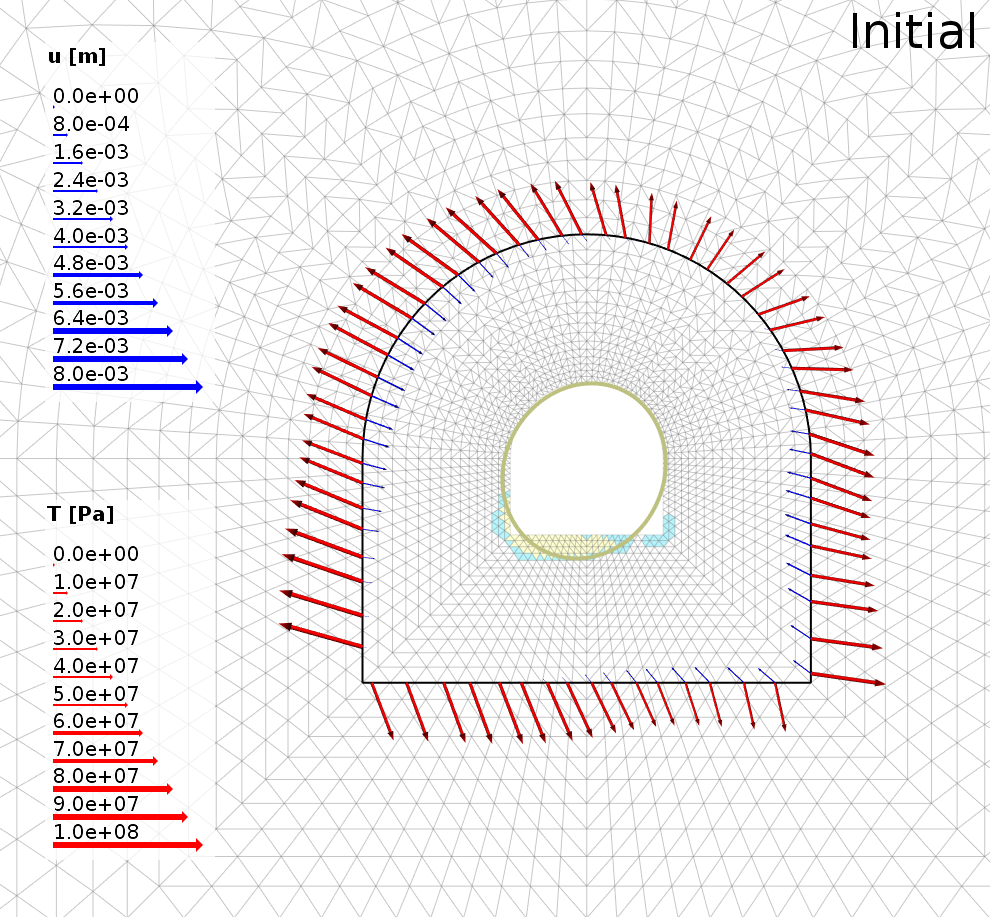}\hspace{0.02\textwidth}\includegraphics[width=0.3\textwidth]{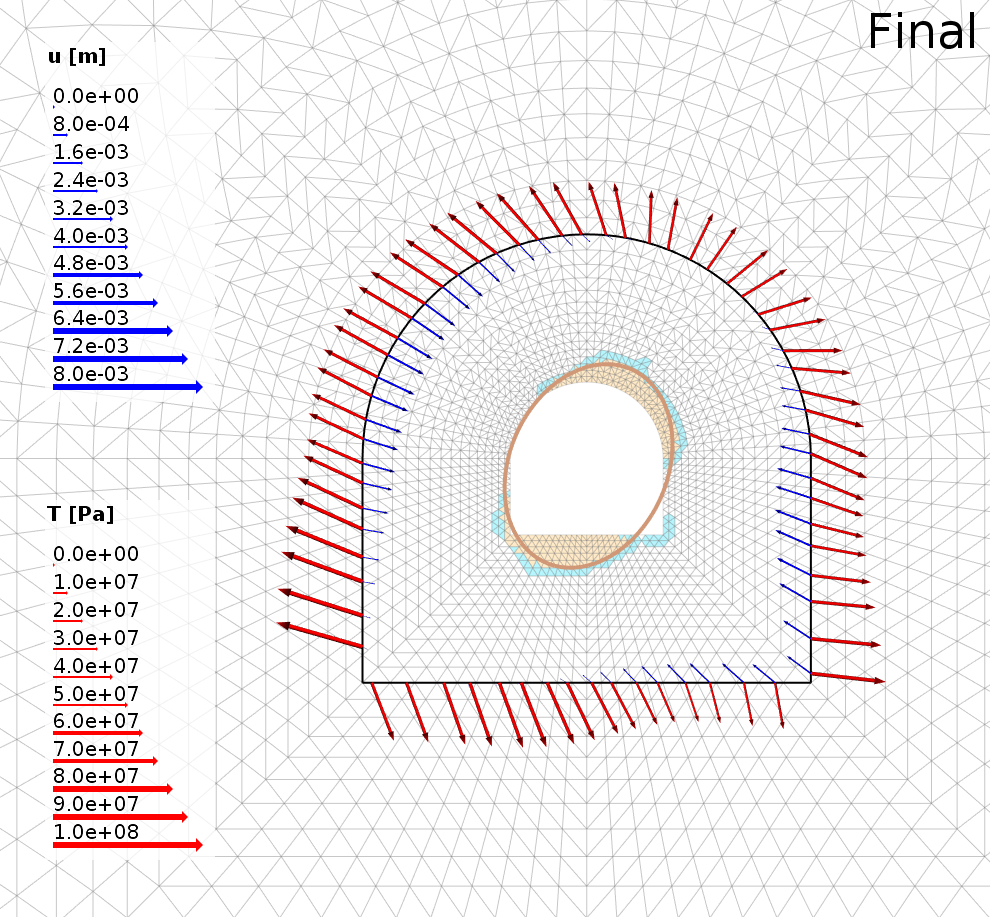}\hspace{0.02\textwidth}\includegraphics[width=0.3\textwidth]{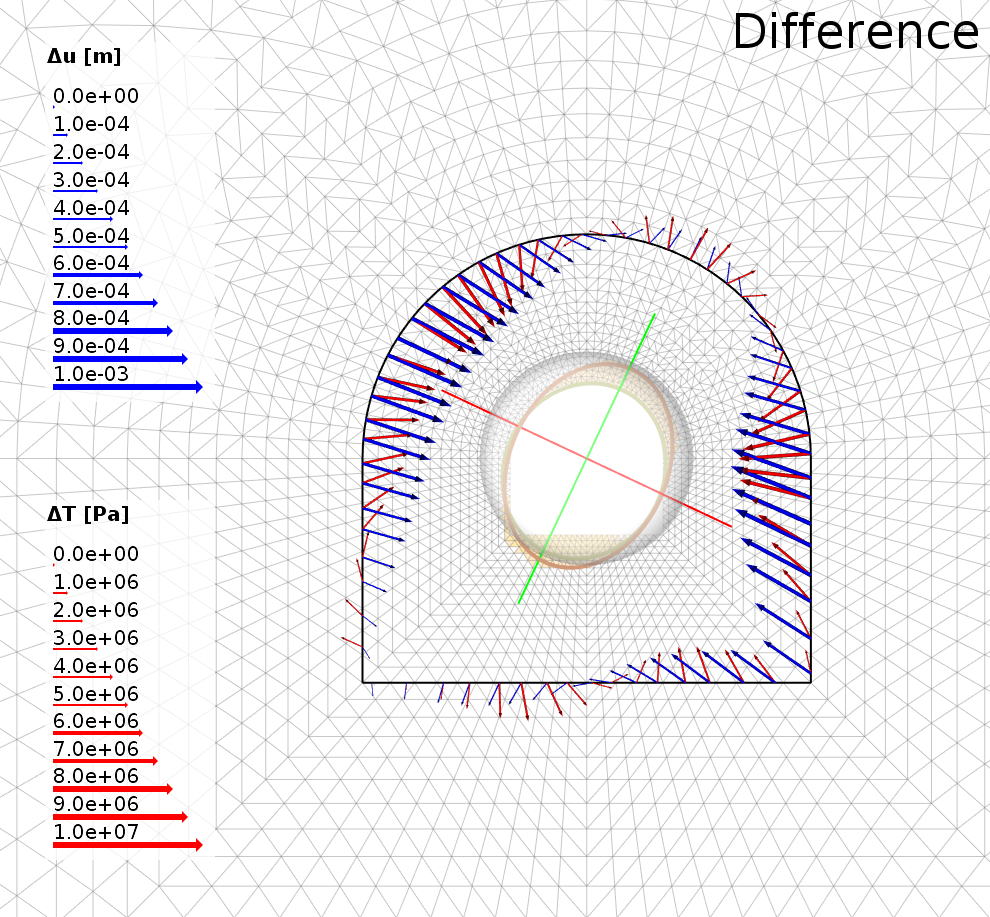}
\par\end{centering}
\caption{\label{fig:Approximation-ellipses-case3}Illustration of the seismic
mechanism calculation for Case 3 using fitted ellipses and Equation
(\ref{eq:kirchhoff}). Descriptions are provided in the text.}
\end{figure}

For the six cases considered in Section \ref{sec:Modelling}, we select
approximating ellipses based on the depth of failure. In practice,
the depth of failure can be evaluated through visual observation in
boreholes or through the use of empirical relations to the magnitudes
of principal stresses and the uniaxial compressive strength of the
rockmass \cite{Cai-Kaiser-2018,Kaiser-Cai-2021}. We follow \cite{perras2016predicting}
in taking the depth of failure to be defined by the boundary between
the inner and outer excavation damage zones, which is identified by
the transition of tensile to compressive volumetric strain in the
failed region. These tensile and compressive zones are shown in yellow/orange
and blue, respectively, in Figure \ref{fig:Approximation-ellipses-case3}
for Case 3. The area defined by the tunnel's profile and the tensile
failed region is used to determine an ellipse based on moments of
inertia \cite{Jain-1989}. In particular, the center of the ellipse
coincides with the area's center of mass, the orientation is taken
to be the angle minimizing the area's moment of inertia, and the major
and minor semi-axes are selected such that the area and the ellipse
have the same second moments of inertia. The ellipses derived for
Case 3 are overlaid in Figure \ref{fig:Approximation-ellipses-case3}
along with displacement and traction computed on the same contour
used in Section \ref{subsec:Results}, which are adopted in Equation
\ref{eq:kirchhoff} to determine the mechanism shown.

Plots summarizing the elliptical approximation of the expansion of
the damaged region are given in Appendix \ref{sec:Appendix-1:-Details}
and results of the corresponding seismic moment calculations are summarized
in Tables \ref{tab:Results-M0} and \ref{tab:Results-Hudson+stereo}
(listed as $N$). Comparison with the respective Kirchhoff-type solutions
of Subsection \ref{subsec:Results} indicates that the suggested approximation
as the expansion of an elliptical cavity is reasonable:
\begin{itemize}
\item The seismic moment ratios range from $1.20$ (Case 3) to $1.41$ (Case
5), which corresponds to Hanks-Kanamori moment magnitude differences
of $0.05$ and $0.10$, respectively. This slight overestimation is
possibly a result of the approach taken to selecting approximating
ellipses.
\item In most case, the source types for the suggested approximations deviate
less from the Kirchhoff-type solution ($K$), than the adjusted conventional
solution ($C$) does. The difference does not exceed $0.03$ and $0.11$
for the $k$ (strength of the isotropic component) and $T$ (deviation
of the deviatoric component from a double-couple model) parameters,
respectively.
\item The maximum difference in the orientation of the P-axis from the Kirchhoff-type
solutions does not exceed $2.8^{\circ}$. There is much larger variation
in B- and T-axis orientation; however, this is to be expected given
the previously discussed near degeneracy of the corresponding eigenvalues.
\end{itemize}
These deviations from the Kirchhoff-type solution are all within the
typical uncertainty of source mechanisms evaluated from real data.

\subsection{\label{subsec:Analytical-approximation}Analytical approximation}

The results presented in Subsection \ref{subsec:Numerical-approximation}
demonstrate that instances of dynamic stress fracturing around tunnels
can be modelled as the expansion of an elliptical cavity. However,
the ellipses used were derived from the results of numerical modelling
of stresses and strains, and the calculation of the corresponding
source mechanisms required numerical integration. In this subsection,
we present a parameterization of the elliptical cavity expansion and
a corresponding analytical expression for approximating the source
mechanism.

Dynamic stress fracturing around a tunnel can be represented as a
sudden increase $\Delta d_{f}$ in the depth of failure. As shown
in Figure \ref{fig:Cavity-expansion-and-parametrisation}, the original
and expanded elliptical cavities can be parameterized in terms of
this increase:
\begin{itemize}
\item We assume that the original and expanded cavities have minor axes
aligned with the direction of loading. This is consistent with stress
being concentrated in the tunnel sides orthogonal to loading, which
facilitates growth of the damaged zone.
\item The lengths of the major and minor axes of the initial ellipse are
denoted as $L_{A}$ and $L_{B}$, respectively. These represent the
effective dimensions of the tunnel including its pre-existing damaged
zone.
\item The lengths of the post-expansion axes are $L_{A}+\Delta d_{f}^{A}$
and $L_{B}+\Delta d_{f}^{B}$ , where $\Delta d_{f}^{A}=\Delta d_{f}^{A1}+\Delta d_{f}^{A2}$
and $\Delta d_{f}^{B}=\Delta d_{f}^{B1}+\Delta d_{f}^{B2}$ are the
increases in depth of failure in the directions of the major and minor
axes, respectively.
\item The extension of both the pre- and post-expansion cavities along the
tunnel's axis is $L_{3}$.
\end{itemize}
In terms of the parameterization outlined above, it is possible to
derive an approximate analytical expression for the source mechanism
resulting from an elliptical cavity expansion. This is detailed in
Appendix \ref{sec:Appendix-2:-Transformations}, with the resulting
expression being
\begin{equation}
M_{ij}=\begin{cases}
C_{M}\left[\frac{\pi}{2}\nu+(1-\nu)C_{1}-C_{2}\right] & \mathrm{if\,}i,j=1,\\
C_{M}\left[\frac{\pi}{2}(1-\nu)+\nu C_{1}+C_{2}\right] & \mathrm{if\,}i,j=2,\\
C_{M}\left[\frac{\pi}{2}\nu+\nu C_{1}\right] & \mathrm{if\,}i,j=3,\\
0 & \mathrm{otherwise},
\end{cases}\label{eq:Definition=0000233d}
\end{equation}
where
\begin{align*}
C_{M}= & 2\frac{1-\nu}{1-2\nu}\sigma_{\max}L_{3}\overline{L_{A}}\triangle d_{f}^{A},\\
C_{1}= & \frac{\pi}{2}\frac{1}{k_{\sigma}}\frac{\triangle d_{f}^{B}}{\triangle d_{f}^{A}}\frac{\overline{L_{B}}}{\overline{L_{A}}},\\
C_{2}= & \frac{\pi}{8}(1-2\nu)(1-\frac{1}{k_{\sigma}})(\frac{\overline{L_{B}}}{\overline{L_{A}}}+\frac{\triangle d_{f}^{B}}{\triangle d_{f}^{A}}),
\end{align*}

\noindent $\overline{L_{A}}=L_{A}+\triangle d_{f}^{A}/2$ and $\overline{L_{B}}=L_{B}+\triangle d_{f}^{B}/2$
are the mean of pre- and post-event effective tunnel dimensions, $\sigma_{\mathrm{max}}$
and $\sigma_{\mathrm{min}}$ are the maximum and minimum principal
stresses orthogonal to the tunnel's axis, $\nu$ is Poisson's ratio
for the rockmass, and $k_{\sigma}=\sigma_{\mathrm{max}}/\sigma_{\mathrm{min}}$.
Note that this expression is in a coordinate system defined by the
tunnel and its loading, with $\hat{\boldsymbol{x}}_{1}$ in the direction
of $\sigma_{\mathrm{min}}$, $\hat{\boldsymbol{x}}_{2}$ in the direction
of $\sigma_{\mathrm{max}}$, and $\hat{\boldsymbol{x}}_{3}$ directed
along the tunnel's axis.

To verify the analytical approximations of Equation (\ref{eq:Definition=0000233d}),
we have parameterized the ellipses fitted in Subsection \ref{subsec:Numerical-approximation}
in terms of $L_{A}$, $L_{B}$, $\Delta d_{f}^{A}$, and $\Delta d_{f}^{B}$
as listed in Table \ref{tab:Ellipses-param}. The remaining input
parameters have been given in Section \ref{subsec:Cases}. The approximate
seismic moments calculated using Equation (\ref{eq:Definition=0000233d})
are summarized in Tables \ref{tab:Results-M0} and \ref{tab:Results-Hudson+stereo}.
There is a similar degree of agreement with the Kirchhoff-type solutions
of Subsection \ref{subsec:Results} to that observed for the cavity
expansion solutions relying on numerical integration presented in
Subsection \ref{subsec:Numerical-approximation}:
\begin{itemize}
\item The seismic moment ratios range from $1.21$ (Case 2) to $1.40$ (Case
4), which corresponds to Hanks-Kanamori moment magnitude differences
of $0.05$ and $0.10$, respectively.
\item In most cases, the source types for the suggested approximations do
not deviate significantly further for the Kirchhoff-type solution
($K$) than the numerical solutions ($N$) do.
\item The principal axes for the analytical solutions are fixed by the loading
and tunnel orientation. The maximum difference in the orientation
of the P- axis from the Kirchhoff-type solutions does not exceed $18^{\circ}$.
\end{itemize}
Again, these deviations from the Kirchhoff-type solution are all within
the typical uncertainty of source mechanisms evaluated from real data.

\textcolor{black}{}
\begin{table}[H]
\centering{}\textcolor{black}{\caption{\textcolor{black}{\label{tab:Ellipses-param}Geometric parameters
of effective elliptical cavities fitted in Subsection }\ref{subsec:Numerical-approximation}\textcolor{black}{.}}
\medskip{}
}%
\begin{tabular}{ccccccc}
\hline 
\textcolor{black}{\small{}Parameter} & \textcolor{black}{\small{}Case 1} & \textcolor{black}{\small{}Case 2} & \textcolor{black}{\small{}Case 3} & \textcolor{black}{\small{}Case 4} & \textcolor{black}{\small{}Case 5} & \textcolor{black}{\small{}Case 6}\tabularnewline
\hline 
{\small{}$\unit[L_{A}]{[m]}$} & {\small{}5.54} & {\small{}6.10} & {\small{}5.85} & {\small{}7.19} & {\small{}6.65} & 5.85\tabularnewline
{\small{}$\unit[L_{B}]{[m]}$} & {\small{}5.34} & {\small{}5.41} & {\small{}5.21} & {\small{}5.40} & {\small{}5.17} & 5.21\tabularnewline
{\small{}$\unit[\Delta d_{f}^{A}]{[m]}$} & {\small{}2.08} & {\small{}1.53} & {\small{}1.07} & {\small{}0.39} & {\small{}0.36} & 1.08\tabularnewline
{\small{}$\unit[\Delta d_{f}^{B}]{[m]}$} & {\small{}-0.16} & {\small{}0.02} & {\small{}-0.05} & {\small{}-0.03} & {\small{}0.05} & 0.03\tabularnewline
\hline 
\end{tabular}
\end{table}

\begin{figure}[H]
\begin{centering}
\includegraphics[width=0.8\textwidth]{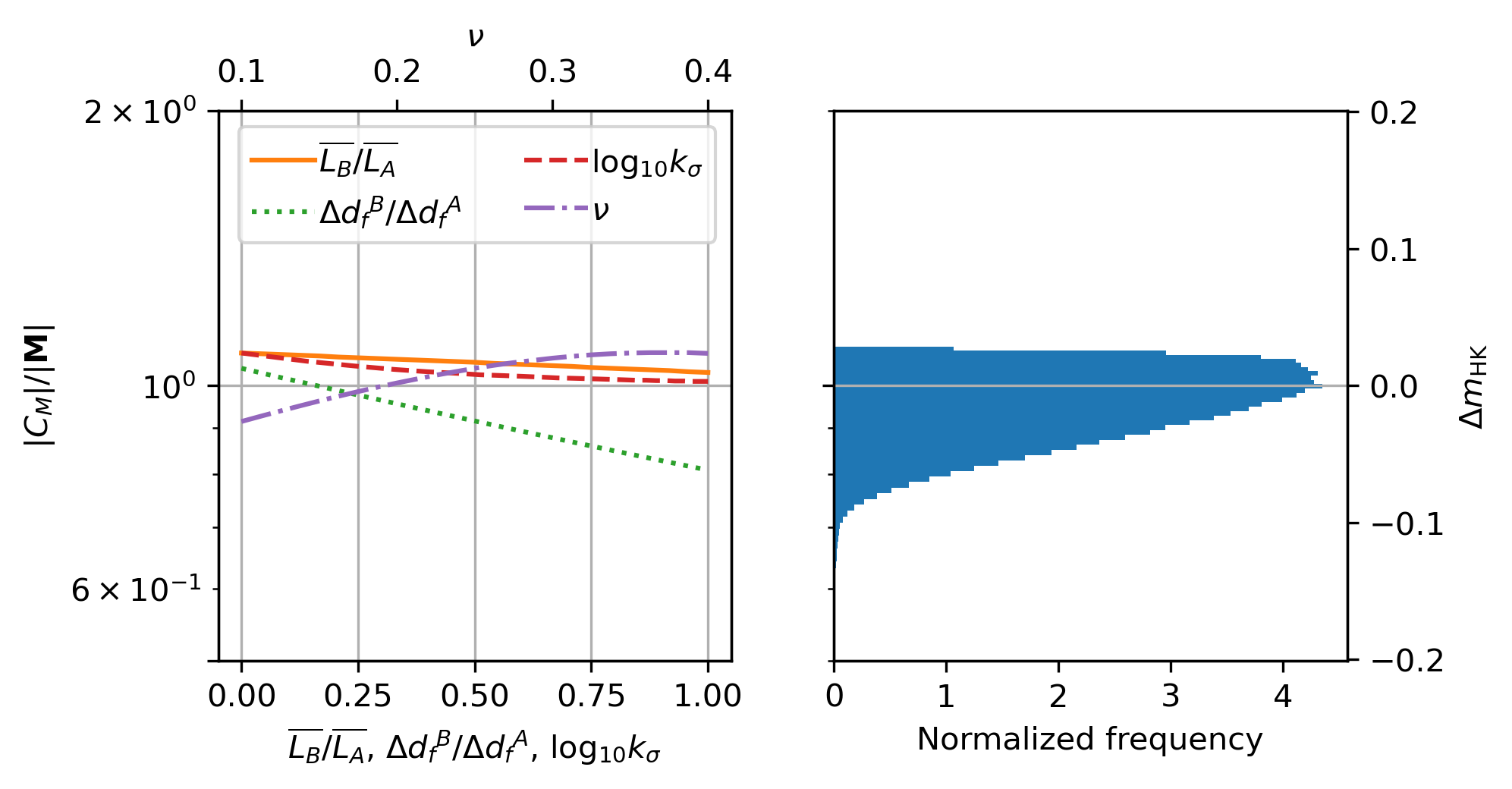}
\par\end{centering}
\caption{\label{fig:M0-approx}Approximation of scalar moment $\left|\mathbf{M}\right|$
for the mechanism described in (\ref{eq:Definition=0000233d}) by
the parameter $C_{M}$. The effect of varying one of $\overline{L_{B}}/\overline{L_{A}}$,
$\triangle d_{f}^{B}/\triangle d_{f}^{A}$, $k_{\sigma}$, or $\nu$
with respect to the reference mechanism ($\overline{L_{B}}/\overline{L_{A}}=0.8$,
$\triangle d_{f}^{B}/\triangle d_{f}^{A}=0$, $k_{\sigma}=2$, and
$\nu=0.25$) on the ratio $|C_{M}|/\left|\mathbf{M}\right|$ \textit{(left)}.
Histogram of $|C_{M}|/\left|\mathbf{M}\right|$ values resulting from
sampling $\overline{L_{B}}/\overline{L_{A}}$, $\triangle d_{f}^{B}/\triangle d_{f}^{A}$,
$\log_{10}k_{\sigma}$, and $\nu$ uniformly over the same ranges
\textit{(right)}. The right vertical axis shows the difference between
$\left|\mathbf{M}\right|$ and $|C_{M}|$ expressed in terms of Hanks-Kanamori
moment magnitude.}
\end{figure}

The scalar seismic moment $\left|\mathbf{M}\right|=\sqrt{(M_{11}^{2}+M_{22}^{2}+M_{33}^{2})/2}$
for the expression of Equation (\ref{eq:Definition=0000233d}) depends
on $C_{M}$, ratios of geometric parameters ($\overline{L_{B}}/\overline{L_{A}}$
and $\triangle d_{f}^{B}/\triangle d_{f}^{A}$), the ratio of principal
stresses orthogonal to the tunnel's axis ($k_{\sigma}$), and the
material's elastic properties ($\nu$). However, it can be seen in
Table \ref{tab:Results-M0} that $\left|\mathbf{M}\right|\approx|C_{M}|$
across all six cases. To show that this agreement holds across a wide
range of reasonable inputs, we consider variants on a ``reference''
mechanism with $\overline{L_{B}}/\overline{L_{A}}=0.8$, $\triangle d_{f}^{B}/\triangle d_{f}^{A}=0$,
$k_{\sigma}=2$, and $\nu=0.25$. As is shown in the left of Figure
\ref{fig:M0-approx}, individually varying any of these four parameters
over a wide range has little effect on the ratio $|C_{M}|/\left|\mathbf{M}\right|$
or difference in moment magnitude $\Delta m_{\mathrm{HK}}$. These
ranges and the corresponding bounds on $|C_{M}|/\left|\mathbf{M}\right|$
and $\Delta m_{\mathrm{HK}}$ are listed in Table \ref{tab:Maximum/minimum-ratios-}.
The effect of varying multiple parameters has also been investigated
by uniformly sampling $\overline{L_{B}}/\overline{L_{A}}$, $\triangle d_{f}^{B}/\triangle d_{f}^{A}$,
$\log_{10}k_{\sigma}$, and $\nu$ uniformly over the same ranges
to produce the histogram on the right of Figure \ref{fig:M0-approx}.
As listed in Table \ref{tab:Maximum/minimum-ratios-}, this still
results in relatively stable values of $0.55<|C_{M}|/\left|\mathbf{M}\right|<1.10$
and $-0.17<\Delta m_{\mathrm{HK}}<0.03$. As such, 
\begin{equation}
\left|\mathbf{M}\right|\approx2\frac{1-\nu}{1-2\nu}|\sigma_{\max}|L_{3}\overline{L_{A}}\triangle d_{f}^{A}\label{eq:M0-approx}
\end{equation}
can be used as a simple yet quite accurate approximation to the scalar
moment that can be derived from the full moment tensor solution of
Equation (\ref{eq:Definition=0000233d}). In the case of $\nu\approx0.25$,
this reduces to the even simpler $\left|\mathbf{M}\right|\approx3\left|\sigma_{\max}\right|L_{3}\overline{L_{A}}\triangle d_{f}^{A}$.

\begin{table}[H]
\centering{}\caption{\label{tab:Maximum/minimum-ratios-}Maximum/minimum ratios $|C_{M}|/\left|\mathbf{M}\right|$
and moment magnitude differences $\Delta m_{\mathrm{HK}}$ for mechanisms
varying from the reference case ($\overline{L_{B}}/\overline{L_{A}}=0.8$,
$\triangle d_{f}^{B}/\triangle d_{f}^{A}=0$, $k_{\sigma}=2$, and
$\nu=0.25$) by the listed parameter(s).}
\textcolor{black}{\medskip{}
}%
\begin{tabular}{cccccc}
\toprule 
Parameter varied & $\overline{L_{B}}/\overline{L_{A}}$ & $\triangle d_{f}^{B}/\triangle d_{f}^{A}$ & $k_{\sigma}$ & $\nu$ & All\tabularnewline
\midrule
Parameter minimum & 0 & 0 & 1 & 0.15 & -\tabularnewline
Parameter maximum & 1 & 1 & 10 & 0.4 & -\tabularnewline
Minimum $|C_{M}|/\left|\mathbf{M}\right|$ & 1.03 & 0.81 & 1.01 & 0.91 & 0.55\tabularnewline
Maximum $|C_{M}|/\left|\mathbf{M}\right|$ & 1.08 & 1.05 & 1.09 & 1.09 & 1.10\tabularnewline
Minimum $\Delta m_{\mathrm{HK}}$ & 0.01 & -0.06 & 0.00 & -0.03 & -0.17\tabularnewline
Maximum $\Delta m_{\mathrm{HK}}$ & 0.02 & 0.01 & 0.02 & 0.02 & 0.03\tabularnewline
\bottomrule
\end{tabular}
\end{table}

\subsection{Depth of failure inversion}

Rearranging Equation (\ref{eq:M0-approx}) yields
\begin{equation}
\Delta d_{f}^{A}=\sqrt{L_{A}^{2}+\frac{1-2\nu}{1-\nu}\frac{\left|\mathbf{M}\right|}{|\sigma_{\max}|L_{3}}}-L_{A}.\label{eq:dof-increase}
\end{equation}
In practice, this equation can be used to infer the depth of failure
increase $\Delta d_{f}^{A}$ from the observed seismic moment $\left|\mathbf{M}\right|$
assuming knowledge (or reasonable estimates) of $L_{A}$, $\nu$,
$|\sigma_{\max}|$, and $L_{3}$. As a demonstration of this process
and to give an idea of the uncertainties involved, we have done this
for the six modelled cases considered. In each case, we take $\left|\mathbf{M}\right|$
as listed in the Kirchhoff column of Table \ref{tab:Results-M0} with
an uncertainty of $\pm40\%$ (corresponding to roughly $\pm0.1$ $m_{\mathrm{HK}}$),
$\unit[L_{A}=6]{m\,\pm20\%}$, $v=0.25\pm10\%$, $\unit[L_{A}=5]{m\,\pm20\%}$,
and $|\sigma_{\max}|=k|\sigma_{\mathrm{min}}|\pm10\%$. The inferred
values obtained from Equation \ref{eq:dof-increase} are listed in
Table \ref{tab:Inferred-and-measured}. For comparison, ``measured''
depth of failure increases are also listed in Table \ref{tab:Inferred-and-measured},
which have been determined based on the depth of failure definition
discussed in Section \ref{subsec:Numerical-approximation}. In particular,
the depth of failure increase has been averaged within a $45^{\circ}$
window centered at the direction of minor loading (that is, $22.5^{\circ}$
either side). This procedure has been chosen eliminate any impact
from the method of selecting approximating ellipses and to more closely
match the methodology of depth of failure measurement using boreholes
(we note, however, that the values do not vary significantly from
those derived from approximating ellipses as listed in Table \ref{tab:Maximum/minimum-ratios-}).
In general, it can be seen that the inferred and measured values are
in reasonable agreement given their respective uncertainties.
\begin{table}[H]
\centering{}\caption{\label{tab:Inferred-and-measured}Inferred and measured depth of failure
increases for the six cases modelled.}
\begin{tabular}{ccc}
\toprule 
Case & $\Delta d_{f}^{A}$ inferred {[}m{]} & $\Delta d_{f}^{A}$ measured {[}m{]}\tabularnewline
\midrule
1 & $1.43\pm0.64$ & $1.95\pm0.24$\tabularnewline
2 & $1.26\pm0.57$ & $1.65\pm0.07$\tabularnewline
3 & $0.81\pm0.38$ & $0.93\pm0.29$\tabularnewline
4 & $0.32\pm0.15$ & $0.31\pm0.13$\tabularnewline
5 & $0.30\pm0.15$ & $0.32\pm0.22$\tabularnewline
6 & $0.79\pm0.37$ & $0.91\pm0.30$\tabularnewline
\bottomrule
\end{tabular}
\end{table}

\section{\label{sec:Examples}Examples of Seismic Events from a Mine}

In this section, the source model presented in Subsection \ref{subsec:Analytical-approximation}
is applied to real data, which comes from a seismic monitoring system
operating in a deep underground hard-rock mine in Western Australia.
This system is composed of geophones with natural frequencies of $\unit[4.5]{Hz}$
and $\unit[14]{Hz}$ that are installed and grouted in boreholes away
from excavations. The acquired seismic data is of high quality: many
waveforms are recorded by sites within line-of-sight of the source
(that is, not shielded by excavations) and consist of P- and S-wave
pulses with little coda. As a result, the use of an elastodynamic
Green's function $\mathbf{G}$ for homogeneous isotropic space is
appropriate. Given the system's good coverage of sources and the simplicity
of the recorded waveforms, source mechanisms of events can be analyzed
in detail.

\subsection{Overview\label{subsec:Overview}}

A significant seismic event ($m_{\mathrm{HK}}=2.3$) was recorded
in the deep levels of the mine on 8 June 2019. The event was followed
by a strong aftershock sequence, which included $m_{\mathrm{HK}}=2.0$
and $m_{\mathrm{HK}}=1.7$ events within three minutes of the mainshock
and a $m_{\mathrm{HK}}=1.9$ event approximately 27 hours later. There
was no direct spatial or temporal association of the mainshock with
the excavation of large volumes of rock, with the largest nearby stope
blasting occurring more than four days prior to the event. 

\begin{figure}[H]
\begin{centering}
\includegraphics[width=0.49\textwidth]{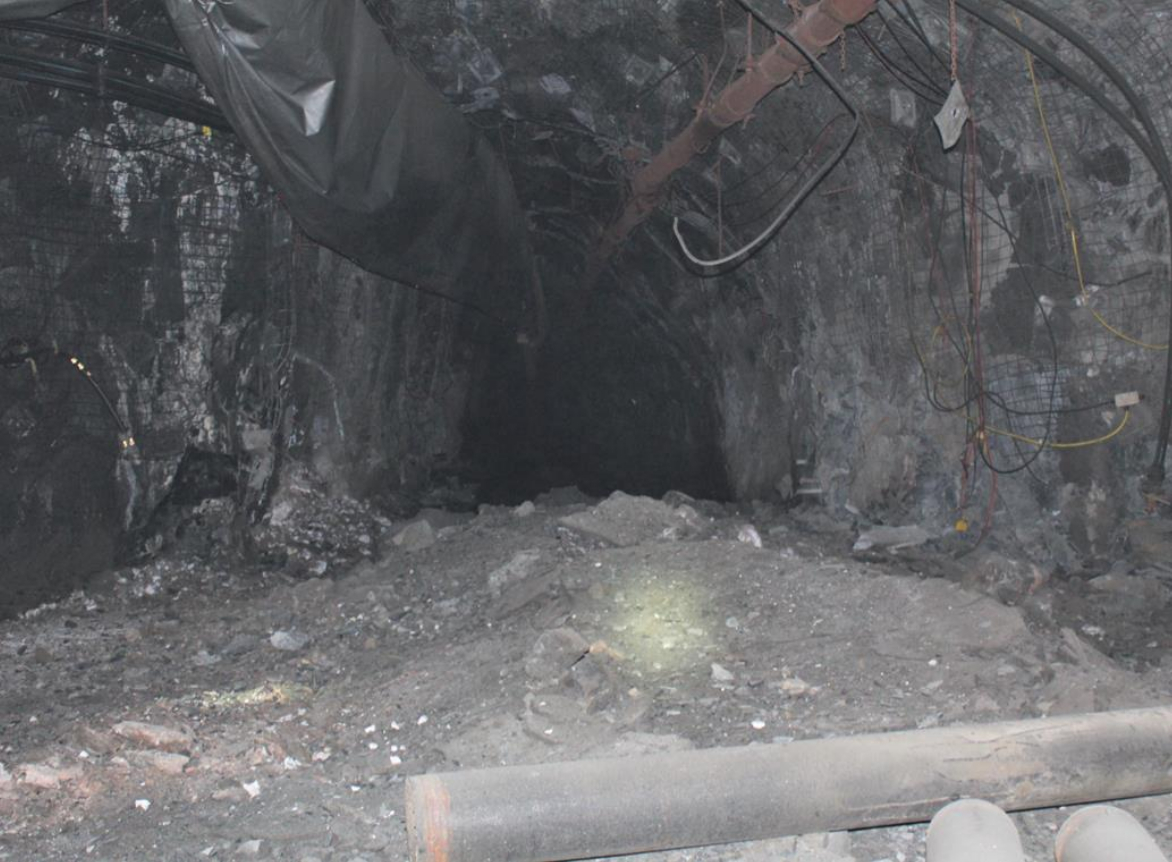}\hfill{}\includegraphics[width=0.49\textwidth]{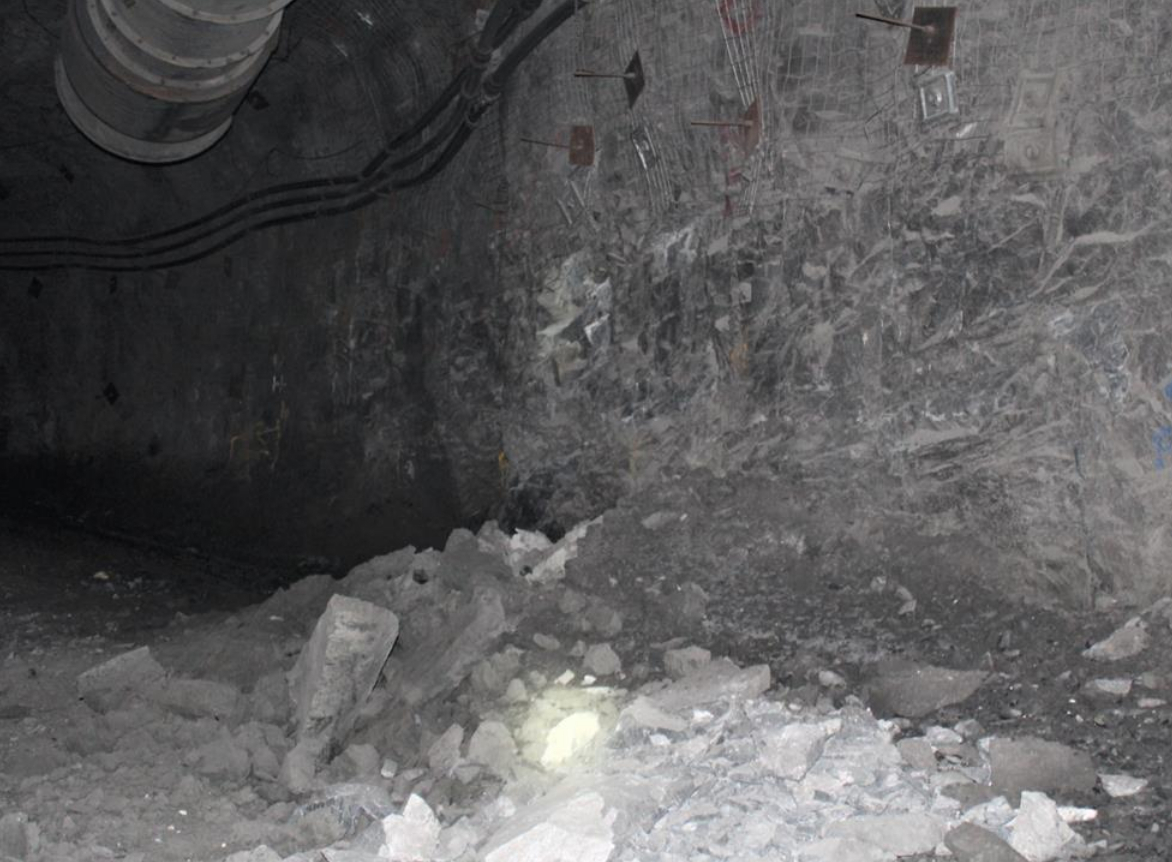}\smallskip{}
\par\end{centering}
\begin{centering}
\includegraphics[width=0.49\textwidth]{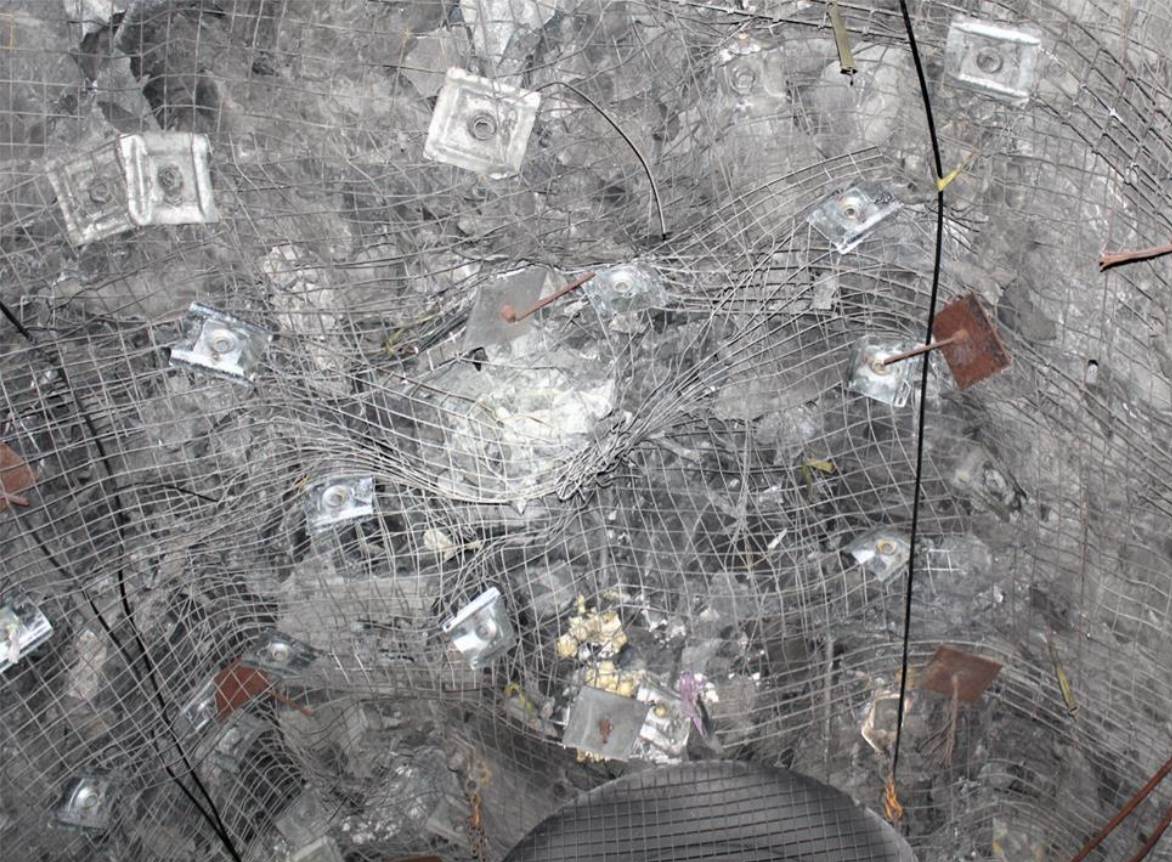}\hfill{}\includegraphics[width=0.49\textwidth]{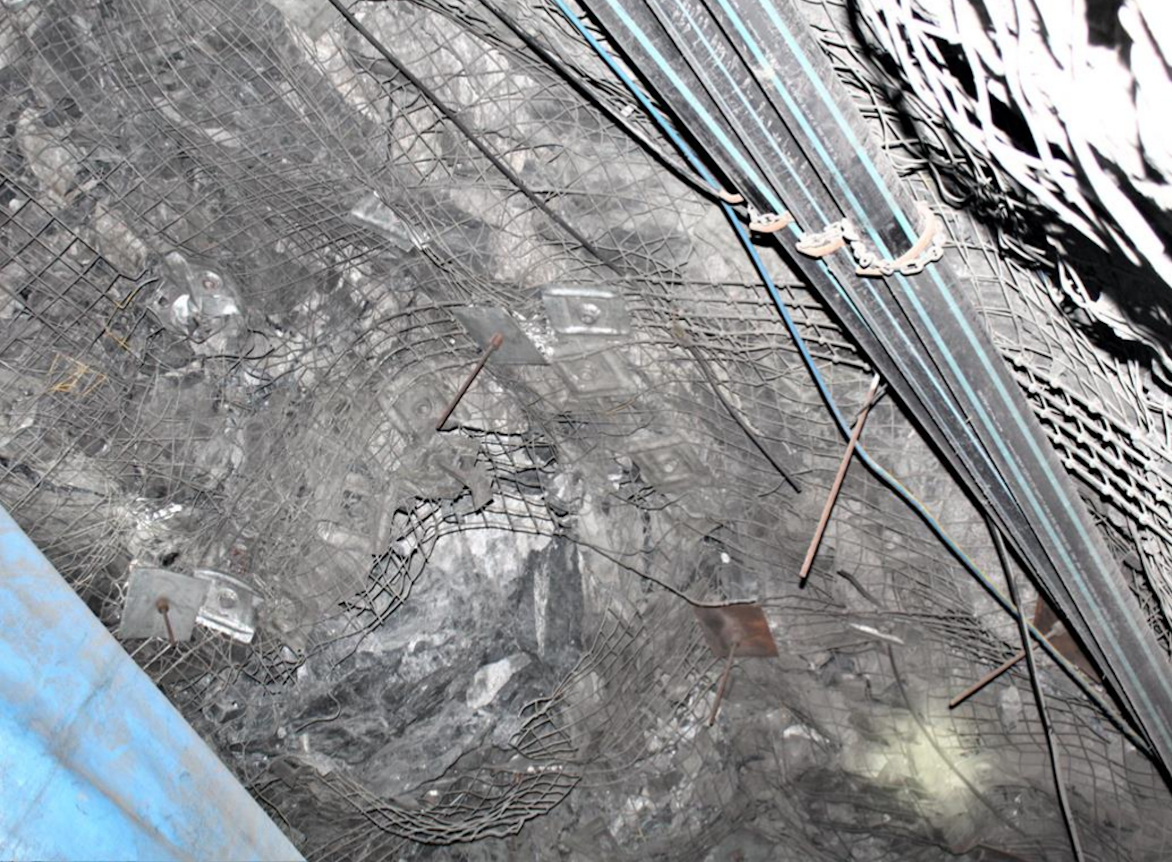}
\par\end{centering}
\caption{\label{fig:Example-damage}Damage observed underground: floor heave
\textit{(top)} and deformation of the supported roof \textit{(bottom)}.}
\end{figure}

After the aftershock activity decayed, the tunnels were inspected.
Damage was observed at four levels, manifesting primarily in the form
of floor heave, an example of which is shown in Figure \ref{fig:Example-damage}.
There were also isolated cases of support deformation and failure
in the roof (back) of the tunnels (bottom of Figure \ref{fig:Example-damage})
and shakedown of rocks from the lower unsupported parts of tunnels.
This association of major damage with floor and backs is consistent
with expectation given that maximum and intermediate (compressional)
principal stresses are subhorizontal.

\begin{figure}[H]
\begin{centering}
\includegraphics[width=1\textwidth]{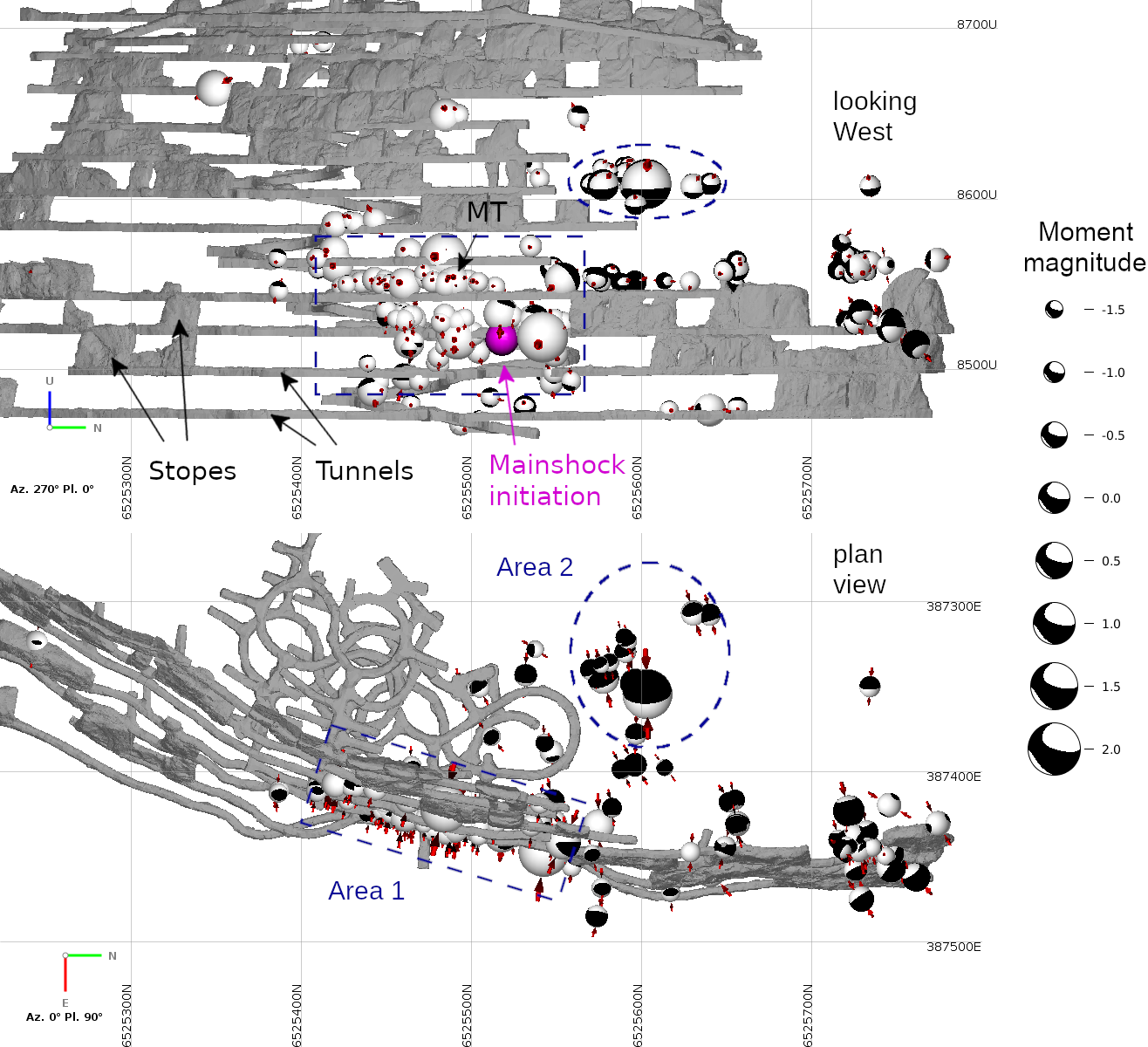}
\par\end{centering}
\caption{\label{fig:Example-mechanisms-view}Analyzed seismicity presented
in a form of beachballs sized according to $m_{\mathrm{HK}}$. Red
dipoles represent the P-axes of source mechanisms. The gray wireframes
describe excavations (tunnels and stopes from which ore has been extracted)
at the time of the analyzed seismic event. The two contoured areas
(1 and 2) are described in the text. The mainshock is represented
by a magenta sphere located at the inferred point of initiation (not
sized by $m_{\mathrm{HK}}$).}
\end{figure}

\begin{figure}[H]
\begin{centering}
\includegraphics[width=1\textwidth]{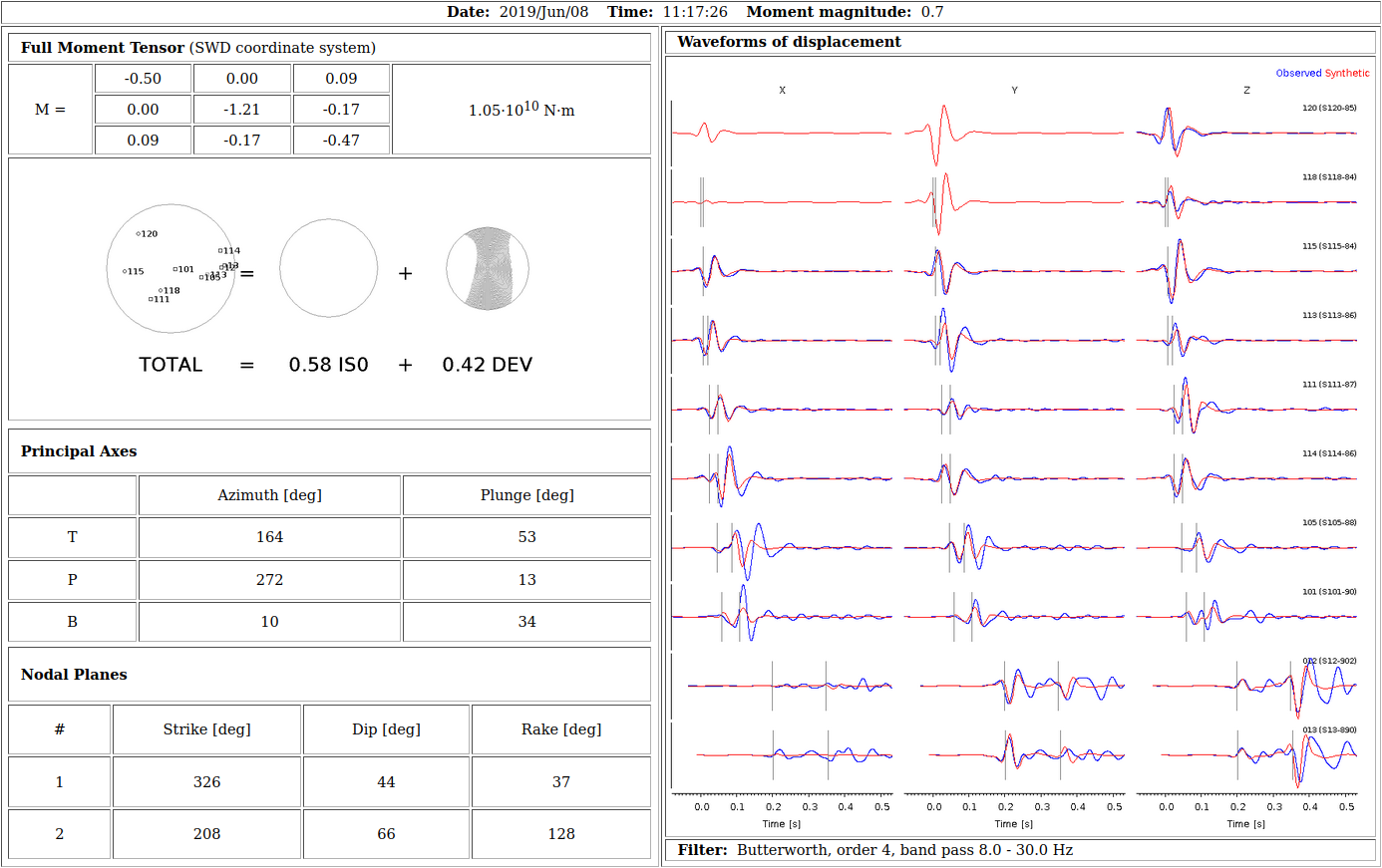}
\par\end{centering}
\caption{\label{fig:Example-MT-report}Results of moment tensor inversion for
a medium-size seismic event marked as ``MT'' in Figure \ref{fig:Example-mechanisms-view}.
Inverted source mechanism of the event \textit{(left)}. Observed waveforms
of displacement are compared with the waveforms modelled for the inverted
moment tensor \textit{(right)}. Sensor response is taken into account
in the synthetic waveforms. A 4th-order Butterworth bandpass 8-30
Hz filter is applied to both observed and synthetic waveforms. Vertical
gray lines mark the arrivals of direct P- and S-waves picked from
unfiltered waveforms.}
\end{figure}

Source mechanisms were calculated for 185 out of 220 events recorded
within 48 hours of the mainshock (including all 19 events with $m_{\mathrm{HK}}>0.0$).
These mechanisms, with the mainshock excluded, are shown in Figure
\ref{fig:Example-mechanisms-view}. The results of full-waveform inversion
for one of these events is summarized in Figure \ref{fig:Example-MT-report},
which demonstrates a good fit between observed and synthetic waveforms.

\begin{figure}[H]
\begin{centering}
\includegraphics[width=1\textwidth]{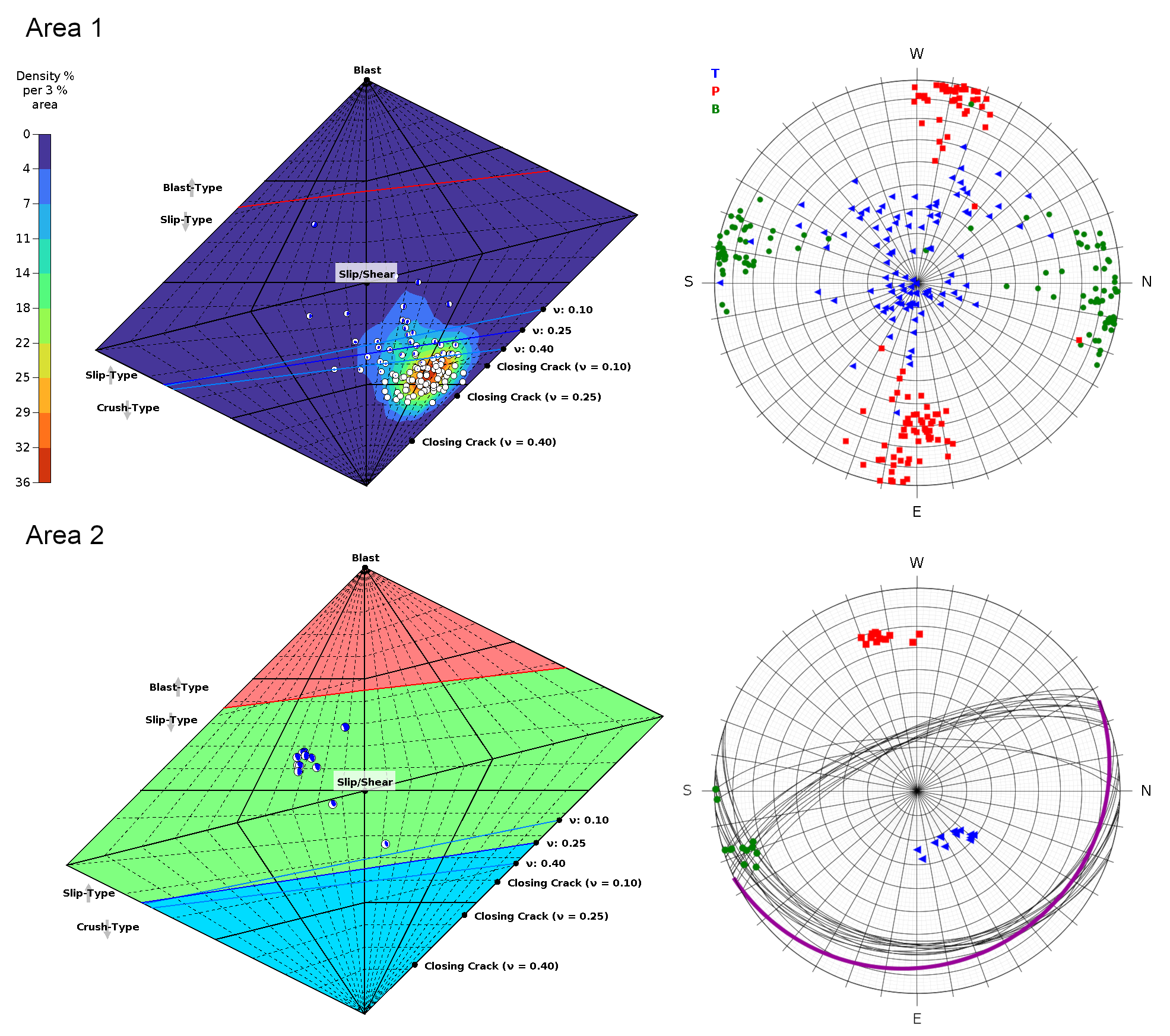}
\par\end{centering}
\caption{\label{fig:Mechanism-characteristics-2-areas}Characteristics of source
mechanisms of events recorded in two areas contoured in Figure \ref{fig:Example-mechanisms-view}.
Note that the stereonets are rotated 90° clockwise to match the view
in Figure \ref{fig:Example-mechanisms-view}. The orientation of the
common location plane of events in the Area 2 is shown in magenta
in the associated stereonet.}
\end{figure}

Spatially, there are several groups of events in the dataset analyzed.
Those defined by Areas 1 and 2 as contoured in Figure \ref{fig:Example-mechanisms-view}
are considered:
\begin{itemize}
\item The events in Area 1, which includes the mainshock, are clustered
around tunnels (mainly ore drives), and the majority of them (80\%)
occurred within 24 hours of the mainshock. In total, 103 mechanisms
were determined for events in this area, which includes the mainshock
and previously mentioned $m_{\mathrm{HK}}=2.0$ and $m_{\mathrm{HK}}=1.7$
aftershocks. As shown in Figure \ref{fig:Mechanism-characteristics-2-areas},
the majority of these mechanisms have a significant implosive component,
a deviatoric component ranging from double couple to pancake-shaped
CLVD, and a principal axis oriented approximately orthogonal to the
direction of the ore drives. The locations and mechanisms of these
events suggest interpretation in terms of the processes discussed
in previous sections; that is, as episodes of dynamic stress fracturing
around tunnels and the associated convergence of surrounding rockmass.
In particular, the loading orientation and distribution of damage
resemble Case 1 as considered in Sections \ref{sec:Modelling} and
\ref{sec:Approximation} (and presented graphically in Appendix \ref{sec:Appendix-1:-Details}).
\item The events in Area 2 locate away from excavations in the footwall
of the orebody and occurred more than 26 hours after the mainshock.
Mechanisms were determined for 12 events in this area, including the
$m_{\mathrm{HK}}=1.9$ event, and all have a significant double-couple
component as shown in Figure \ref{fig:Mechanism-characteristics-2-areas}.
The locations of these sources lie approximately on a common plane
with a dip of $15^{\circ}$ and dip direction of $65^{\circ}$ (strike
of $335^{\circ}$), which is also consistent with the inferred nodal
planes as shown in a stereonet in Figure \ref{fig:Mechanism-characteristics-2-areas}.
It follows that it is appropriate to interpret this cluster of events
as episodes of slip (reverse faulting) along such a plane.
\end{itemize}

\subsection{Analysis of the mainshock\label{subsec:Analysis-of-the}}

\begin{figure}[H]
\begin{centering}
\includegraphics[width=1\textwidth]{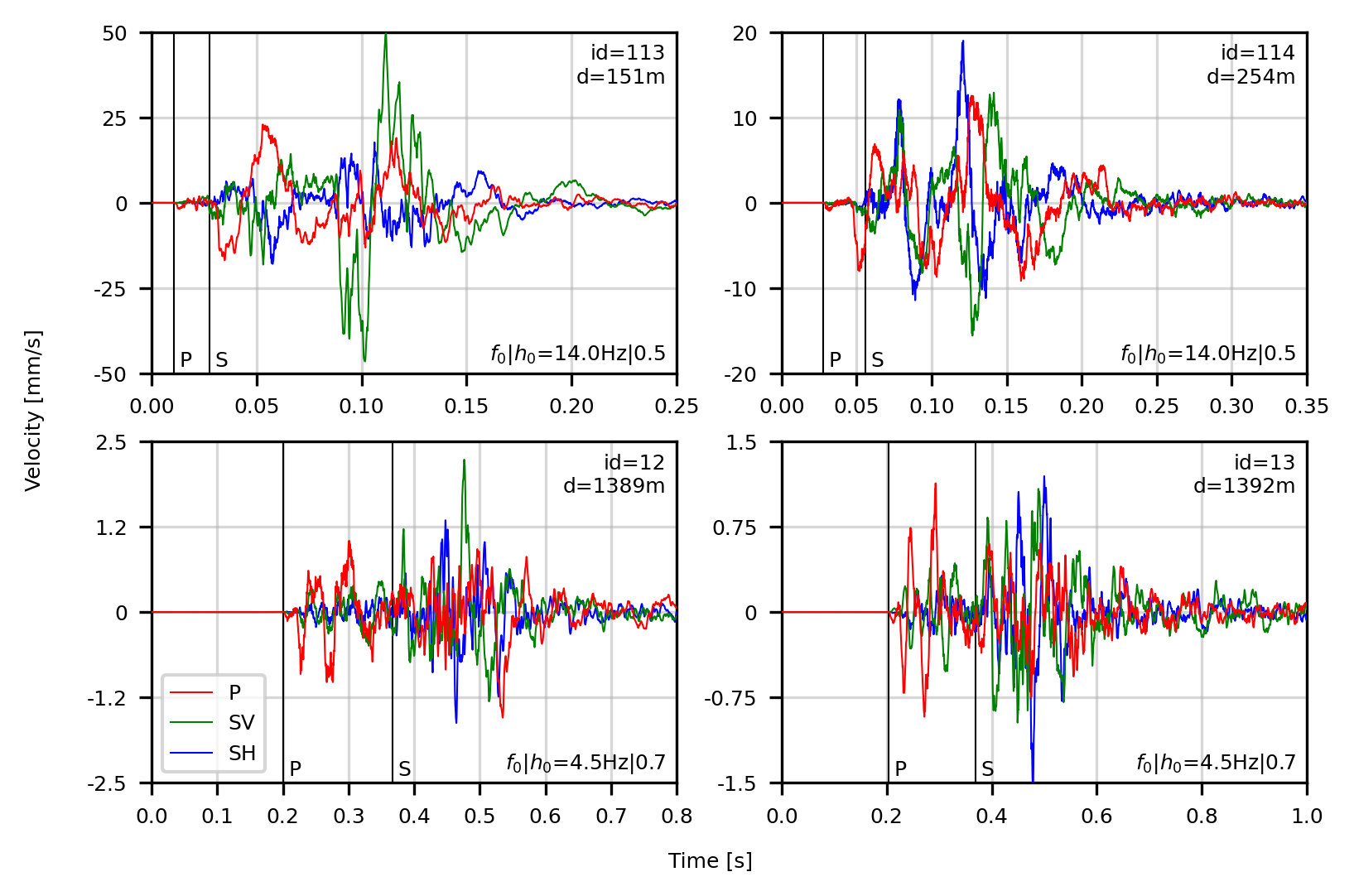}
\par\end{centering}
\caption{\label{fig:Example-mainshock-waveforms}Examples of the mainshock
waveforms as recorded at four representative sites: two located close
to the source \textit{(top)} and two far from the source\textit{ (bottom)}.
Each seismogram is rotated into a local P-SV-SH coordinate system
of site, where the P-axis corresponds to the direction of strongest
motion after the selected P-wave arrival (shown by a vertical line),
while SV- and SH-axes are selected to be orthogonal to P-axis with
the added constraint of the SH-axis being horizontal. Listed in the
top right of each seismogram are site index ($\mathrm{id}$) and distance
$d$ from the point of initiation. The natural frequency $f_{0}$
and damping coefficient $h_{0}$ of the geophones are listed in the
bottom right.}
\end{figure}

As shown in Figure \ref{fig:Example-mainshock-waveforms}, the mainshock
waveforms are complex. This cannot be fully explained simply by the
effects of wave propagation through an inhomogenious medium as smaller
events in the same area yield relatively simple waveforms, which are
dominated by direct P- and S-waves. This indicates that the complexity
results from the mainshock source consisting of multiple episodes
of failure distributed in time (and likely space). Waveforms observed
at distance sites (such as those with indices 12 and 13 in Figure
\ref{fig:Example-mainshock-waveforms}) support this view, with at
least three P-wave pulses visible: an initial wave, which was picked
for location; a stronger one approximately $\unit[35]{ms}$ later;
and one with even greater amplitude approximately $\unit[80]{ms}$
after the initial arrival. Although less clear, similar patterns can
also be seen in the S-waves. At sites closer to the source (such as
those with indices 113 and 114 in Figure \ref{fig:Example-mainshock-waveforms}),
there is greater overlap of P- and S-waves, which makes identification
of individual pulses more difficult.

Interpretation of the mainshock source and its relation to observed
damage is not straightforward. In particular, interpretation in terms
of slip along a plane is not supported by any clear planar structure
in the immediate aftershocks in Area 1 (the inferred weak plane in
Area 2 lies a significant distance from the mainshock's point of initiation,
and only became active 26 hours later; other clusters of aftershocks
dominated by double-couple mechanisms are similarly distant in terms
of space and time). As noted in Subsection \ref{subsec:Overview},
the mechanisms of aftershocks in Area 1 are consistent with dynamic
stress fracturing around tunnels as discussed in Sections \ref{sec:Modelling}
and \ref{sec:Approximation} (significant implosive and pancake-shaped
CLVD components and a P-axis approximately orthogonal to the direction
of the tunnels). This suggests the hypothesis that the mainshock source
is composed of cascading stress fracturing around tunnels, in which
damage around one section of tunnel rapidly transfers load to neighboring
sections (either on the same or adjacent levels), resulting in expansion
of the damaged region. 

\begin{figure}[H]
\begin{centering}
\includegraphics[width=0.9\textwidth]{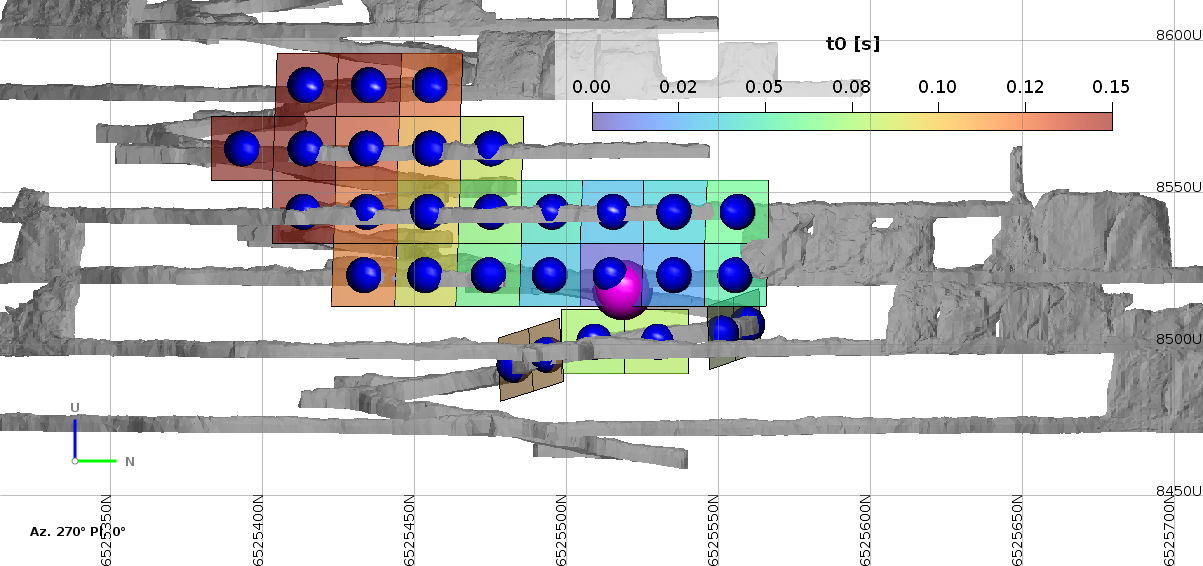}
\par\end{centering}
\caption{\label{fig:Example-mainshock-finite-source-model}Parametrization
of the mainshock finite source. Blue spheres describe 29 point sources
distributed along the tunnels where damage and aftershocks were observed.
Each of these sources describes the convergence and ride of a corresponding
$\mathrm{21\times\unit[21]{m}}$ element. These elements are colored
by delay $t_{0}$ relative to initiation, which occurred at the point
marked by the magenta sphere.}
\end{figure}

To test this hypothesis of cascading stress fracturing, we employ
a variant of the procedure of \cite{Ide-2007} by constructing a distributed
source model based on the locations of damage and immediate aftershocks,
with parameters of the subsources being inverted from velocity waveforms
and interpreted in terms of the model proposed in Section \ref{sec:Approximation}.
The details of the distribution of 29 employed subsources is shown
in Figure \ref{fig:Example-mainshock-finite-source-model}, each of
which corresponds to a $\mathrm{21\times\unit[21]{m}}$ square element
normal to the expected orientation of the maximum principal stress
orthogonal to the tunnel's axis $\sigma_{\mathrm{max}}$. Also shown
is the assumed time evolution of the source: radial propagation from
the point of initiation at an apparent velocity of $\unit[750]{ms^{-1}}$.

\begin{figure}[H]
\begin{centering}
\includegraphics[width=1\textwidth]{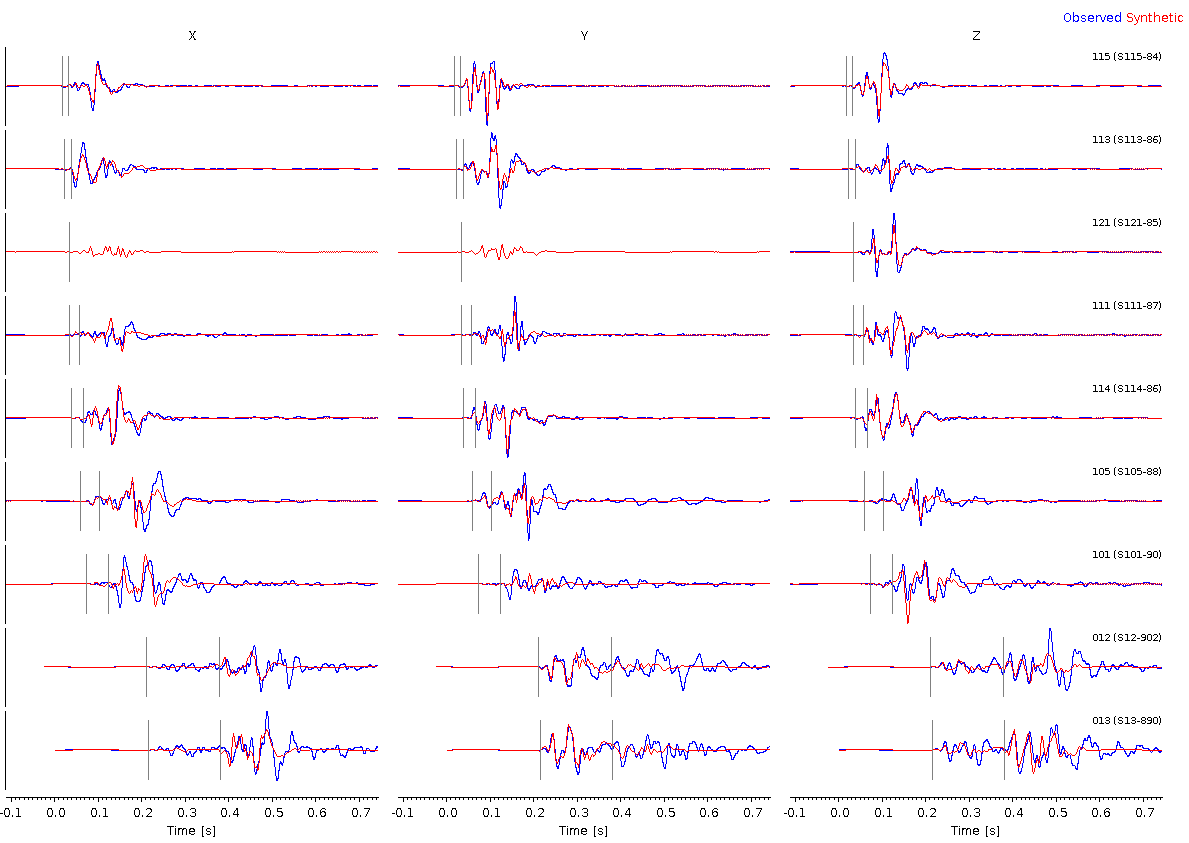}
\par\end{centering}
\caption{\label{fig:Example-mainshock-waveform-fit}Fit of velocity waveforms
obtained from finite-source inversion. Note that for each site, the
waveforms have been normalized by maximum observed amplitude.}
\end{figure}

Inversion of subsource parameters was performed using a variant of
the slip inversion framework outlined in \cite{Olson-Apsel-1982}.
In the original framework, a pure double-couple subsource is assumed,
with inversion being performed for slip in two mutually orthogonal
directions in the plane of the corresponding square element. In our
approach, we assume a subsource decomposition in terms of two different
parameters. The first of these is the magnitude of a mechanism composed
of three orthogonal compressional vector dipoles with an amplitude
ratio of $1:1:(1/\nu-1)$, where the latter dipole is oriented orthogonal
to the square element {[}this mechanism approximates the point-source
model suggested in Equation (\ref{eq:Definition=0000233d}){]}. The
second parameter is the amplitude of slip along the square element
in a prescribed direction, which we take to be vertical (corresponding
to reverse faulting). The technical aspects of inverting for the subsource
parameters by matching observed and synthetic waveforms are summarized
below.
\begin{itemize}
\item The elastodynamic Green's function for homogeneous isotropic space
was used.
\item Matching was performed using velocity waveforms filtered below $\unit[100]{Hz}$
(as opposed to the displacement waveforms used in the point-source
inversions discussed in Section \ref{subsec:Overview}).
\item To permit variation around the prescribed radial evolution, the time
history of each subsource was parameterized using the multi-time-window
method of \cite{Olson-Apsel-1982,Ide-2007}. In particular, $15$
time steps of $\unit[2.5]{ms}$ centered around the prescribed time
$t_{0}$ were used, leading to $870=29\times2\times15$ unknowns for
which to invert ($29$ sources, two components, and $15$ time steps).
\item Nonnegative least square inversion was employed \cite{Olson-Apsel-1982}
to ensure that deformation at the source (convergence and slip) is
unidirectional (that is, it does not change in sign). 
\item The total scalar seismic moment of all subsources was constrained
by the low-frequency plateau of the source spectra ($\unit[3.2\times10^{12}]{N\cdot m}$,
which corresponds to $m_{\mathrm{HK}}=2.3$). 
\end{itemize}
The observed and synthetic waveforms resulting from inversion are
compared in Figure \ref{fig:Example-mainshock-waveform-fit}. A reasonable
fit can be seen between the two, particularly in the initial and middle
sections of the waveforms. There is some discrepancy in the tail of
waveforms at more distance sites; however, this may be related to
imperfections introduced by the adopted Green's function, and so we
have not attempted to reduce the misfit by increasing the model's
complexity (that is, adding more subsources or extending their duration).

\begin{figure}[H]
\begin{centering}
\includegraphics[width=0.48\textwidth]{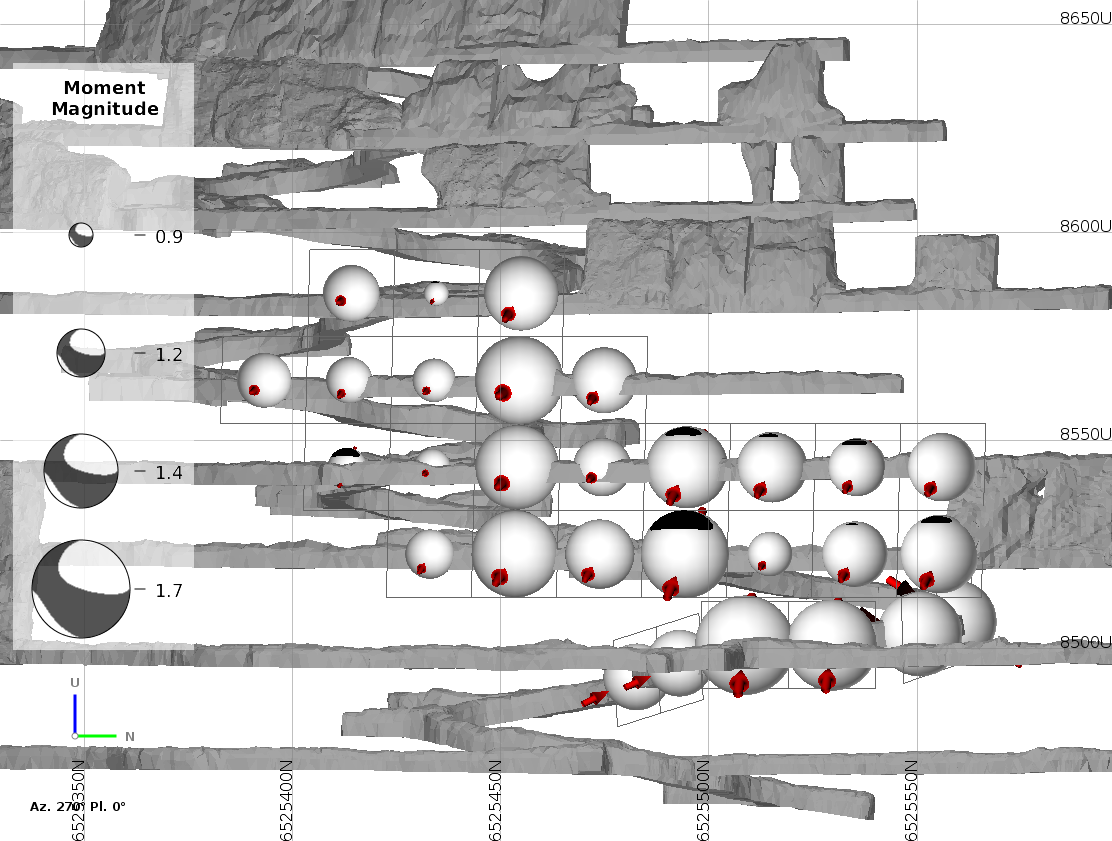}\hfill{}\includegraphics[width=0.48\textwidth]{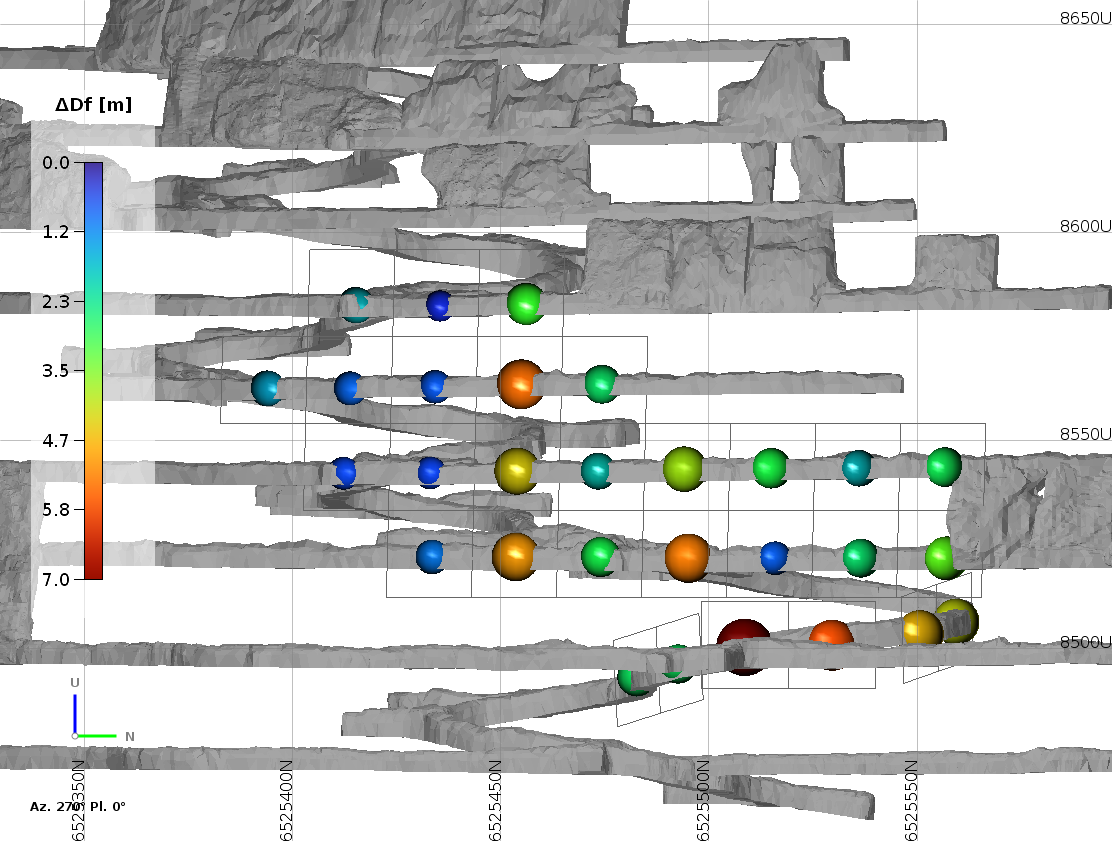}
\par\end{centering}
\caption{\label{fig:Example-mainshock-finite-source-results}Results of inversion
and interpretation of finite source parameters. The mechanisms of
subsources are shown as beachballs sized according to $m_{\mathrm{HK}}$
and P-axes depicted by red dipoles \textit{(left)}. The evaluated
increase in the depth of failure is shown by spheres colored by $\triangle d_{f}^{A}$
and with diameter of $L_{A}+\triangle d_{f}^{A}$\textit{ (right)}.}
\end{figure}

The left of Figure \ref{fig:Example-mainshock-finite-source-results}
shows the mechanisms of the subsources summed over the 15 time steps.
The majority of these are dominated by implosive and pancake-shaped
CLVD content, with double-couple content being less significant. As
such, it is appropriate to interpret their scalar moments in terms
of the model of Equation (\ref{eq:M0-approx}), which permits evaluation
of a corresponding increase in the depth of failure $\triangle d_{f}^{A}$
using Equation \ref{eq:dof-increase}. Other required parameters have
been selected as follows:
\begin{itemize}
\item Poisson's ratio of $\nu=\unit{0.23}$ (inferred from seismic data).
\item Maximum stress orthogonal to the tunnel's axis of $\sigma_{\mathrm{max}}=\unit[-90]{MPa}$
(based on elastic stress modelling results that take the \textit{in
situ} stress field and stress induced by stopes into account).
\item Extent along the tunnel's axis of $L_{3}=\unit[21]{m}$ (according
to the discretization of the finite source).
\item Effective tunnel dimension of $L_{A}=\unit[7.0]{m}$ (which incorporates
the design tunnel height and presumed pre-event depth of failure of
$\unit[1]{m}$ in both the roof and floor).
\end{itemize}
The obtained values of $\triangle d_{f}^{A}$ are shown in the right
of Figure \ref{fig:Example-mainshock-finite-source-results} and have
a maximum of $\unit[6.7]{m}$, mean of $\unit[2.8]{m}$, and median
of $\unit[2.4]{m}$. These values are not inconsistent with the observed
damage, from which it follows that the suggested hypothesis of cascading
stress fracturing around tunnels is plausible.

\subsection{Analysis of smaller-scale seismicity}

\begin{figure}[H]
\begin{centering}
\includegraphics[width=1\textwidth]{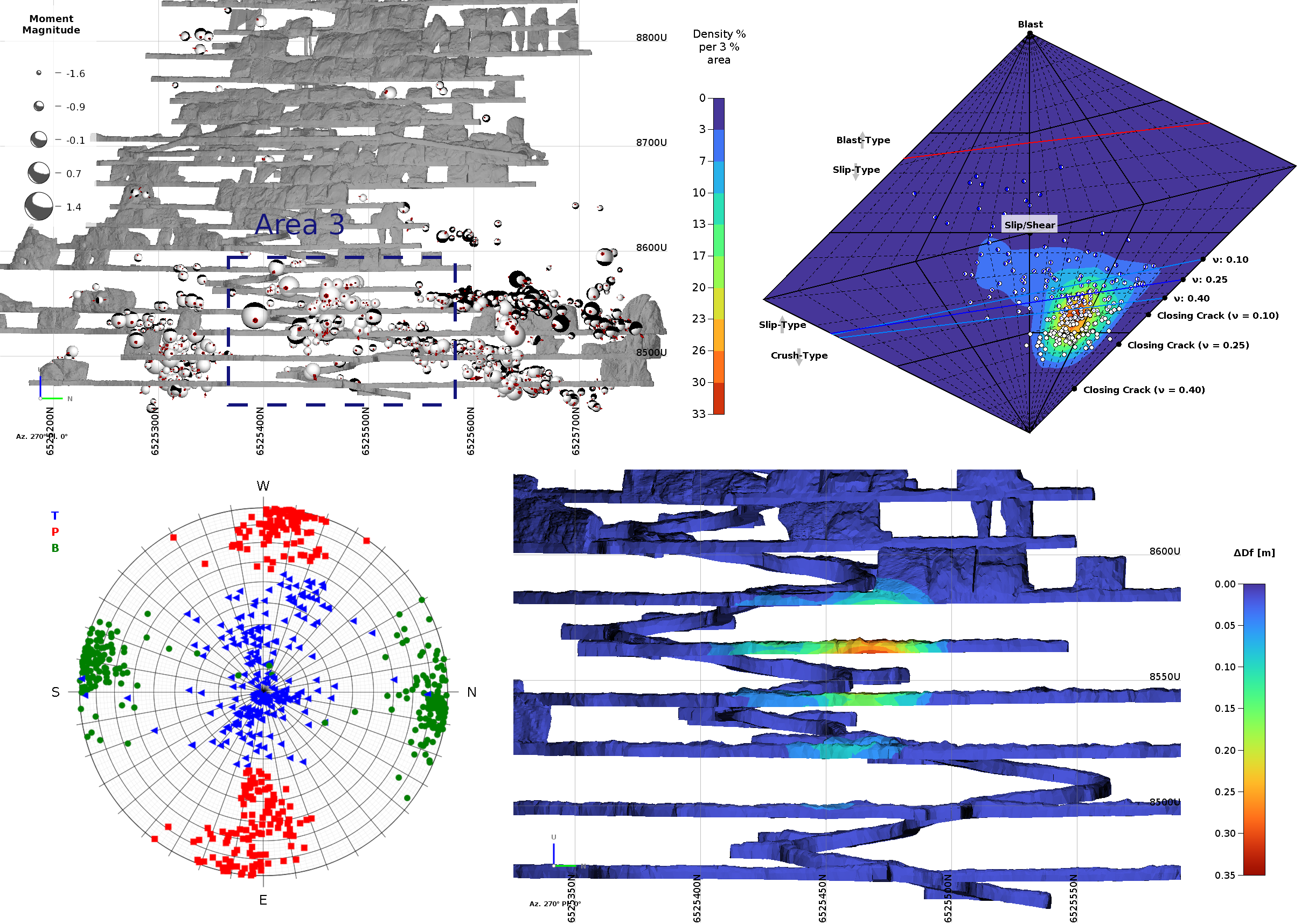}
\par\end{centering}
\caption{\label{fig:Example-previous-seismicity}Seismic events recorded from
1 January to 8 June 2019 \textit{(top left)}. The visualization of
seismic events is the same as in Figure \ref{fig:Example-mechanisms-view}.
Gray wireframes describe excavations as of the beginning of 2019.
The mechanisms of events in Area 3 are visualized in a source-type
plot \textit{(top right)} and stereonet of principal axes (\textit{bottom
left)}. The stereonet is rotated 90° clockwise to match the view in
Figure \ref{fig:Example-mechanisms-view}. Contours of increase in
depth of failure $\Delta d_{f}^{A}$ inferred from seismic events
\textit{(bottom right)}.}
\end{figure}

Figure \ref{fig:Example-previous-seismicity} shows seismicity observed
over a six-month period (1 January to 8 June 2019) prior to the large
event considered in Subsection \ref{subsec:Analysis-of-the}. In total,
1047 events were recorded, of which mechanisms were determined for
877 (including all events with $m_{\mathrm{HK}}>0.0$). This seismicity
can also be divided into several spatial groups, with those events
in Area 3 (as contoured in Figure \ref{fig:Example-previous-seismicity})
being our focus here. As shown in Figure \ref{fig:Example-previous-seismicity},
these events have similar characteristics to the mainshock and its
aftershocks in Area 1: clustering around the orebody in the proximity
of the tunnels (ore drives), significant implosive and pancake-shape
CLVD components, and P-axes approximately orthogonal to the direction
of the drives. Again, these features suggest interpretation in terms
of episodes of sudden stress fracturing in the roof and/or floor of
the ore drives accompanied by horizontal elastic convergence of the
surrounding rockmass.

Beyond this qualitative interpretation, an estimate of the increase
in depth of failure can be made by mapping the source parameters of
seismic events to the tunnels and determining $\Delta d_{f}^{A}$
from scalar moment using Equation (\ref{eq:dof-increase}). The direction
of this increase is controlled by the P-axis orientations of the projected
mechanisms (we omit further details of this mapping process for the
sake of brevity). The results of this procedure are presented in Figure
\ref{fig:Example-previous-seismicity}, where it can be seen that
the largest increase in depth of failure is expected to be in the
roof and floor (due to the subhorizontal loading) of the tunnels in
the central part of Area 3. In comparison to the values determined
for the mainshock, these increases are relatively minor, not exceeding
$\unit[0.35]{m}$.

This example serves to demonstrate the possibility of quantitative
monitoring of damage zone evolution around tunnels using seismic data.
However, the suggested approach requires testing (including the comparison
with borehole measurements), and this deserves a separate study. An
important aspect of the employed mapping procedure is the careful
selection of only those events associated with tunnels. While in the
case considered here, many seismic events outside of Area 3 have mechanisms
with significant implosive and pancake-shape CLVD components, they
are likely related to a sudden increase in the depth of failure around
stopes. The reduction to two-dimensional geometry in the derivation
of Equation (\ref{eq:M0-approx}) is not valid for such excavation
geometries, and so we have not attempted to infer the depth of failure
increase using it.

\section{\label{sec:Discussion-and-Conclusions}Discussion}

\subsection{\label{subsec:Classification}Classification}

In the context of underground mining operations, it has been suggested
to discriminate between ``crush-type'' and ``slip-type'' seismic
events \cite{Ryder-1988}; the former being described as an ``unstable
crushing of volume of rock in close proximity to mining void'' and
the latter as an ``unstable release of shear stress by slip over
a planar area (plane of weakness, including 'intact' rock).'' This
discrimination can be achieved by measuring the similarity of a given
event to seismic point sources for ``ideal'' crush- and slip-type
events. However, no particular source model is proposed in \cite{Ryder-1988}
to describe crush-type events. Given this, we suggest the use of the
closing-crack model \cite{Day-McLaughlin-1991,Bowers-Walter-2002},
which is composed of three orthogonal compressional vector dipoles
with an amplitude ratio of $1:1:(1/\nu-1)$, for an ideal crush-type
event. The closing crack model with a sub-vertically oriented largest
dipole has previously been used to describe processes such as near-surface
spalling during underground nuclear explosions and massive collapse
in mines with sub-horizontally oriented ore bodies. The point sources
obtained in Sections \ref{sec:Modelling} and \ref{sec:Approximation}
also resemble a closing-crack source with the largest dipole co-oriented
with the tunnel's loading. As this direction is arbitrary in general,
it is sensible for any discrimination procedure between crush- and
slip-type events to be orientation invariant. Graphically, this can
be achieved by assessing ``closeness'' to closing-crack and double-couple
point sources on source-type plots such as those presented in \cite{Tape-Tape-2012,Vavrycuk-2015,Hudson-1989}.
For the plots of \cite{Hudson-1989} employed in this paper (see,
for example, Table \ref{tab:Results-Hudson+stereo}, Figures \ref{fig:Mechanism-characteristics-2-areas}
and \ref{fig:Example-previous-seismicity}), we have marked the location
of closing-crack point sources for three variants of Poisson's ratio
($\nu=0.1$, $0.25$, and $0.4$). Given the proximity of the modelled
sources in Sections \ref{sec:Modelling} and \ref{sec:Approximation}
(as well as the majority of observed sources in Section \ref{sec:Examples})
to these points, we consider it appropriate to classify them as crush
type.

This classification process can be made quantitative through the use
of a distance metric to assess the ``closeness'' of sources. We
take the distance between moment tensors $M$ and $N$ as the angle
$\omega$ between the vectors $\mathbf{m}=(m_{1},m_{2},m_{3})$ and
$\mathbf{n}=(n_{1},n_{2},n_{3})$, where $m_{i}$ and $n_{i}$ are
the $i$th largest eigenvalues of $M$ and $N$, respectively; that
is, $\omega=\arccos\left(\mathbf{m}\cdot\mathbf{n}/\left\Vert \mathbf{m}\right\Vert \left\Vert \mathbf{n}\right\Vert \right).$
We also extend the classification of \cite{Ryder-1988} by including
the category of ``blast-type'' events, for which the ideal point
source represents a combination of three orthogonal extensional vector
dipoles of equal amplitude. Using the described metric, the source-type
plot of \cite{Hudson-1989} can be separated into three regions by
a line of equal proximity to ideal blast- and slip-type sources 
\[
k=\begin{cases}
\frac{T+4}{\sqrt{2/3}T+2(\sqrt{6}+2)} & \mathrm{for\,}-1\leq T\leq0,\\
\frac{T-4}{T-2(\sqrt{6}+2)} & \mathrm{for\,}0\leq T\leq1,
\end{cases}
\]
and a line of equal proximity to ideal slip- and crush-type sources
\[
k=\begin{cases}
-\frac{2(\sqrt{2}q+2-1/\nu)+(q/\sqrt{2}+2-1/\nu)T}{2(\sqrt{2}q+3)+(q/\sqrt{2}+2-1/\nu)T} & \mathrm{for\,}-1\leq T\leq0,\\
-\frac{2(\sqrt{2}q+2-1/\nu)-qT/\sqrt{2}}{2(\sqrt{2}q+3)-qT/\sqrt{2}} & \mathrm{for\,}0\leq T\leq1,
\end{cases}
\]

\noindent where $\nu$ is the Poisson's ratio and $q=\sqrt{3+1/\nu^{2}-2/\nu}$.
These lines are shown on the source-type plots presented in this paper
(see, for example, Table \ref{tab:Results-Hudson+stereo}, Figures
\ref{fig:Mechanism-characteristics-2-areas} and \ref{fig:Example-previous-seismicity}).
That all the modelled sources in Sections \ref{sec:Modelling} and
\ref{sec:Approximation} fall below the line separating slip and crush-type
sources quantitatively justifies their being classified as crush-type.
The same applies for the majority of real events presented in Section
\ref{sec:Examples}.

\subsection{Interpretation}

We note that crush-type sources should not be considered as exotic:
there are numerous observations in mines that report source mechanisms
with significant implosive components \cite{McGarr-1992,Stickney-Sprenke-1993,Julia-2009}.
Although examples from a single mine are presented in Section \ref{sec:Examples},
the authors have observed crush-type sources at a number of other
mines. In fact, at some underground mines, seismic events of this
type constitute the majority of recorded seismicity (both in terms
of number of events and even cumulative seismic moment).

Quite often, seismic data recorded in mines is processed and interpreted
assuming a double-couple source model. The nodal planes of the inferred
mechanisms are then compared with mapped geological structures (such
as faults, contacts of lithological units, etc.) or considered as
possible surfaces of shear rupture of intact rock. The results presented
in previous sections suggest that such interpretation is not always
meaningful. Although the modelled source mechanisms shown in Table
\ref{tab:Results-Hudson+stereo} have non-zero double-couple components,
the corresponding nodal planes do not indicate the presence of slip
or shear rupture surfaces. If the sources of seismic events are located
close to tunnels (or other excavations, as discussed below) and have
crush-type mechanisms, then it can be more useful to attempt interpretation
in terms of the model of Equation (\ref{eq:Definition=0000233d}). 

\subsection{Limitations}

The numerical modelling presented in Section \ref{sec:Modelling}
and the approximations outlined in Section \ref{sec:Approximation}
have the following known limitations:
\begin{itemize}
\item Tunnels were considered two-dimensionally under the assumption of
plane strain, and the extent of failure along the tunnel was taken
into account by means of a simple length coefficient $L_{3}$ (as
discussed previously, this is equivalent to taking a finite slice
of an infinite-length tunnel). In reality, damage is not distributed
uniformly along the tunnel, and full three-dimensional modelling of
this distribution would be preferable. Furthermore, approximation
in terms of ellipsoidal cavity expansion rather than the elliptical
cavity expansion considered in Section \ref{sec:Approximation} may
be more appropriate; this requires further exploration in future work.
\item The proposed model is focused on the case of isolated tunnel in homogeneous
stress field. However, in mining environments, tunnels are typically
surrounded by other excavations (such as caves, stopes and other tunnels),
which make the loading conditions softer. This means that the same
increment in the depth of failure $\Delta d_{f}^{A}$ for a tunnel
may produce different elastic convergence of the surrounding rock
mass depending on the proximity to and configuration of other excavations.
In this sense, Equation (\ref{eq:M0-approx}) provides the lower-bound
estimate of seismic moment $\left|\mathbf{M}\right|$. It is expected
that softer loading conditions (due to the presence of neighboring
excavations) will results in larger seismic moment $\left|\mathbf{M}\right|$
for the same $\Delta d_{f}^{A}$. Analysis of this would be an interesting
avenue for future research.
\end{itemize}

\subsection{Applications}

The suggested source mechanism model of Equation (\ref{eq:Definition=0000233d})
has several applications:
\begin{itemize}
\item Forensic analysis of damaging events: If a seismic event locates close
to a damaged tunnel, then it makes sense to test the hypothesis that
the source of the event and the source of damage are the same. A modelled
source mechanism can be determined using Equation (\ref{eq:Definition=0000233d})
and compared with one inverted from recorded seismic waveforms. If
these mechanisms have reasonable agreement in terms of scalar moment,
direction of principal axis, and source type, then it makes sense
to conclude that dynamic stress fracturing around the tunnel was the
source of the seismic event. Alternatively, if for example, the modelled
scalar moment is smaller than that derived from observation, it implies
that the rock fracturing associated with damage is not substantial
enough to explain the entire recorded seismic radiation. In such cases,
it can be useful to detect ``fingerprints'' of damage within the
recorded waveforms. This can be done by calculating synthetic waveforms
for the modelled mechanism (corresponding to damage) and correlating
these with high-frequency pulses in the recorded waveforms. If a match
is found, then the temporal and spatial relation between the damage
and larger-scale deformation of the rockmass (which is responsible
for the majority of recorded seismic radiation) can be established.
\item Monitoring of deformation of tunnels: Identification of crush-type
events around tunnels and interpreting them in terms of Equation (\ref{eq:Definition=0000233d})
makes it possible to quantify the growth of damage zone around tunnels.
This can be used in assessing the consumption of capacity of ground
support systems installed in tunnels.
\item Assessment of dynamic loading to ground support of tunnels: If sources
of seismic event in mines are located away from the tunnel of interest,
then dynamic loading is assessed in terms of intensity of ground motion
(shaking). For seismic sources considered in this work, a better characteristics
of loading will be associated with the amount or rate of deformation
within the stress-fracturing rock (for example, the amplitude or rate
of tangential straining on the perimeter of excavation).
\item Other excavations: While the focus of this work has been on events
associated with tunnels, the suggested model {[}described by Equation
(\ref{eq:Definition=0000233d}){]} can be applied to other types of
excavations, such as ore passes, shafts, stopes, or cave\textcolor{black}{s,
provided that their geometrical characteristics can be approximated
by a two-dimensional elliptical cavity. If the shape of excavations
is more complex, then Equations (\ref{eq:kirchhoff}) or (\ref{eq:adjusted-conventional})
need to be used. The model is also potentially applicable to underground
excavations associated with non-mining applications (for example,
nuclear waste storage, hydro-electric stations, etc.). The key requirement
in all of these cases is that the seismic wavelengths of interest
must be larger than the size of the excavation; for example, at} wavelengths
of hundreds of meters, the model may be applicable for stopes (dimensions
of $\unit[20-50]{m}$) but not for caves (dimensions of several hundred
meters).
\end{itemize}

\section*{Data and Resources}

\noindent Permission is required to obtain the data used in Section
\ref{sec:Examples}, which are property of the mine. The supplemental
material describes the verification tests of the Material Point Method
and constitutive relations used for modelling presented in Section
\ref{sec:Modelling}. Also included are animations showing the dynamics
of the sources for Cases 1-6.

\section*{Acknowledgments}

The model considered in this work was motivated by the presentations
of Prof. Peter Kaiser at Institute of Mine Seismology (IMS) seminars.
The advice of Prof. Mark Diederichs and Prof. Peter Kaiser regarding
the modelling parameters is highly appreciated. We are thankful to
the mine, which provided permission to use their data in Section \ref{sec:Examples}.
Gys Basson (IMS) advised and assisted in numerical modelling. Martin
Gal (IMS) has helped to validate the expressions presented in Section
\ref{sec:Discussion-and-Conclusions}.

\bibliographystyle{plain}
\bibliography{Seismic-sources}

\appendix

\part*{Appendices}

\section{Details of modelling of seismic sources\label{sec:Appendix-1:-Details}}

\begin{figure}[H]
\begin{centering}
\includegraphics[width=0.75\textwidth]{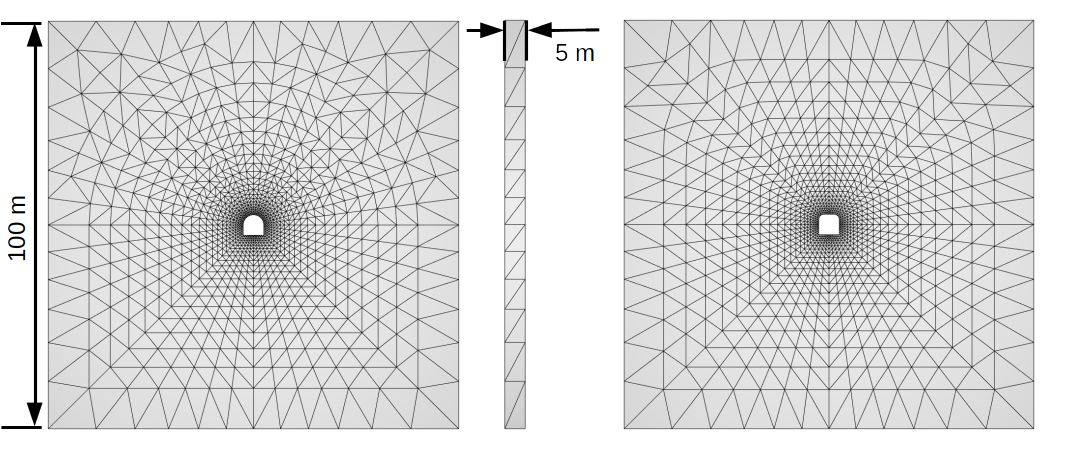}
\par\end{centering}
\caption{\label{fig:Meshing}\textcolor{black}{North views of a simplified
version of the tetrahedral mesh used for Cases 1-4, and 6 }\textit{\textcolor{black}{(left)}}\textcolor{black}{{}
and for Case 5 }\textit{\textcolor{black}{(right)}}\textcolor{black}{.
East view is shown between them }\textit{\textcolor{black}{(middle)}}\textcolor{black}{.}}
\end{figure}

Following the procedure of the verification tests available in the
supplemental material, our MPM simulations employ grids based on tetrahedral
meshes produced using \textsc{TetGen} \cite{Si-2015}. An example
of a simplified (coarse) version of the mesh used for Cases 1-4 and
6 is shown in Figure \ref{fig:Meshing}. The meshes are $\unit[100]{m}$
diameter square plates with a thickness of $\unit[5]{m}$ composed
of a single layer of tetrahedra whose vertices lie on contours that,
with allowances for boundary geometry, are concentric to the profile
of the tunnel itself (this reduction to what is essentially a two-dimensional
setup follows the discussion of Section \ref{subsec:Cases}). In practice,
$96$ vertices lie on each of these contours (rather than the $48$
shown in Figure \ref{fig:Meshing}) and are separated by equal angles
as measured from the center of the tunnel. The radial distance between
adjacent contours is approximately equal to the spacing of vertices
in the smaller contour, making the mesh increasingly coarse away from
the tunnel.

Each cell (tetrahedron) is populated with a single particle. The initial
stress of these particles is set according to case-varying loading
described in Section \ref{subsec:Cases} {[}stress in the direction
of the tunnel is set to $\nu(\sigma_{\max}+\sigma_{\min})$ to satisfy
the plane strain assumption{]}. Fixed stress conditions are applied
at the top, bottom, east, and west boundaries; fixed normal displacement
(roller) conditions are applied at the north and south boundaries.

As noted in Section \ref{subsec:Cases}, a CWFS material is used.
Brittle failure at low confinement is implemented using an elastic-brittle-plastic
Mohr-Coulomb constitutive relation with tension cutoff based on the
formulation of \cite{FLAC3D-manual-2009}. Strain hardening at high
confinement is implemented based on the strain-softening Hoek-Brown
material constituitive model detailed in \cite{sorensen2015finite}.
Verification of these two constitutive relations can be found in the
supplemental material.

The initial simulation is performed quasi-statically using damping
as described in \cite{FLAC3D-manual-2009,Wang-2016}. Expansion of
the failed region is simulated dynamically by removing damping within
a 50m of the tunnel's center (we preserve damping near the boundaries
to limit unwanted reflections) prior to applying one of the perturbations
outlined in Section \ref{subsec:Cases}.

Calculating the surface integrals in Subsection \ref{subsec:Results}
requires values of displacement and traction at faces of the tetrahedral
mesh. In the case of displacement, this is achieved by first mapping
particle displacements to the nodes of their containing cells (tetrahedra)
via the shape function. The displacement of any given face is then
taken to be the average of the displacement of its constituent nodes.
Similarly, stress at faces is determined in the same way, allowing
traction to be obtained using the faces' normal vectors.

\begin{figure}
\centering{}\includegraphics[width=0.45\textwidth]{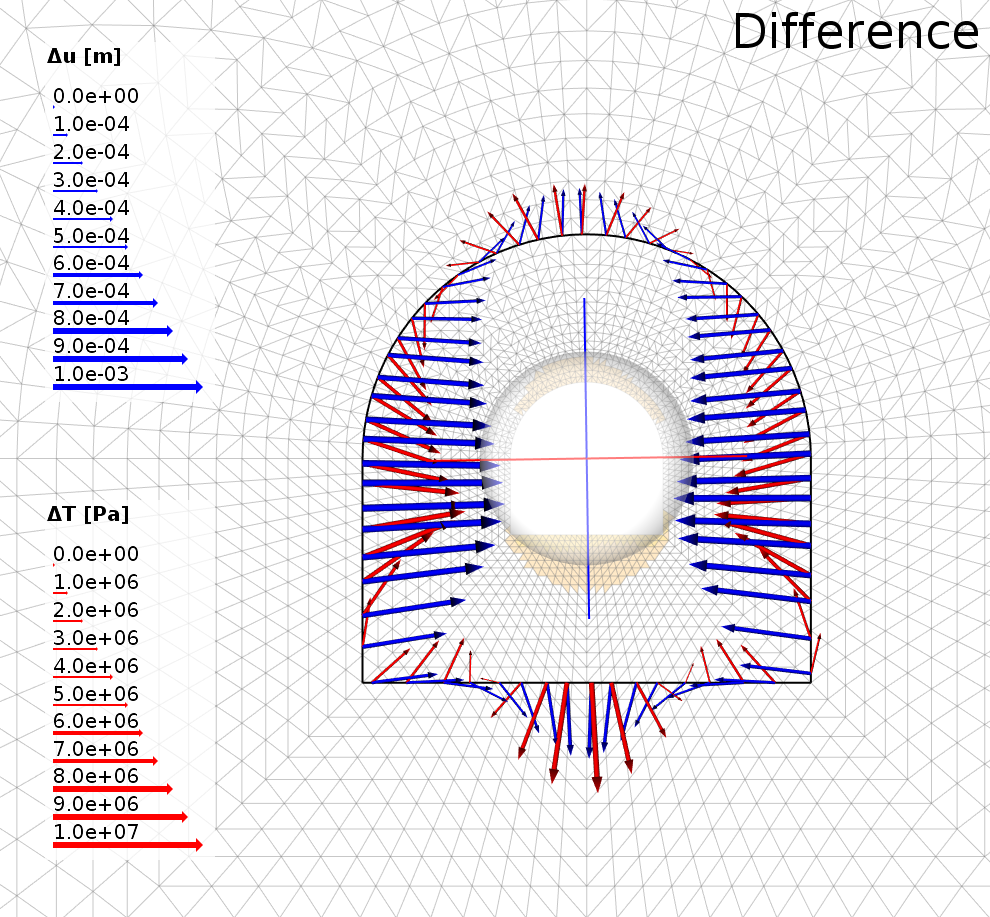}\hspace{0.02\textwidth}\includegraphics[width=0.45\textwidth]{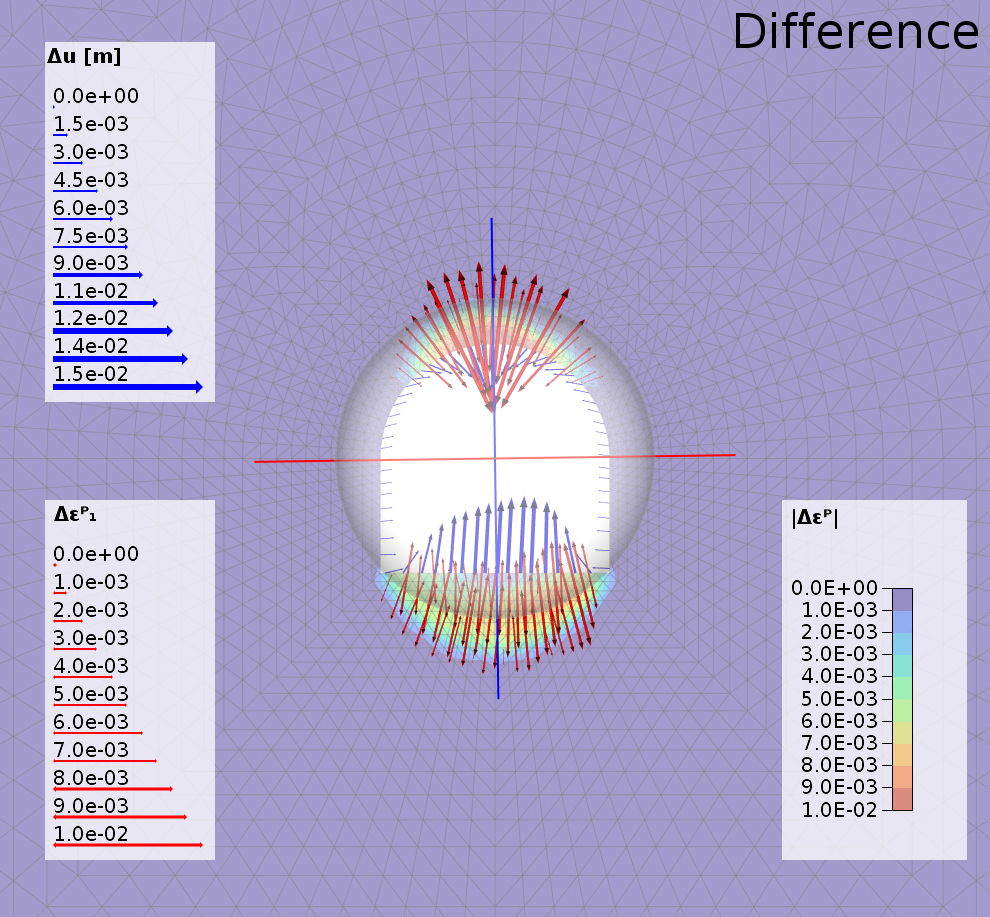}\medskip{}
\\
\includegraphics[width=0.45\textwidth]{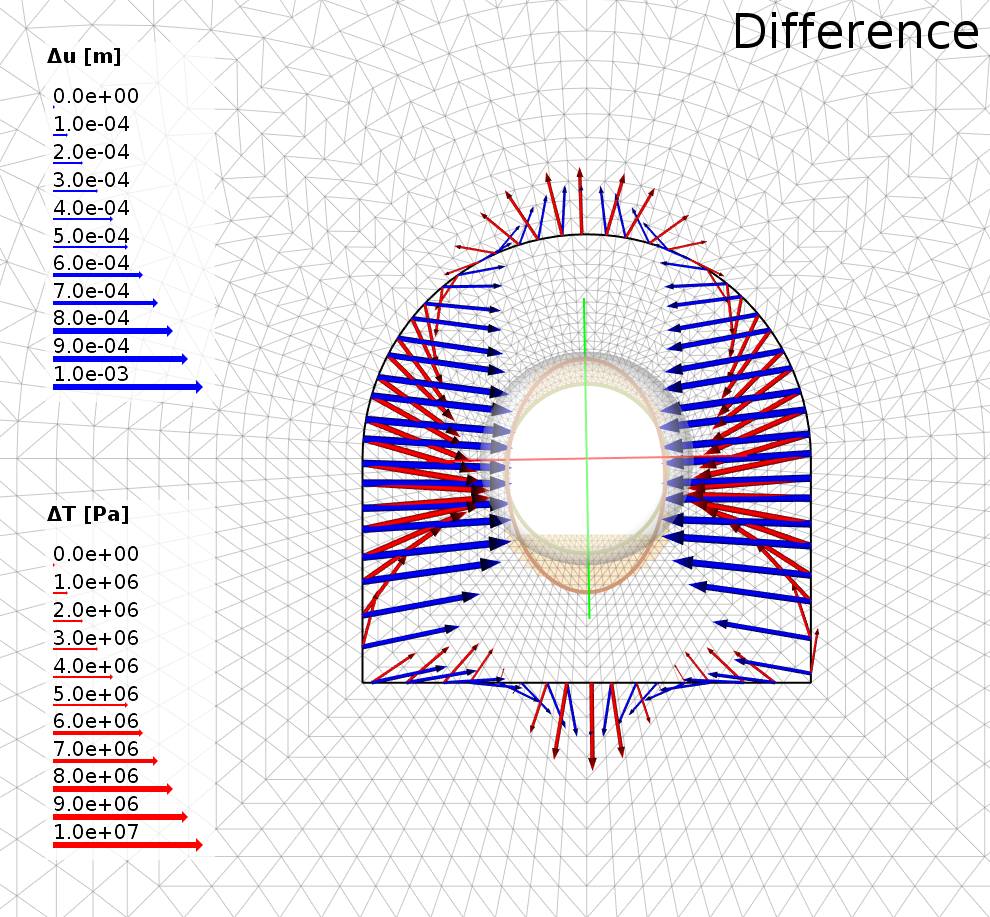}\hspace{0.02\textwidth}\includegraphics[width=0.45\textwidth]{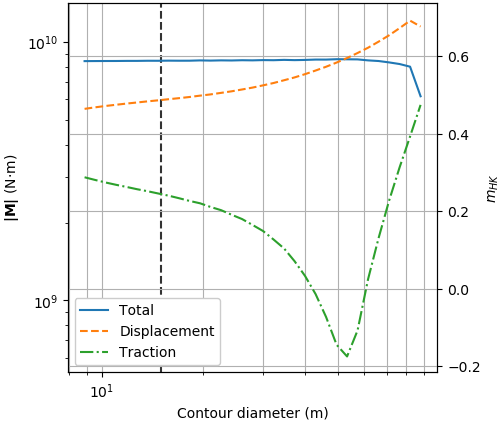}\caption{Summaries for Case 1 (horizontal loading, stress increase, failure
in the roof and floor) of Kirchhoff \textit{(top left)}, adjusted
conventional \textit{(top right)}, and elliptical \textit{(bottom
left)} moment tensor calculations. Demonstration of contour invariance
for Kirchhoff-type calculation \textit{(bottom right)}. See text for
more details.\label{fig:Summaries-for-Case-1}}
\end{figure}

\begin{figure}
\centering{}\includegraphics[width=0.45\textwidth]{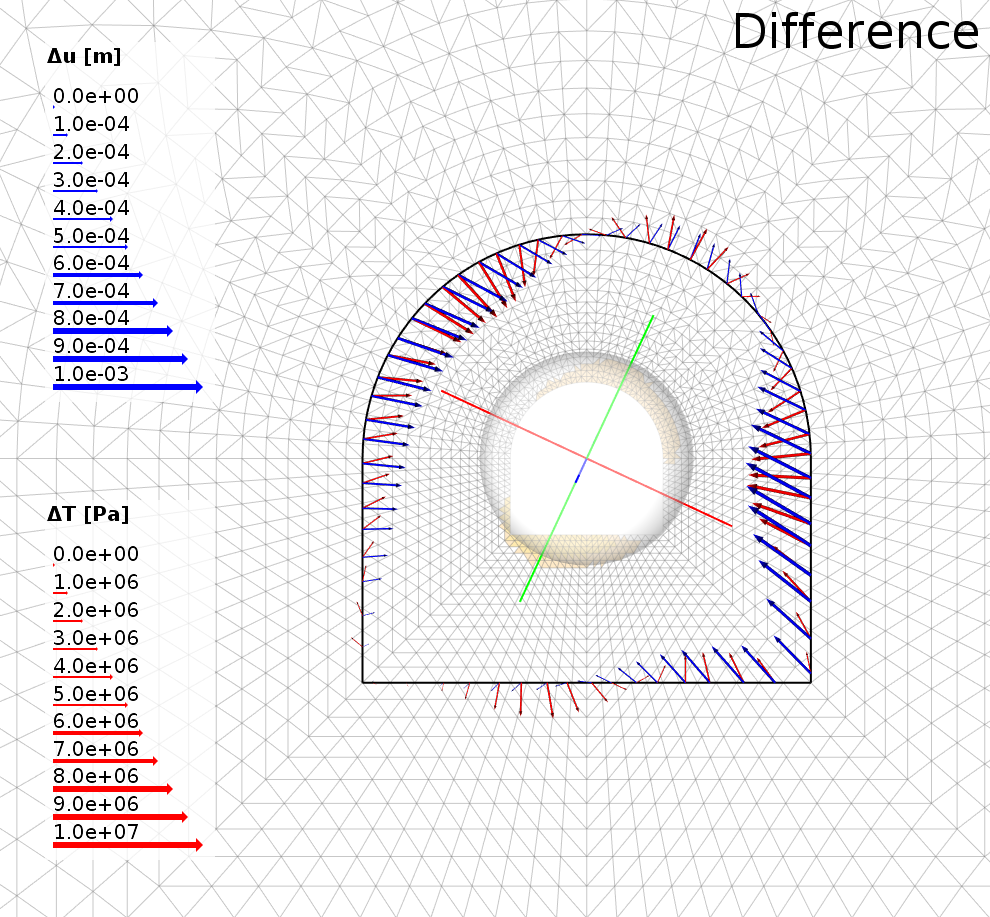}\hspace{0.02\textwidth}\includegraphics[width=0.45\textwidth]{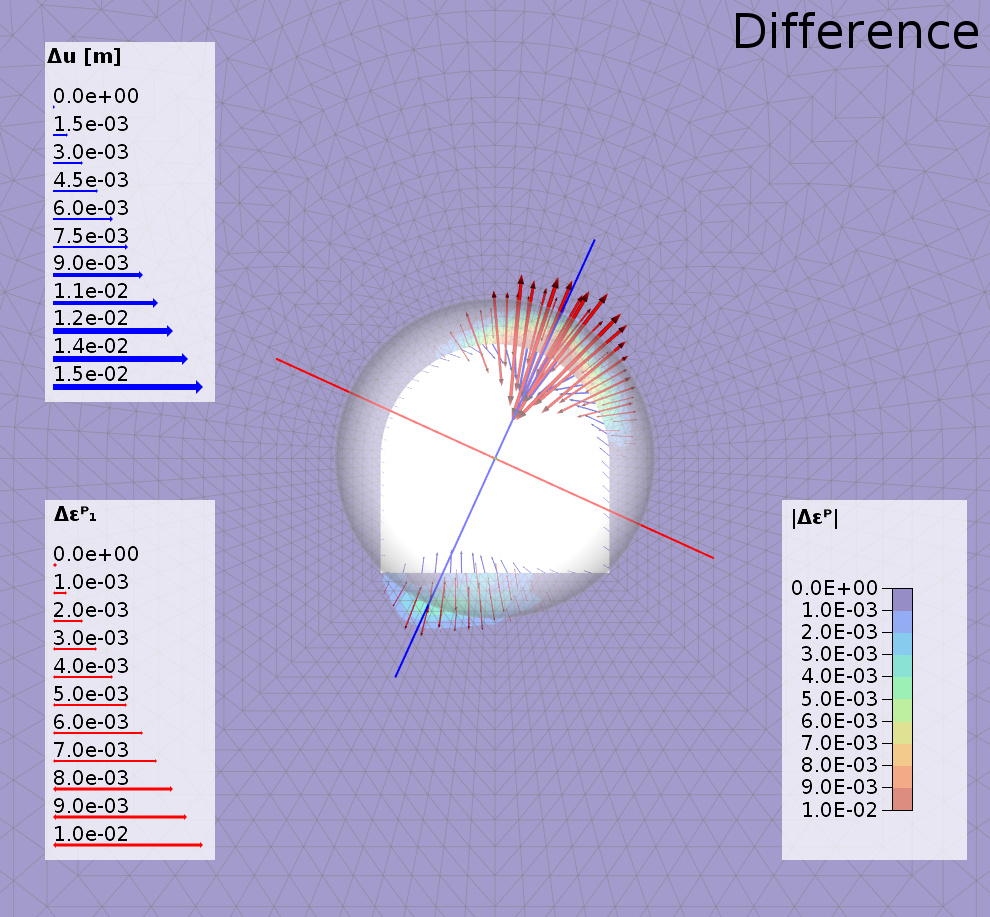}\medskip{}
\\
\includegraphics[width=0.45\textwidth]{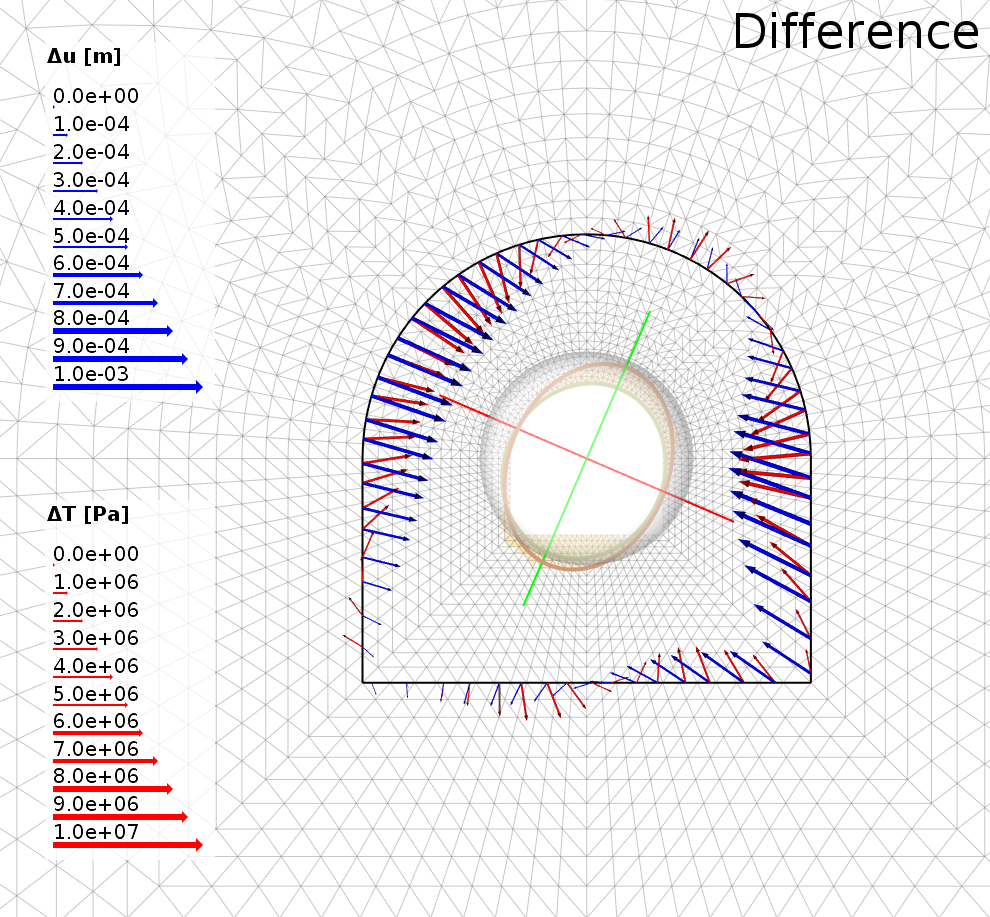}\hspace{0.02\textwidth}\includegraphics[width=0.45\textwidth]{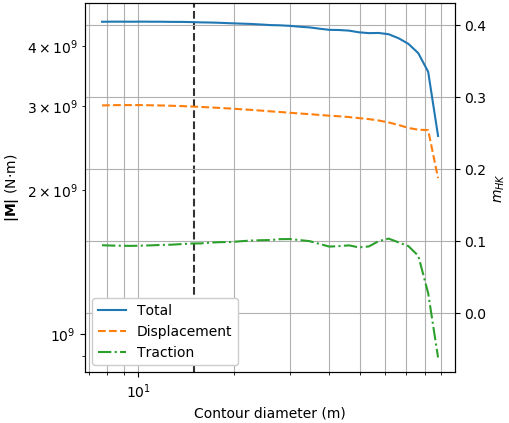}\caption{Summaries for Case 6 (eastward plunging loading, stress wave, failure
in the eastern shoulder and western bottom corner) of Kirchhoff \textit{(top
left)}, adjusted conventional \textit{(top right)}, and elliptical
\textit{(bottom left)} moment tensor calculations. Demonstration of
contour invariance for Kirchhoff-type calculation \textit{(bottom
right)}. See text for more details.\label{fig:Summaries-for-Case-6}}
\end{figure}

Figures \ref{fig:Summaries-for-Case-1} to \ref{fig:Summaries-for-Case-6}
show details of the source mechanism calculations for the five modelled
cases:
\begin{itemize}
\item The top row shows plots matching those presented in the bottom row
of Figure \ref{fig:Conv-vs-Kirch-case3} for Case 3. The left plot
shows the inputs to Equation (\ref{eq:kirchhoff}): differential displacement
$\mathbf{\triangle u}$ and traction $\mathbf{\triangle T}$ along
a surface $S$ in the elastic region, which is contoured in black.
The right plot shows the inputs to Equation (\ref{eq:adjusted-conventional}):
differential plastic strain $\boldsymbol{\triangle\varepsilon^{p}}$
and displacement $\mathbf{\triangle u}$ along the tunnel surface
$\varSigma$.
\item The bottom left plot matches that presented on the right of Figure
\ref{fig:Approximation-ellipses-case3} for Case 3. Pre- and post-expansion
approximating ellipses are shown along with $\mathbf{\triangle u}$
and $\mathbf{\triangle T}$ along $S$.
\item The bottom right plot shows the relative invariance in the surface
$S$ selected on the Kirchhoff-type source mechanism calculated from
Equation (\ref{eq:kirchhoff}). In particular, it shows the variation
in the scalar moments $\left|\mathbf{M^{T}}+\mathbf{M^{U}}\right|$,
$\left|\mathbf{M^{U}}\right|$, and $\left|\mathbf{M^{T}}\right|$
for surfaces of varying diameter. The dashed vertical line marks the
surface diameter of $\unit[15]{m}$ used in calculating the mechanisms
presented in Sections \ref{sec:Modelling} and \ref{sec:Approximation}.
\end{itemize}

\begin{figure}
\centering{}\includegraphics[width=0.45\textwidth]{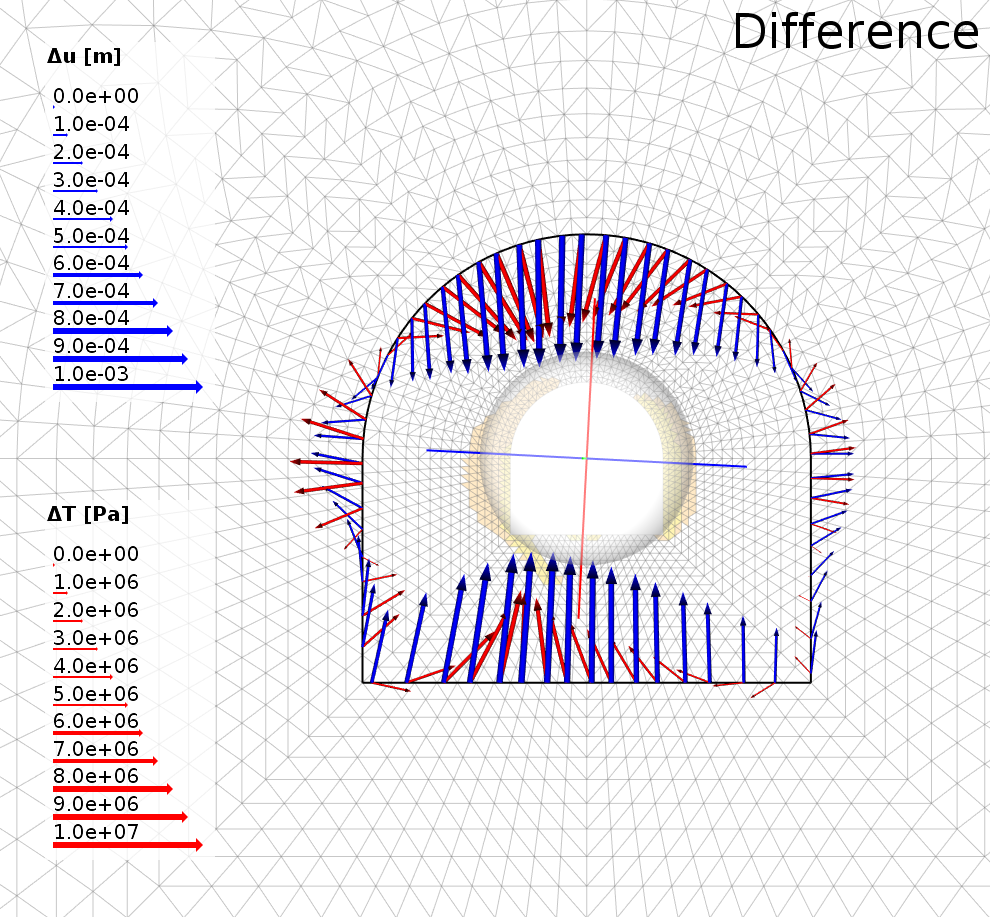}\hspace{0.02\textwidth}\includegraphics[width=0.45\textwidth]{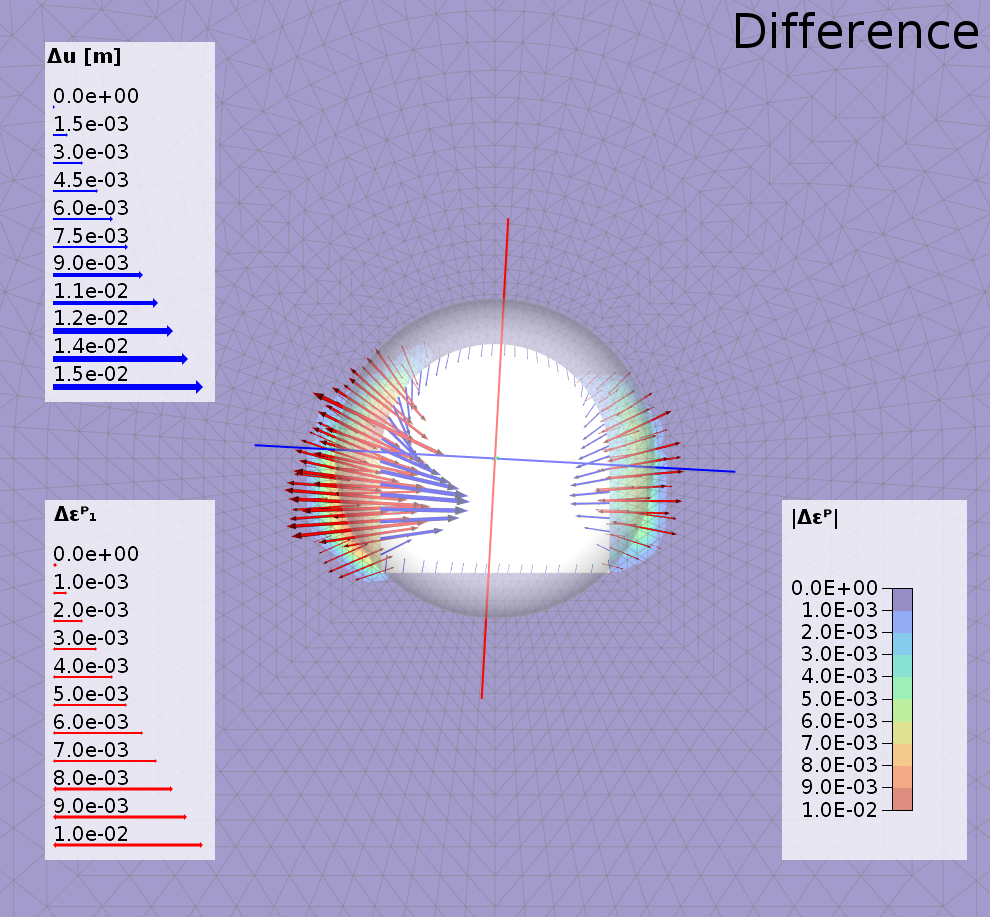}\medskip{}
\\
\includegraphics[width=0.45\textwidth]{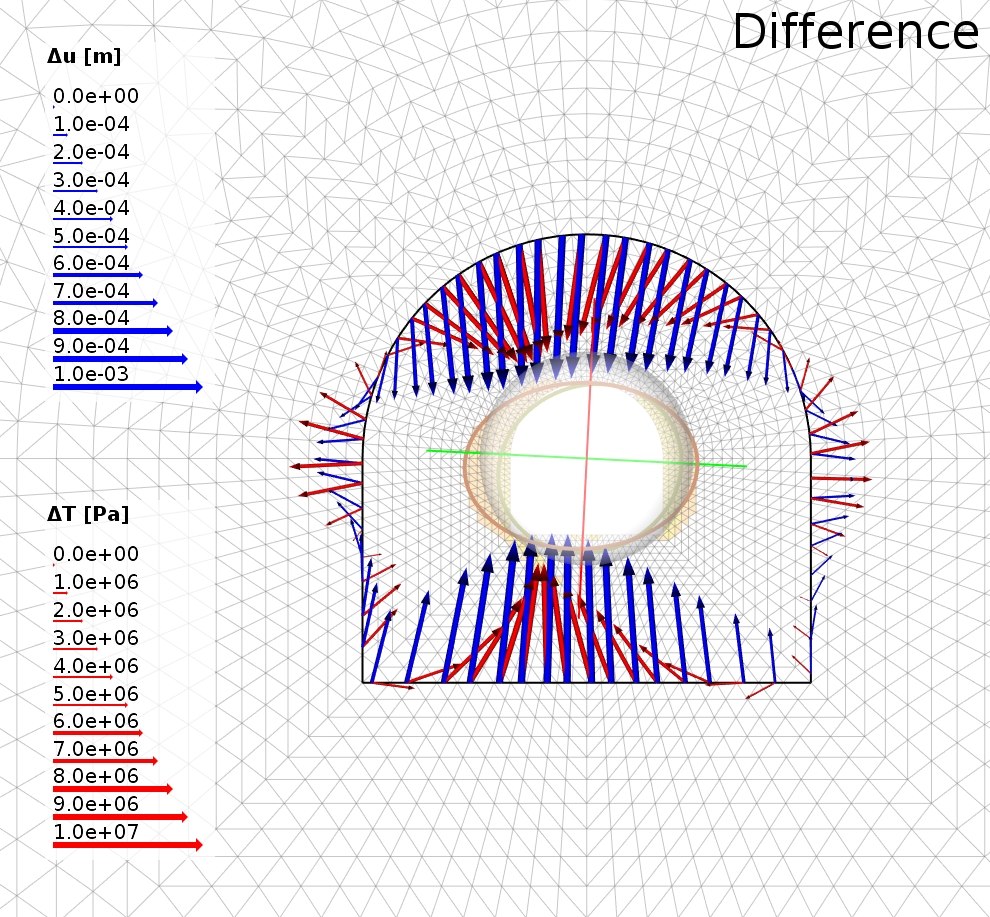}\hspace{0.02\textwidth}\includegraphics[width=0.45\textwidth]{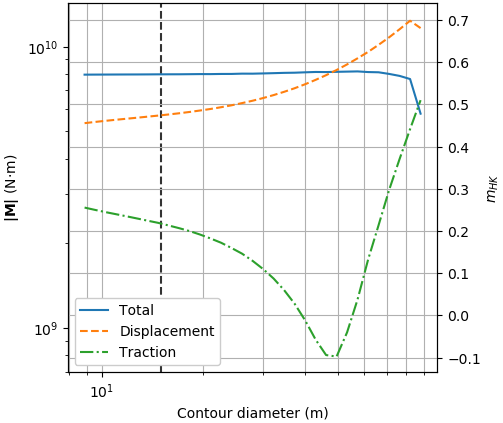}\caption{Summaries for Case 2 (vertical loading, stress increase, failure in
the walls) of Kirchhoff \textit{(top left)}, adjusted conventional
\textit{(top right)}, and elliptical \textit{(bottom left)} moment
tensor calculations. Demonstration of contour invariance for Kirchhoff-type
calculation \textit{(bottom right)}. See text for more details.}
\end{figure}

\begin{figure}
\centering{}\includegraphics[width=0.45\textwidth]{case-3-kirchhoff-diff}\hspace{0.02\textwidth}\includegraphics[width=0.45\textwidth]{case-3-conventional-diff}\medskip{}
\\
\includegraphics[width=0.45\textwidth]{case-3-ellipse-diff}\hspace{0.02\textwidth}\includegraphics[width=0.45\textwidth]{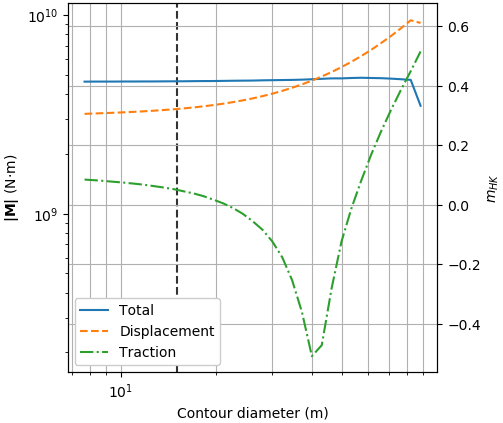}\caption{Summaries for Case 3 (eastward plunging loading, stress increase,
failure in the eastern shoulder and western bottom corner) of Kirchhoff
\textit{(top left)}, adjusted conventional \textit{(top right)}, and
elliptical \textit{(bottom left)} moment tensor calculations. Demonstration
of contour invariance for Kirchhoff-type calculation \textit{(bottom
right)}. See text for more details.}
\end{figure}

\begin{figure}
\centering{}\includegraphics[width=0.45\textwidth]{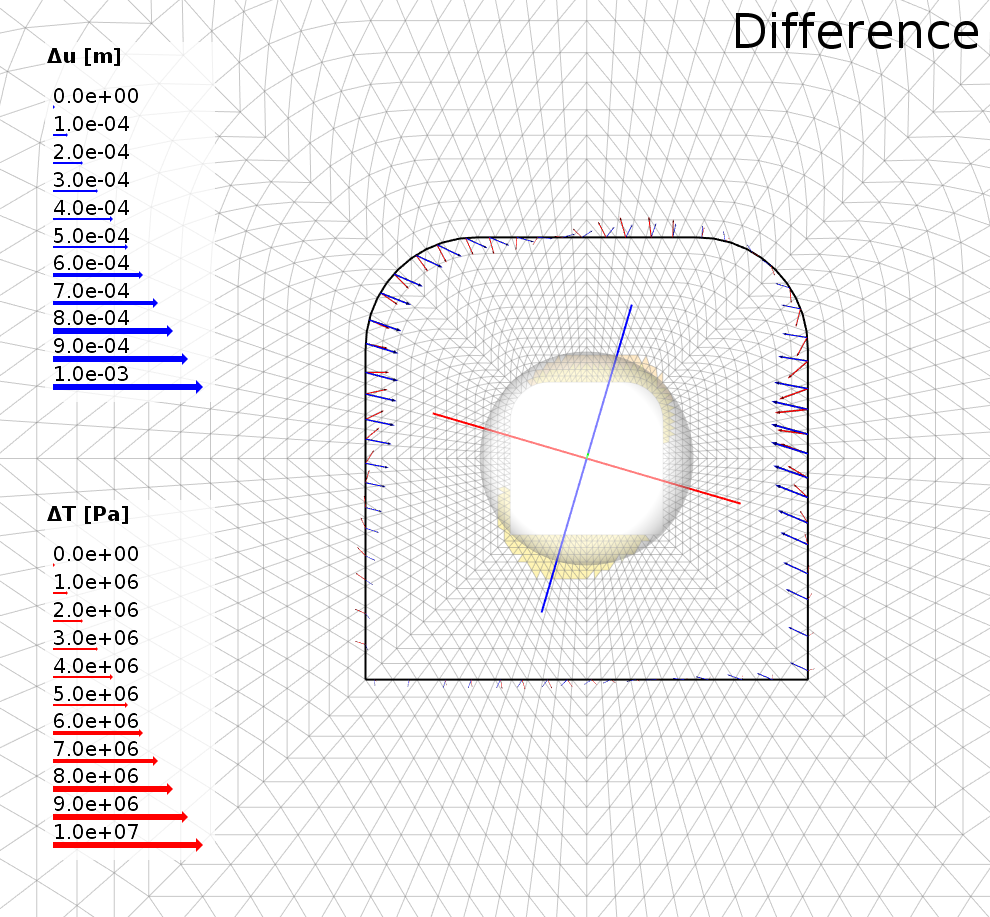}\hspace{0.02\textwidth}\includegraphics[width=0.45\textwidth]{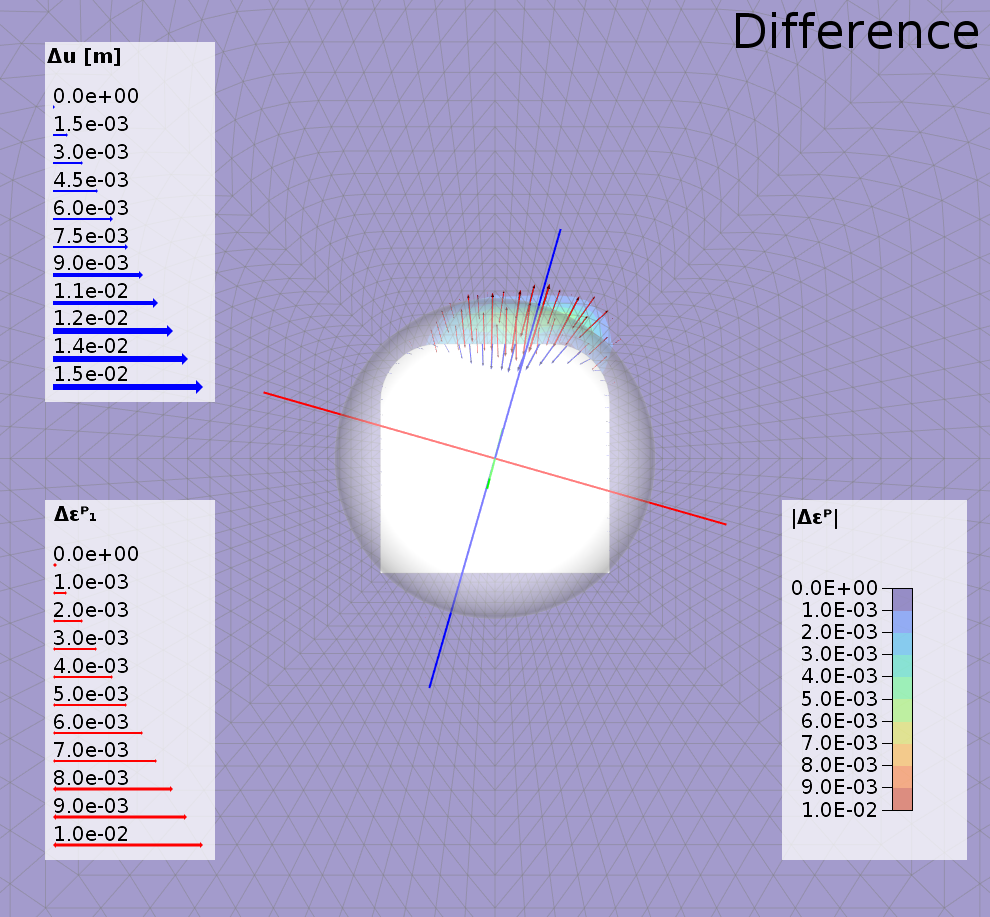}\medskip{}
\\
\includegraphics[width=0.45\textwidth]{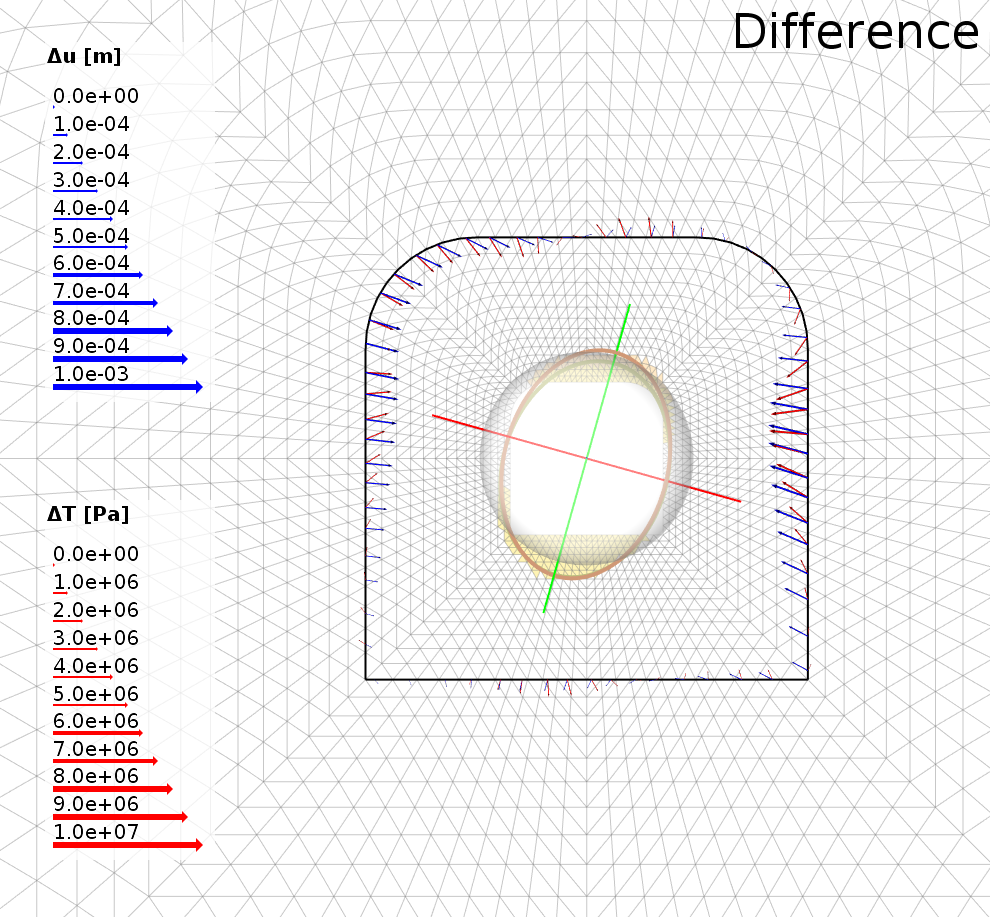}\hspace{0.02\textwidth}\includegraphics[width=0.45\textwidth]{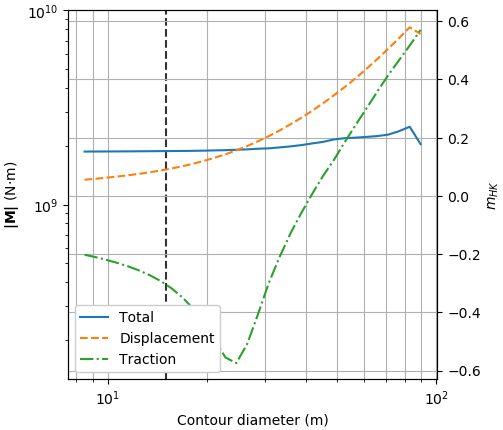}\caption{Summaries for Case 4 (eastward plunging loading, stress increase,
different profile, failure in the eastern shoulder) of Kirchhoff \textit{(top
left)}, adjusted conventional \textit{(top right)}, and elliptical
\textit{(bottom left)} moment tensor calculations. Demonstration of
contour invariance for Kirchhoff-type calculation \textit{(bottom
right)}. See text for more details.}
\end{figure}

\begin{figure}
\centering{}\includegraphics[width=0.45\textwidth]{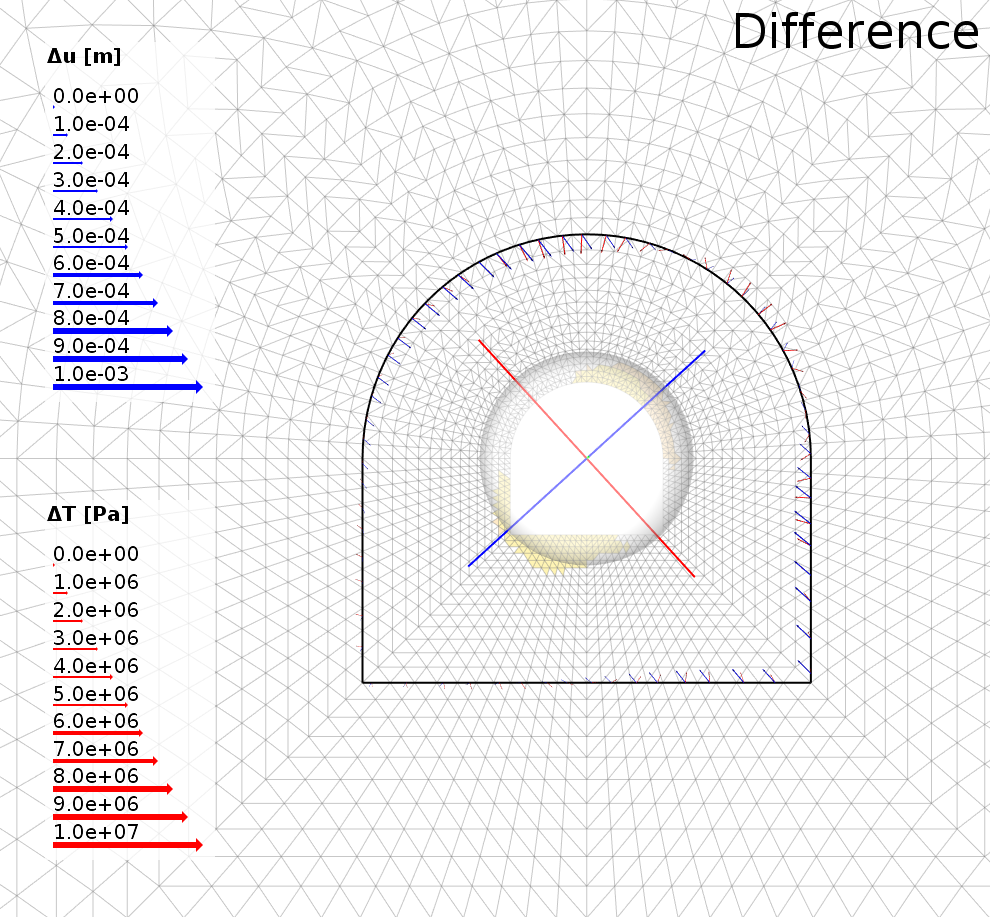}\hspace{0.02\textwidth}\includegraphics[width=0.45\textwidth]{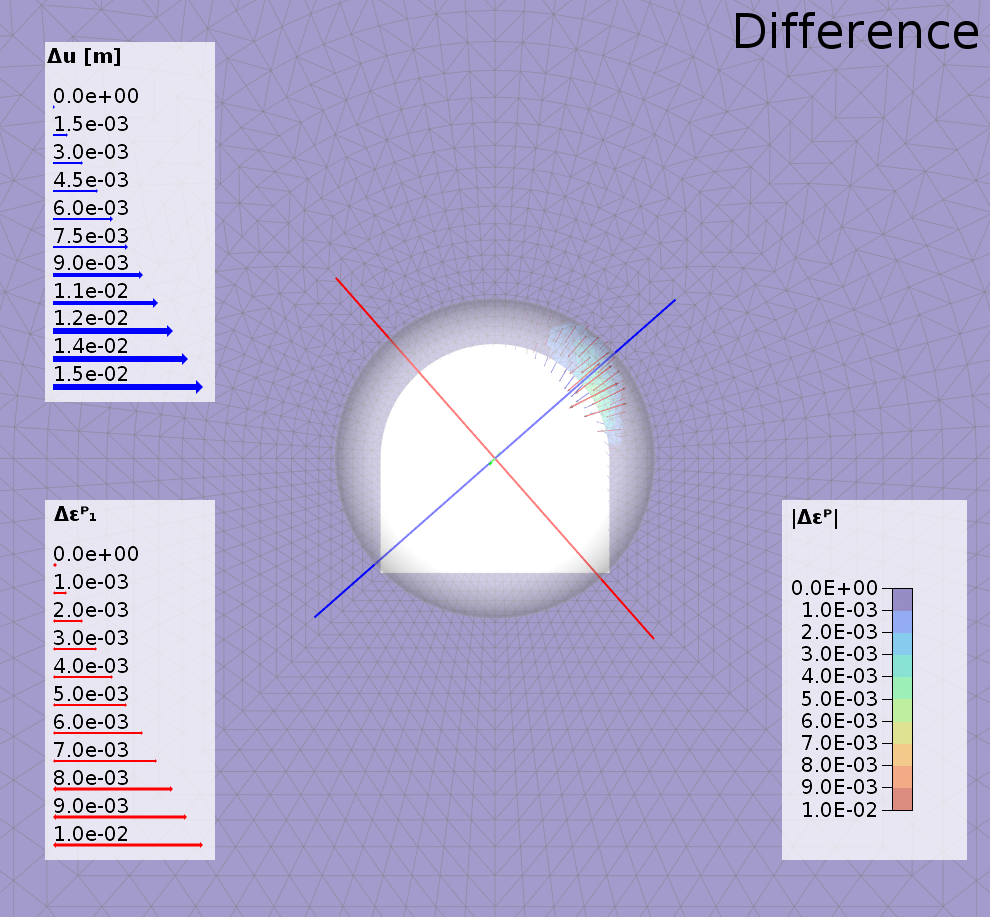}\medskip{}
\\
\includegraphics[width=0.45\textwidth]{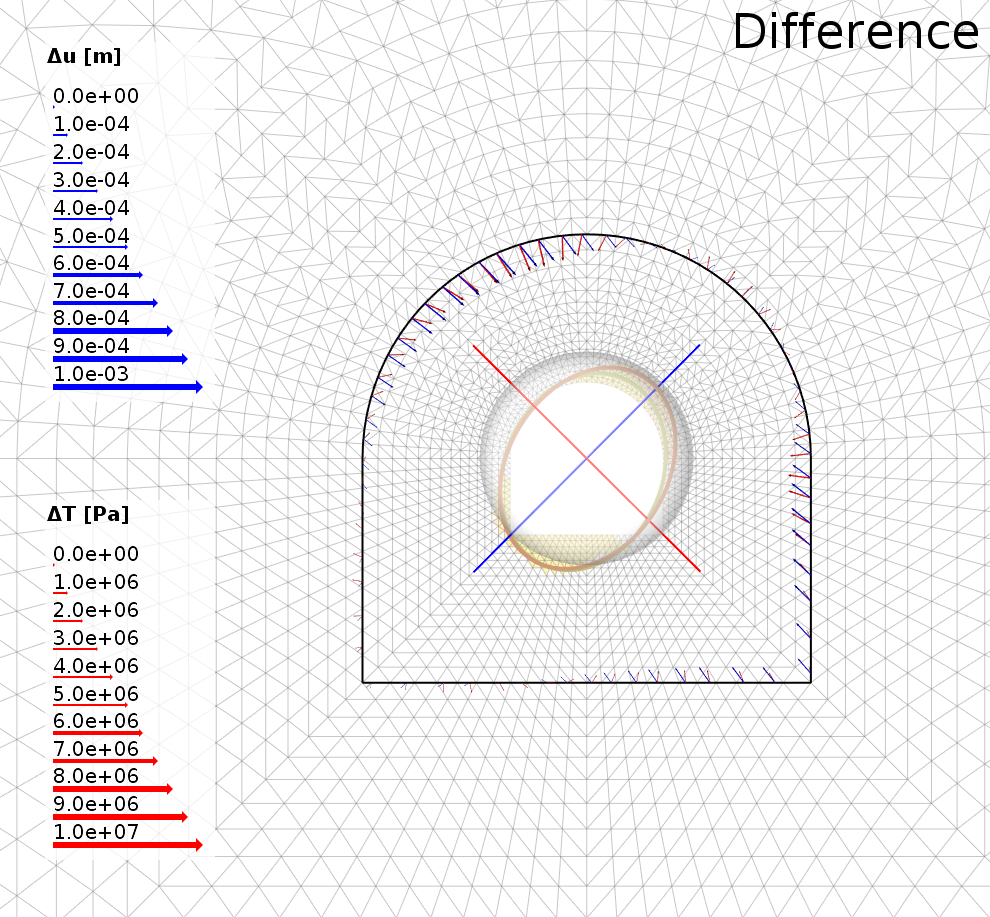}\hspace{0.02\textwidth}\includegraphics[width=0.45\textwidth]{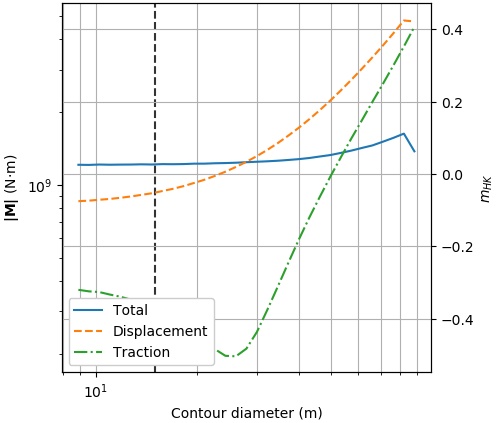}\caption{Summaries for Case 5 (eastward plunging loading, stress increase,
weaker material, failure in the eastern corner) of Kirchhoff \textit{(top
left)}, adjusted conventional \textit{(top right)}, and elliptical
\textit{(bottom left)} moment tensor calculations. Demonstration of
contour invariance for Kirchhoff-type calculation \textit{(bottom
right)}. See text for more details.}
\end{figure}

\pagebreak{}

\section{Transformations of Equation (\ref{eq:kirchhoff}) for an expanding
elliptical cavity\label{sec:Appendix-2:-Transformations}}

\begin{singlespace}
When applied to the traction-free surface of the expanded elliptical
cavity $S_{2}$, Equation (\ref{eq:kirchhoff}) has the form
\begin{equation}
M_{ij}=-\iintop_{S_{2}}\left\{ T_{i}^{(\mathrm{before})}(\mathbf{\mathbf{x}})(x_{j}-x_{j}^{(0)})+c_{ijkl}(\mathbf{x})\left[u_{k}^{\mathrm{(after)}}(\mathbf{x})-u_{k}^{\mathrm{(before)}}(\mathbf{x})\right]n_{l}(\mathbf{x})\right\} dS(\mathbf{x}).\label{eq:Definition=0000233b}
\end{equation}
A simple analytical expression is known for the displacement $\mathbf{u}^{\mathrm{(after})}$
on $S_{2}$ \cite{Maugis-1992}, and integration of that component
is relatively straightforward. Expressions for $\mathbf{T}^{\mathrm{(before})}$
and $\mathbf{u}^{\mathrm{(before})}$ on $S_{2}$ are also available,
but their increased complexity makes their integration less straightforward.
In this section, we demonstrate a procedure for transforming Equation
\ref{eq:Definition=0000233b} into a form that makes this integration
possible. In the derivation of this transformed expression, we will
deal with the deformation and stress fields of the pre-expansion state
only, so to simplify notation, we will omit ``before'' superscripts
(that is, $\boldsymbol{\sigma}\equiv\boldsymbol{\sigma}^{\mathrm{(before)}}$,
$\mathbf{u}\equiv\mathbf{u}^{\mathrm{(before)}}$, etc.).

\begin{figure}[H]
\begin{centering}
\includegraphics[width=0.8\textwidth]{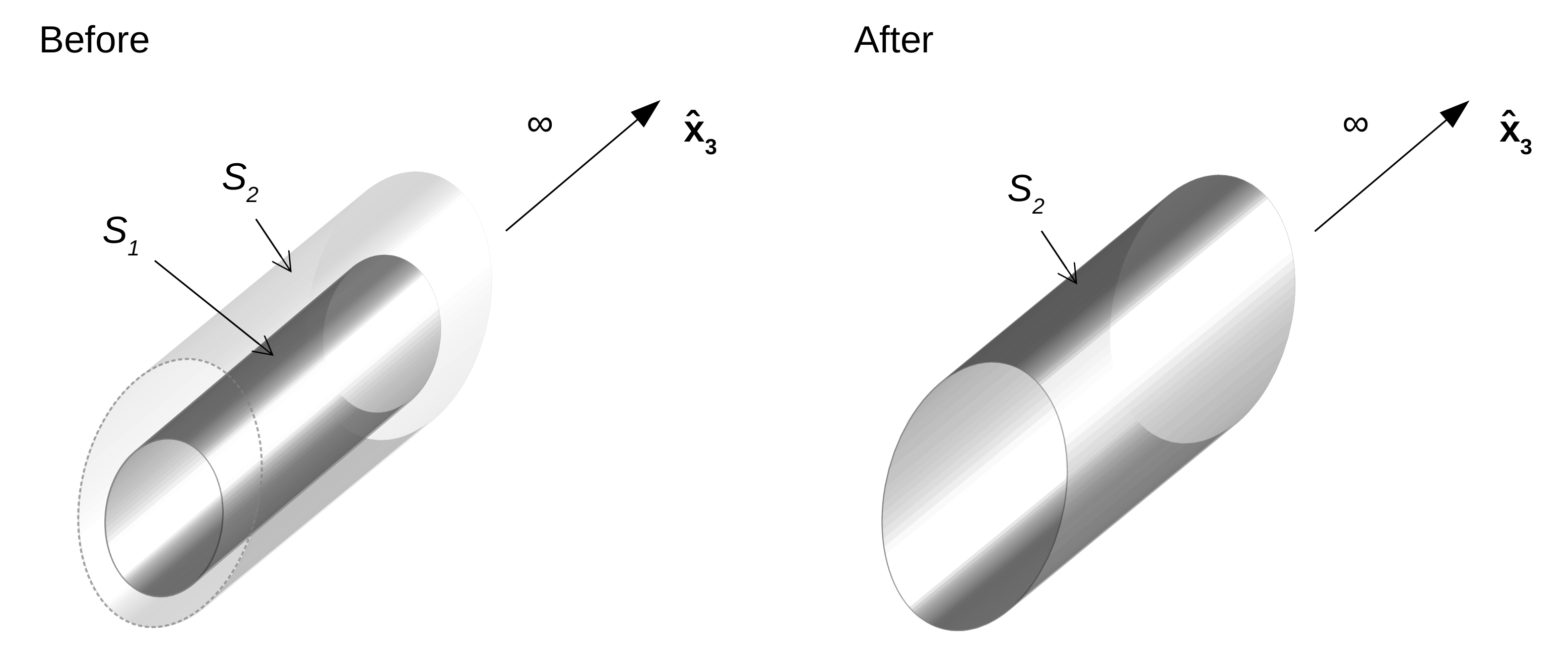}
\par\end{centering}
\caption{\label{fig:Elliptical-cylinders}Before \textit{(left)} and after
\textit{(right)} the expansion of an infinite extent elliptical cavity
oriented in the direction $\hat{\mathbf{x}}_{3}$. Only a limited
section is shown and used in the source mechanism calculations. The
pre and post-expansion surfaces $S_{1}$ and $S_{2}$ are shown in
dark gray in their respective plots ($S_{2}$ is also superimposed
on the pre-expansion $S_{1}$ to indicate their relative position).
Plane strain is assumed in both cases.}
\end{figure}

\end{singlespace}

Taking $\hat{\mathbf{x}}_{3}$ to be oriented along the cavity's axis
as shown in Figure \ref{fig:Elliptical-cylinders}, the plane strain
condition dictates that displacement in this direction is zero; that
is, $u_{3}(\mathbf{x})=0$. It also implies that $\sigma_{13}(\mathbf{x})=\sigma_{23}(\mathbf{x})=\sigma_{31}(\mathbf{x})=\sigma_{32}(\mathbf{x})=0$
and $\sigma_{33}(\mathbf{x})=\nu[\sigma_{11}(\mathbf{x})+\sigma_{22}(\mathbf{x})]$,
where $\nu$ is Poisson's ratio. Consider the toroidal volume $V$
confined by the surfaces $S_{1}$ and $S_{2}$ radially, and by the
surfaces $S_{3+}$ and $S_{3-}$ in the direction of $\hat{\mathbf{x}}_{3}$.
At equilibrium, 
\[
\frac{\partial\sigma_{im}(\mathbf{x})}{\partial x_{m}}=0
\]
for all $\mathbf{x}\in V$. Multiplying this by the $j$th component
of $\mathbf{x}$ relative to an arbitrary location $\mathbf{x}^{(0)}$
gives 
\[
\frac{\partial\sigma_{im}(\mathbf{x})}{\partial x_{m}}(x_{j}-x_{j}^{(0)})=0.
\]
Following integration over $V$, applying the product rule, and the
divergence theorem, this becomes
\[
\iintop_{\partial V}T_{i}^{V}(\mathbf{x})(x_{j}-x_{j}^{(0)})dS(\mathbf{x})-\iiintop_{V}\sigma_{ij}(\mathbf{x})dV(\mathbf{x})=0,
\]
where stress is presented in terms of the the traction $T_{i}^{V}(\mathbf{x})=\sigma_{im}(\mathbf{x})n_{m}^{V}(\mathbf{x})$,
and the surface $\partial V$ is composed of $S_{1}$, $S_{2}$, $S_{3+}$,
and $S_{3-}$. Note that this normal vector $\mathbf{n}^{V}$ is oriented
outwards from the volume $V$, meaning that it is not the same as
$\mathbf{n}$ in Equation (\ref{eq:Definition=0000233b}): $\mathbf{n}^{V}(\mathbf{x})=\mathbf{n}(\mathbf{x})$
for $\mathbf{x}\in S_{2}$ but $\mathbf{n}^{V}(\mathbf{x})=-\mathbf{n}(\mathbf{x})$
for $\mathbf{x}\in S_{1}$. Substituting $\sigma_{ij}(\mathbf{x})=c_{ijkl}\partial u_{k}(\mathbf{x})/\partial x_{l}$
and applying the divergence theorem to the second integral yields\footnote{For example, $\iiintop_{V}\partial u_{1}(\mathbf{x})/\partial x_{2}dV(\mathbf{x})=\iintop_{\partial V}u_{1}(\mathbf{x})n_{2}(\mathbf{x})dS(\mathbf{x})$
is obtained by applying the divergence theorem to the vector field
$(0,u_{1}(\mathbf{x}),0)$. This equation also justifies the invariance
of Kirchhoff-type definition for moment tensor of Equation (\ref{eq:kirchhoff})
in regards to selection of the integration surface $S$ \textcolor{black}{(taking
the effect of side surfaces $S_{3+}$ and $S_{3-}$ into account).}}
\[
\iintop_{\partial V}T_{i}^{V}(\mathbf{x})(x_{j}-x_{j}^{(0)})dS(\mathbf{x})-\iintop_{\partial V}c_{ijkl}u_{k}(\mathbf{x})n_{l}^{V}(\mathbf{x})dS(\mathbf{x})=0.
\]
For $i=1$ or $2$, we have

\[
\iintop_{S_{2}}T_{i}^{V}(\mathbf{x})(x_{j}-x_{j}^{(0)})dS(\mathbf{x})-\iintop_{S_{2}}c_{ijkl}u_{k}(\mathbf{x})n_{l}^{V}(\mathbf{x})dS(\mathbf{x})=\iintop_{S_{1}}c_{ijkl}u_{k}(\mathbf{x})n_{l}^{V}(\mathbf{x})dS(\mathbf{x}),
\]
and for $i=3$, we have

\[
\iintop_{S_{3+}\cup S_{3-}}T_{3}^{V}(\mathbf{x})(x_{3}-x_{3}^{(0)})dS(\mathbf{x})=\iintop_{S_{1}\cup S_{2}}\lambda[u_{1}(\mathbf{x})n_{1}^{V}(\mathbf{x})+u_{2}(\mathbf{x})n_{2}^{V}(\mathbf{x})]dS(\mathbf{x}).
\]
In terms of notations used in Equation (\ref{eq:Definition=0000233b}),
these equations become

\[
-\iintop_{S_{2}}T_{i}^{\mathrm{(before)}}(\mathbf{x})(x_{j}-x_{j}^{(0)})dS(\mathbf{x})+\iintop_{S_{2}}c_{ijkl}u_{k}^{\mathrm{(before)}}(\mathbf{x})n_{l}(\mathbf{x})dS(\mathbf{x})=\iintop_{S_{1}}c_{ijkl}u_{k}^{\mathrm{(before)}}(\mathbf{x})n_{l}(\mathbf{x})dS(\mathbf{x})
\]
and

\[
\begin{array}{c}
\iintop_{S_{3+}\cup S_{3-}}T_{3}^{\mathrm{(before)}}(\mathbf{x})(x_{3}-x_{3}^{(0)})dS(\mathbf{x})+\iintop_{S_{1}}\lambda[u_{1}^{\mathrm{(before)}}(\mathbf{x})n_{1}(\mathbf{x})+u_{2}^{\mathrm{(before)}}(\mathbf{x})n_{2}(\mathbf{x})]dS(\mathbf{x})=\\
\iintop_{S_{2}}\lambda[u_{1}^{\mathrm{(before)}}(\mathbf{x})n_{1}(\mathbf{x})+u_{2}^{\mathrm{(before)}}(\mathbf{x})n_{2}(\mathbf{x})]dS(\mathbf{x}),
\end{array}
\]
respectively. These equations allow Equation (\ref{eq:Definition=0000233b})
to be re-expressed as

\begin{equation}
\begin{cases}
M_{ij}=\iintop_{S_{1}}c_{ijkl}(\mathbf{x})u_{k}^{\mathrm{(before)}}(\mathbf{x})n_{l}(\mathbf{x})dS(\mathbf{x})-\iintop_{S_{2}}c_{ijkl}(\mathbf{x})u_{k}^{\mathrm{(after)}}(\mathbf{x})n_{l}(\mathbf{x})dS(\mathbf{x}) & \mathrm{if\,}i,j=1\mathrm{\,or\,2},\\
\begin{aligned}M_{33}= & \iintop_{S_{1}}\lambda[u_{1}^{\mathrm{(before)}}(\mathbf{x})n_{1}(\mathbf{x})+u_{2}^{\mathrm{(before)}}(\mathbf{x})n_{2}(\mathbf{x})]dS(\mathbf{x})+\\
 & \iintop_{S_{3+}\cup S_{3-}}T_{3}^{\mathrm{(before)}}(\mathbf{x})(x_{3}-x_{3}^{(0)})dS(\mathbf{x})-\\
 & \iintop_{S_{2}}\lambda[u_{1}^{\mathrm{(after)}}(\mathbf{x})n_{1}(\mathbf{x})+u_{2}^{\mathrm{(after)}}(\mathbf{x})n_{2}(\mathbf{x})]dS(\mathbf{x})
\end{aligned}
 & ,\\
M_{ij}=0 & \mathrm{otherwise}.
\end{cases}\label{eq:Definition=0000233c}
\end{equation}

The next step is to express Equation (\ref{eq:Definition=0000233c})
in terms of the characteristics of loading and the geometric parameters
of the ellipses, which will require two approximations. The first
of these is to assume that the minor semi-axes of both ellipses are
aligned with the direction of maximum in-plane compressive stress
(marked as ``Load'' in Figure \ref{fig:Cavity-expansion-and-parametrisation}).
This is a reasonable assumption as stress concentration and damage
tend to occur in this direction as illustrated by the modelling of
Section \ref{sec:Modelling}. The second approximation is to neglect
the traction term in the expression for $M_{33}$, allowing the source
mechanism to be defined entirely by displacements on the original
and final cavities. Denoting the major and minor semi-axes of the
original ellipse as $a^{\mathrm{(before)}}$ and $b^{\mathrm{(before)}}$,
respectively, and those of the final ellipse as and $a^{\mathrm{(after)}}$
and $b^{\mathrm{(after)}}$, respectively, the components of the moment
tensor in a coordinate system with $\hat{\mathbf{x}}_{1}$ and $\hat{\mathbf{x}}_{2}$
oriented along the major and minor semi-axes are

\begin{equation}
M_{ij}=\begin{cases}
K\left[\frac{\nu}{1-2\nu}+\frac{1-\nu}{1-2\nu}\frac{1}{k_{\sigma}}\frac{b}{a}\frac{\triangle b}{\triangle a}-\frac{k_{\sigma}-1}{4k_{\sigma}}(\frac{b}{a}+\frac{\triangle b}{\triangle a})+\frac{\nu}{1-2\nu}\frac{1}{2}\frac{\triangle a}{a}-\frac{k_{\sigma}-1}{4k_{\sigma}}\frac{\triangle b}{a}+\frac{1-\nu}{1-2\nu}\frac{1}{2k_{\sigma}}\frac{\triangle b}{a}\frac{\triangle b}{\triangle a}\right] & \mathrm{if\,}i,j=1,\\
K\left[\frac{1-\nu}{1-2\nu}+\frac{\nu}{1-2\nu}\frac{1}{k_{\sigma}}\frac{b}{a}\frac{\triangle b}{\triangle a}+\frac{k_{\sigma}-1}{4k_{\sigma}}(\frac{b}{a}+\frac{\triangle b}{\triangle a})+\frac{1-\nu}{1-2\nu}\frac{1}{2}\frac{\triangle a}{a}+\frac{k_{\sigma}-1}{4k_{\sigma}}\frac{\triangle b}{a}+\frac{\nu}{1-2\nu}\frac{1}{2k_{\sigma}}\frac{\triangle b}{a}\frac{\triangle b}{\triangle a}\right] & \mathrm{if\,}i,j=2,\\
K\frac{\nu}{1-2\nu}\left[1+\frac{1}{k_{\sigma}}\frac{b}{a}\frac{\triangle b}{\triangle a}+\frac{1}{2}\frac{\triangle a}{a}+\frac{1}{2k_{\sigma}}\frac{\triangle b}{a}\frac{\triangle b}{\triangle a}\right] & \mathrm{if\,}i,j=3,\\
0 & \mathrm{otherwise},
\end{cases}\label{eq:Definition=0000233c-1-1}
\end{equation}

\noindent where $K=4\pi k_{\sigma}\sigma(1-\nu)a\triangle aL$, $\sigma$
is the loading along the major semi-axis, $k_{\sigma}\sigma$ is the
loading along the minor semi-axis, $\nu$ is the rockmass' Poisson's
ratio, $L$ is the tunnel's extent in $\hat{\mathbf{x}}_{3}$, $\Delta a\equiv a^{\mathrm{(after)}}-a^{\mathrm{(before)}}$,
and $\Delta b\equiv b^{\mathrm{(after)}}-b^{\mathrm{(before)}}$.
\end{document}